\documentclass[a4paper,11pt]{book}
\usepackage[numbers,sort&compress]{natbib}
\usepackage[utf8]{inputenc}
\usepackage[printonlyused]{acronym}
\usepackage{hyperref}
\hypersetup{
    colorlinks=true,
    linkcolor=blue,
    filecolor=green,      
    urlcolor=magenta,
    citecolor=red,
}
\usepackage[symbol]{footmisc}
\usepackage{amsmath,amsfonts,amssymb}
\usepackage{graphicx}
\usepackage{xfrac}   
\usepackage{comment}
\usepackage{orcidlink}
\usepackage{authblk}

\usepackage{url}

\usepackage[margin=1in,footskip=0.25in]{geometry}

\usepackage{soul}

\usepackage{orcidlink}

\usepackage{color}

\date{\today}
\title{A White Paper on \\ The Multi-Messenger Science Landscape in India\footnote{Chief Editors: Suvodip Mukherjee and Amol Dighe}}

\author[1]{Samsuzzaman Afroz\orcidlink{0009-0004-4459-2981}}
\author[2,3]{Sanjib Kumar Agarwalla\orcidlink{0000-0002-9714-8866}}
\author[4]{Dipankar Bhattacharya\orcidlink{0000-0003-3352-3142}}
\author[5]{Soumya Bhattacharya\orcidlink{0000-0003-2540-7504}}
\author[6]{Subir Bhattacharyya}
\author[7]{Varun Bhalerao\orcidlink{0000-0002-6112-7609}}
\author[8]{Debanjan Bose}
\author[6]{Chinmay Borwanker}
\author[9]{Ishwara Chandra C. H. \orcidlink{0000-0001-5356-1221}}
\author[1]{Aniruddha Chakraborty \orcidlink{0009-0004-4937-4633}}
\author[7]{Indranil Chakraborty\orcidlink{0000-0002-9812-2166}}
\author[10]{Sovan Chakraborty\orcidlink{0000-0002-1458-8517}}
\author[11]{Debarati Chatterjee \orcidlink{0000‐0002‐0995‐2329}}
\author[1]{Varsha Chitnis\orcidlink{0000-0001-5046-7504}}
\author[12]{Moon Moon Devi \orcidlink{0000-0001-6569-0792}}
\author[11]{Sanjeev Dhurandhar}
\author[1]{Amol Dighe\orcidlink{0000-0001-6639-0951}}
\author[6]{Bitan Ghosal\orcidlink{0000-0002-7963-0907}}
\author[1]{Sourendu Gupta}
\author[13]{Arpan Hait\orcidlink{0000-0001-7540-0111}}
\author[14]{Md Emanuel Hoque}
\author[14]{Pratik Majumdar\orcidlink{0000-0002-5481-5040}}
\author[1]{Nilmani Mathur}
\author[1]{Harsh Mehta \orcidlink{0009-0007-4664-4820}}
\author[13]{Subhendra Mohanty\orcidlink{0000-0003-0070-6647}}
\author[15]{Reetanjali Moharana \orcidlink{0000-0001-5642-8311}}
\author[14]{Arunava Mukherjee\orcidlink{0000-0003-1274-5846}}
\author[1]{Suvodip Mukherjee \footnote{White Paper Coordinator's email: \href{mailto:suvodip@tifr.res.in}{suvodip@tifr.res.in}}\orcidlink{0000-0002-3373-5236}}
\author[11]{Dhruv Pathak\orcidlink{0000-0001-8129-0473}}
\author[9]{Tirthankar Roy Choudhury\orcidlink{0000-0001-7462-8587}}
\author[1]{Mohit Raj Sah\orcidlink{0009-0005-9881-1788}}
\author[10]{Prantik Sarmah\orcidlink{0000-0003-3159-7148}}
\author[6]{Krishna Kumar Singh\orcidlink{0000-0002-5818-8195}}
\author[1]{Rishi Sharma\orcidlink{0000-0001-8774-3138}}
\author[11]{Swarnim Shirke\orcidlink{0000-0001-8604-5362}}
\author[1, 16]{Shriharsh P. Tendulkar\orcidlink{0000-0003-2548-2926}}
\author[7]{Gaurav Waratkar\orcidlink{0000-0003-3630-9440}}
\author[6]{Kuldeep Yadav}
\affil[1]{Tata Institute of Fundamental Research, 1, Homi Bhabha Road, Mumbai 400005, India}
\affil[2]{Institute of Physics, Sachivalaya Marg, Sainik School Post, Bhubaneswar 751005, India}
\affil[3]{Homi Bhabha National Institute, Training School Complex, Anushakti Nagar, Mumbai 400094, India}
\affil[4]{Department of Physics, Ashoka University, Rai, Sonipat, Haryana 131029, India}
\affil[5]{S. N. Bose National Centre for Basic Sciences, Kolkata 700106, India}
\affil[6]{Astrophysical Sciences Division, Bhabha Atomic Research Centre, Mumbai 400085, India}
\affil[7]{Department of Physics,  Indian Institute of Technology Bombay, Mumbai 400076, India}
\affil[8]{Department of Physics, Central University of Kashmir, Ganderbal 191131, India}
\affil[9]{National Centre for Radio Astrophysics, TIFR,
Post Bag 3, Pune University Campus, Ganeshkhind, Pune 411007, India}
\affil[10]{Indian Institute of Technology Guwahati, Assam 781039, India}
\affil[11]{Inter-University Centre for Astronomy and Astrophysics, Post Bag 4, S P Pune University, Ganeshkhind, Pune 411007, India}
\affil[12]{Department of Physics, Tezpur University, Napam, Assam 784028, India}
\affil[13]{Department of Physics, Indian Institute of Technology Kanpur, Kanpur 208016, India}
\affil[14]{Saha Institute of Nuclear Physics, 1/AF Bidhannagar, Kolkata 700064, India}
\affil[15]{Department of Physics, IIT Jodhpur, Karwar-342037, India}
\affil[16]{CIFAR Azrieli Global Scholars Program, MaRS Centre, Toronto, Ontario, Canada}
\begin{document}

\maketitle

\newpage
  \vspace{1cm}
   \,
\newpage

\begin{figure}[h]
    \centering
    \includegraphics[width=0.76\textwidth]{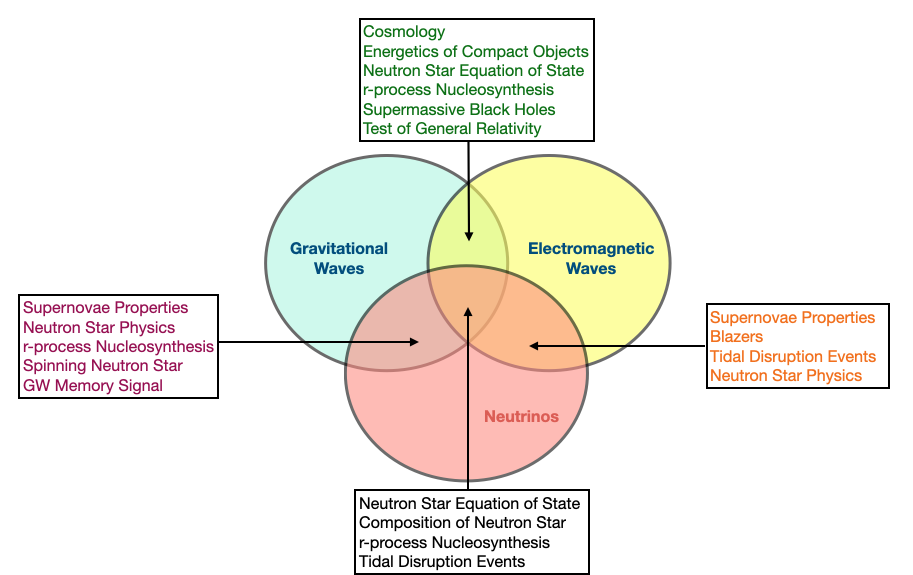}
    \caption{ The key science issues that may be addressed by multi-messenger observations explored in this white paper.}
    \label{fig:venn}
\end{figure}

\begin{center}
\textbf{\Large Executive summary}
\end{center}
The multi-messenger science using different observational windows to the Universe such as Gravitational Waves (GWs), Electromagnetic Waves (EMs), Cosmic Rays (CRs) 
and Neutrinos bring an opportunity to study the Universe from the scale of a neutron star to cosmological scales over a large cosmic time. At the smallest scales, we can study the structure of the neutron star and the different energetics involved in the transition of a pre-merger neutron star to a post-merger neutron star. 
{This will open up a window to study the properties of matter in extreme conditions and a guaranteed discovery space \cite{Lasky:2015uia,Ascenzi:2024wws}.} 
On the other hand, at the largest cosmological scales, multi-messenger observations will make it possible to study the long-standing problems in physical cosmology related to the Hubble constant \cite{Abdalla:2022yfr}, dark matter \cite{RevModPhys.90.045002}, and dark energy \cite{doi:10.1126/science.aaa0980, PhysRevD.108.103519}. 
The multi-messenger observations will be able to map the expansion history of the Universe using GW sources, which are the standard sirens up to high redshift. This will make it possible to determine the value of the Hubble constant and the redshift evolution of the dark energy equation of state and can shed light on the nature of dark matter \cite{Berti:2022wzk}.
From the point of view of astronomy, multi-messenger observations of astrophysical systems will bring deep insights into the physical 
processes associated with compact objects such as white dwarfs, neutron stars, and black holes of all masses, all the way up to a high redshift Universe, that are inaccessible otherwise.
Such multi-messenger observations also open up the opportunity for the discovery of new kinds of compact objects that are currently unknown \cite{Meszaros:2019xej}.
In Fig.~\ref{fig:venn}, we depict the intersection between different areas and science goals that can be addressed by the multi-messenger observations explored in this white paper.    

These vast ranges of discovery space from multi-messenger observations are possible from both isolated or binary compact objects, with the signal being transient or persistent in nature.  These signals can be from a supernova explosion, rotating asymmetric compact objects, stellar-origin or supermassive binary black holes with baryonic matter around them, or exotic compact objects.
The energy spectrum of the transient or persistent signal in multi-messenger observables carries a wealth of information on the source properties which can be used for scientific studies of compact objects as well as 
 for the
inference of cosmology from these compact objects. As a result, it becomes extremely important to measure the multi-messenger signals in both the time domain and the frequency domain to capture the energy spectrum and its variation along with the arrival time and sky position of the signal. 

This kind of scientific study requires different telescopes operating in multiple energy/frequency bands and coordination between different messengers to explore the synergy between them and make it possible to achieve the scientific goals discussed above. Among different messengers, in the field of GWs, the currently ongoing network of GW detectors spans two windows: hecto-hertz and nHz frequency ranges. In the hecto-hertz frequency range, there are currently four major observatories, namely, the two LIGO detectors \cite{KAGRA:2013rdx, LIGOScientific:2014pky}, Virgo \cite{VIRGO:2014yos}, GEO600 \cite{Luck:2010rt}, and KAGRA \cite{PhysRevD.88.043007}, and in the near-future LIGO-Aundha \cite{Unnikrishnan:2013qwa,saleem2021science} is expected to be operational by the end of this decade in this band. In the nHz range, pulsar timing arrays using a network of radio antennae across the globe which includes NanoGrav \cite{NANOGrav:2023gor}, CPTA \cite{Xu:2023wog}, EPTA \cite{EPTA:2023sfo}, InPTA \cite{joshi2018precision}, and PPTA \cite{Reardon:2023gzh} 
that are currently
operational and MeerKat \cite{Miles:2022lkg} as
well as SKA \cite{Janssen:2014dka, ChandraJoshi:2022etw} that are expected to be operational by the end of this decade.
By the end of the next decade, space-based GW detector LISA will be operational and will explore the milli-hertz frequency range of GW signal \cite{2017arXiv170200786A}. In all these bands, transient and persistent sources, which can have signatures from other messengers, 
are expected. 

Multi-messenger observations from  EM and CRs signal are feasible from the gamma-ray band to the radio band with the help of several detectors which include IACTs around the world --  H.E.S.S. \cite{Ohm:2023cgv}, MAGIC \cite{Bigongiari:2005sw},  VERITAS \cite{Acharyya:2023llt}, MACE \cite{Yadav2022}, Fermi \cite{2009ApJ...697.1071A}, The Neil Gehrels Swift Observatory \cite{2004ApJ...611.1005G}, NuStar \cite{NuSTAR:2013yza} AstroSat \cite{2014SPIE_ASTROSATmission}, HST \cite{2017ApJ...848L..17C}, DECam \cite{2015AJ....150..150F}, Las Cumbres Observatory Global Telescope Network \cite{2013PASP..125.1031B}, The South African Large Telescope (SALT) \cite{2006IAUSS...1E...8B,1994assa.symp..111S}, VLA \cite{2011ApJ...739L...1P}, uGMRT \cite{ugmrt1}. In the near future, Rubin-LSST \cite{ivezic2019lsst} 
and Roman Space Telescope \cite{2019arXiv190205569A} will be operational, which will be able to make a paradigm shift in the detection of optical and infrared signals with cadence and sensitivity that are not currently feasible.
 The other important cosmic messengers, neutrinos, are currently detected mainly at IceCube \cite{aartsen2014icecube} and KM3Net \cite{KM3Net:2016zxf}. The upgrades of these detectors \cite{IceCube-Gen2:2020qha,KM3NeT:2024paj} 
will increase their sensitivity and allow them to probe deeper in the Universe.

India is playing a crucial role in multi-messenger science, both through the telescopes built and operated by Indian research institutes and through collaborations with international partners. This has also been identified as a key science area in the Mega Science Vision-2035 Astronomy and Astrophysics document \cite{MSVA-2035} and 
 the Vision Document of the
Astronomical Society of India \cite{ASI-2024}. On the GW side, with the help of observatories such as uGMRT\cite{ugmrt1} and LIGO-Aundha \cite{Unnikrishnan:2013qwa,saleem2021science}, observations in 
{the nHz and hecto-hertz frequency range, respectively, will be possible over the coming years.}
Similarly, 
Indian institutes operate EM detectors such as 
MACE \cite{Yadav2022}, AstroSAT \cite{2014SPIE_ASTROSATmission}, GROWTH-India telescope \cite{Kumar:2022svq}, and uGMRT \cite{ugmrt1} which are useful for the followup of GW signals. Coordinated observations of the GW events for both super-threshold and sub-threshold events with different observatories in the EM band will be required for the multi-messenger studies discussed above. Along with prompt observations, reduction and analysis of data from different observatories will be crucial 
which will require large computational facilities.  Dedicated computing resources and storage facilities, which can perform fast computations to extract the source properties rapidly and trigger the other detectors, will be necessary for this purpose.
In conjunction with a coordinated operation of the existing facilities, planning of the next-generation observatories in EM, GW, and neutrino area will be needed to broaden our horizon of the discovery space. 

A coordinated activity 
for multi-messenger science studies in India will play a vital role in the future of discovering uncharted territories in astrophysics, cosmology, and fundamental physics. 
A dedicated multi-messenger coordination center can facilitate the development of human resources spanning several areas from instrumentation, astronomical observations, data analysis, and theoretical modeling. 
Such a dedicated facility to support multi-messenger science can also play a pivotal role in mentoring undergraduate students, graduate students, and postdocs in this field. 
The development of such a facility, both for coordinating multi-messenger observations from different telescopes around the globe and for the development of instrumentation and computing frontiers in the country, 
will have a strong impact on global science.


\newpage

\tableofcontents
\newpage
\,
\newpage
\thispagestyle{plain}
\begin{center}
    \textbf{\Large Glossaries}
\end{center} 
\begin{acronym}
  \acro{AGN}{Active Galactic Nuclei}
  \acro{BAO}{Baryon Acoustic Oscillation}
  \acro{BH}{Black Hole}
\acro{BBH}{Binary Black Hole}
 \acro{BNS}{Binary Neutron Star}
 \acro{CBM}{Compressed Baryonic Matter}
    \acro{CC-SNe}{Core Collapse Supernovae}
    \acro{CE}{Cosmic Explorer}
    \acro{CGW}{Continuous Gravitational Waves}
    \acro{CMB}{Cosmic Microwave Background}
    \acro{CPTA}{The Chinese PTA}
    \acro{CRA}{Cosmic Ray Anisotropy}
 \acro{CRs}{Cosmic Rays}
    \acro{CSM}{Circum-Stellar Material}
     \acro{CTA}{Cherenkov Telescope Array}
     \acro{CZTI}{Cadmium Zinc Telluride Imager}
     \acro{DESI}{Dark Energy Spectroscopic Instrument}
     \acro{EGMF}{Extra-Galactic Magnetic Field}
  \acro{EM}{Electro-Magnetic}
  \acro{EoS}{Equation of State}
   \newpage
     \thispagestyle{plain}
  \acro{EPTA}{The European PTA} 
   \acro{ERB}{Extra-galactic Radio Background}
 \acro{ET}{Einstein Telescope}
 \acro{FAIR}{Facility for Antiproton and Ion Research}
 \acro{GCN}{General Coordinates Network}
   \acro{GMRT}{Giant Metrewave Radio Telescope}
   \acro{GR}{General Theory of Relativity}
   \acro{GRACEDB}{A Gravitational Wave Candidate Event Database}
    \acro{GRBs}{Gamma-Ray Bursts}
   \acro{GROWTH}{Global Relay of Observatories Watching
Transients Happen}
  \acro{GW}{Gravitational Wave}
   \acro{GWTC-3}{The third Gravitational-Wave Transient Catalog}
\acro{GZK}{Greisen–Zatsepin–Kuzmin}
\acro{HK}{Hyper-Kamiokande}
\acro{IACTs}{Imaging Atmospheric Cherenkov Telescope Arrays }
\acro{InPTA}{The Indian PTA}
   \acro{KAGRA}{Kamioka Gravitational Wave Detector}
   \acro{LAXPC}{Large Area X-ray Proportional Counters}
   \acro{LIGO}{Laser Interferometer Gravitational-Wave Observatory}
 \acro{LISA}{Laser Interferometer Space Antenna}
           \newpage
     \thispagestyle{plain}
 \acro{LMXB}{Low-Mass X-ray Binary}
 \acro{LSST}{Large Synoptic Survey Telescope}
 \acro{LVK}{LIGO-Virgo-KAGRA}
  \acro{MACE}{Major Atmospheric Cerenkov Experiment Telescope} 
     \acro{nHz}{nano hertz}
  \acro{NANOGrav}{North American Nanohertz Observatory for Gravitational Waves}
 \acro{NSBH}{Neutron Star-Black Hole}
 \acro{ORT}{Ooty Radio Telescope}
 \acro{PAO}{Pierre Auger Observatory}
 \acro{PPTA}{The Parkes PTA}
\acro{PTAs}{Pulsar Timing Arrays}
\acro{RIBs}{Radioactive-Ion Beams}
\acro{SBO}{Shock Breakout}
 \acro{SGWB}{Stochastic Gravitational Wave Background}
 \acro{SK}{Super-Kamiokande}
  \acro{SKA}{Square Kilometer Array}
\acro{SMBHs}{Supermassive Black Holes}
\acro{SMBHBs}{Supermassive Black Hole Binaries}
  \acro{SN}{Supernova}
\acro{SXT}{Soft X-ray Telescope}
 \newpage
     \thispagestyle{plain}
\acro{TA}{Telescope Array}
\acro{TDEs}{Tidal Disruption Events}
\acro{ToO}{Target-of-Opportunity} 
\acro{TOV}{Tolman-Oppenheimer-Volkoff}
\acro{uGMRT}{upgraded Giant Metrewave Radio Telescope}
\acro{UHE}{Ultra-High Energy}
 \acro{UHECRs}{Ultra-High Energy Cosmic Rays}
 \acro{UVIT}{Ultraviolet Imaging Telescope}
\acro{WIMPs}{Weakly Interacting Massive Particles}
\acro{ZTF}{Zwicky Transient Facility}
\end{acronym}

\newpage
\thispagestyle{plain}
\newpage
\chapter{Science from Multi-Messenger Observations}
\section{Astrophysics}
\label{astrophysics}

\small {\it Contributors: Chinmay Borwanker, Aniruddha Chakraborty, Sovan Chakraborty, Suvodip Mukherjee, Prantik Sarmah}\\ \normalsize
\small {\it Editors: Varsha Chitnis, Sanjeev Dhurandhar, Amol Dighe, Suvodip Mukherjee, Tirthankar Roy Choudhury, Shriharsh P. Tendulkar} \\ \normalsize

\noindent \textbf{\Large{Short-term science goals:}}

\subsection{Multi-Messenger Astrophysics with Binary Gravitational Wave Sources}
\subsubsection{Introduction}
The discovery of binary compact objects such as \ac{BNS},  \ac{NSBH}, and \ac{BBH} mergers have provided direct probes for the study of the astrophysics of formation of these compact objects in the Universe, and their evolution with cosmic time \cite{GW150914, KAGRA:2021duu}. A multi-messenger observation of these sources would bring complementary information about the system which can give insight into a plethora of scientific aspects such as the constituents of these objects, the host properties of these sources, different physical processes that are associated with these sources, and also the production mechanism of neutron-rich heavy elements in the Universe \cite{GW170817, GW170817_MMA, Siegel:2019mlp}. 

A broad classification of these binary sources can be based on whether one of the sources is a neutron star or not (both are black holes). For sources with at least one neutron star, the signatures of multi-messenger observations are expected from the binaries during different stages, primarily observed through \ac{GW} during the pre-merger phase of BNS and NSBHs, and later in the post-merger phase using different messengers such as GW, \ac{EM}, neutrino, and \ac{CRs}. One such successful observation of the event GW170817 in GW by the \ac{LIGO} and Virgo Collaboration and in multiple EM bands by multiple telescopes has provided the multi-messenger observation from a BNS system \cite{GW170817, GW170817_MMA}. It has brought insights into a vast range of physics questions ranging from astrophysics, cosmology, and fundamental physics. A detailed discussion on cosmology and fundamental physics aspects is presented in chapters \ref{cosmology} and \ref{fundamental-physics}, respectively. Such observations from NSBH events are yet to be made from observations, though theoretical signatures of multi-messenger observations are feasible. 

For BBH systems, the multi-messenger observations covering GW, EM, neutrino, and CRs are possible astrophysically only if there are baryons present \cite{Chen:2023xrm, Bogdanovic:2021aav, Baker:2019nct}. The baryonic presence near BBHs is possible depending on the environment in which they are merging. If the BBHs are merging in the presence of an accretion disc, then a multi-messenger signature is feasible. However, such scenarios are usually possible only for heavier black holes and supermassive black holes \cite{Baker:2019nct}. Stellar-origin black holes without any baryonic matter around, are unlikely to have a multi-messenger signature. The multi-messenger signature from supermassive BBHs can be observable in GW, EM, neutrino, and CRs. Depending on the mechanism of energetics involved due to the dynamics of the baryonic disc with the BBHs, the strength and frequency/energy band of the multi-messenger signal will vary. 

Some candidate multi-messenger observations of astrophysical systems that include black holes are TXS 0506+056 \cite{2018ApJ...863L..10A}, which has a claim to a joint detection of neutrino and gamma-ray, and GW190521 \cite{LIGOScientific:2020iuh} which has a claim on the observation of the EM counterpart from an \ac{AGN} following a GW emission from the merger of BBHs of masses around 80 $M_\odot$ and 60 $M_\odot$ \cite{PhysRevLett.124.251102}.  TXS 0506+056 is a blazer that is fueled by a supermassive black hole situated in a galaxy at a redshift $z=0.33$. A common emission mechanism involving pions is expected to be the source of the signals in multi-messenger observables. For GW190521, one of the popular mechanisms for the multi-messenger signature is the merging BBHs embedded in an AGN disc, and the EM counterpart arising from the AGN \cite{PhysRevD.108.123039}. Though both these mechanisms are yet to be confirmed, the discovery of these events is bringing us a deeper understanding of these astrophysical systems.   

\subsubsection{Current Status:}

On the frontier of multi-messenger observations, recently the \ac{LVK} collaboration \cite{KAGRA:2013rdx} has started the fourth observation run with four detectors LIGO-Hanford, LIGO-Livingston, Virgo, and \ac{KAGRA} with the potential to detect BNSs up to about 160 Mpc from  LIGO-Hanford and LIGO-Livingston, and up to 50 Mpc using Virgo, while KAGRA is currently in its engineering runs. The operation of three detectors with better sensitivities than the third observation run improves the sky-localization area of the GW source and hence enhances the potential for joint detection with other multi-messenger observations. India is a part of the LVK collaboration and is currently planning to build the third LIGO detector, called LIGO-Aundha, in India in collaboration with LIGO. The operation of  LIGO-Aundha will make it possible to reduce the sky localization error significantly and will also improve the network SNR of the detection of a source. 

Along with GW observatories, multiple EM telescopes ranging from gamma rays to radio are operational across the globe which monitor the GW events in these bands. For BNS events, gamma-ray observations are expected immediately within a few seconds after the GW observation, followed by the EM observations at a timescale from minutes to days, in X-ray to radio. The improvement in the sky-localization area with three operational GW detectors, along with prompt detection in gamma-ray, helps in reducing the sky-localization error significantly, which can help in identifying the EM counterparts at lower frequencies. Several telescopes/observatories operated by Indian Institutions, such as Astrosat and \ac{GMRT}, have taken part in the EM follow-up observation of GW170817. These are currently operational and can contribute to the EM follow-ups of BNSs and NSBHs in the near future. Along with these, India is also a part of the \ac{GROWTH} project which contributes to the EM follow-up. 

Along with the EM follow-up of the GW sources, joint neutrino observations of the BNS, NSBH, and BBH systems will be useful in understanding the high-energy emission mechanism. 
Neutrino observatories such as the ANTARES neutrino telescope and the IceCube Neutrino Observatory, jointly with the CR observatory Pierre Auger, performed a low-latency search for the multi-messenger BNS event GW170817 and did not detect any neutrino from the direction of the source \cite{ANTARES:2017bia}. Searches are also being carried out using the unbinned maximum likelihood and Low-Latency Algorithm for Multimessenger Astrophysics pipelines for IceCube for the GW source catalog of \ac{GWTC-3}. However, no joint detections have been made so far. 

\subsubsection{Required Synergy between Messengers}

\textit{GW and EM:} The LVK network of GW detectors issue real-time alerts in response to the modeled signals (such as coalescing compact binary objects like BNSs, NSBHs, and BBHs) and unmodelled searches (such as core-collapse massive stars) using several pipelines such as GstLAL (GStreamer LIGO Scientific Collaboration Algorithm Library) \cite{PhysRevD.109.042008}, PyCBC Live \cite{PhysRevD.98.024050}, MBTA (Multi-Band Template Analysis) \cite{Adams:2015ulm}, RAVEN (Rapid, on-source VOEvent Coincident Monitor) \cite{2023APS..APRH09008A}, and SPIIR (Summed Parallel Infinite Impulse Response) \cite{PhysRevD.105.024023}. The event candidates are listed on \ac{GRACEDB}\cite{gracedb},  and regular alerts are issued using the International Gravitational Wave Network Alert System (igwn-alerts) which are machine-readable JSON files that include the information of the GW sources such as sky map, and initial set of source parameters. 

Following these GW alerts, it is important to carry out a rapid follow-up of the GW sources through EM telescopes operating in India such as the \ac{MACE} in the gamma-rays, Astrosat in X-ray, Growth-India in optical, and GMRT in Radio, along with the international network of EM telescopes. The fast follow-up of the sources for the EM counterparts will be useful for BNS, NS-BH, and BBH events from LVK --- in the future, also from LIGO-Aundha, \ac{CE} \cite{Reitze:2019iox}, and \ac{ET} \cite{Punturo:2010zz} --- and for the supermassive BBHs detectable in the next decade from \ac{LISA}. Furthermore, it will be important to identify the host galaxy of the transient and characterize its properties such as the total stellar mass, the star formation rate, and stellar metallicity. This helps in drawing a connection between the properties of a binary compact object and its host galaxy that can be useful in shedding light on the formation channels of these binary compact objects.  To identify the host galaxy and its properties, a complete galaxy catalog up to high redshift is required with spectroscopic redshift measurements. With the current network of GW detectors, GW sources can be detected with a matched filtering SNR above eight up to a redshift of about $z\sim 0.1$ for BNS, $z\sim0.3$ for NSBH, and $z\sim1$ for stellar origin BBHs. The sub-threshold searches for the GW events (which are below the threshold value of the matched-filtering SNR in a network of GW detectors) can identify potential candidates beyond these redshifts as well. The EM follow-up for these sources is feasible and interesting as it can probe such events up to high redshift and hence enhances the probability of joint detection. 

\subsubsection{Future Prospects: }

The prospect of multi-messenger observation opens up a new window to study the binary compact objects and their formation history and host properties. This avenue is currently limited to stellar-origin compact objects and low-redshift Universe ($z<1$). In the near future, with the operation of {LISA} by the end of the next decade, GW from supermassive BBHs can be detected, which are also going to have EM counterparts. Targeted EM followups of these sources from gamma-ray to radio wavelengths, along with the data from cosmic-ray detectors and neutrino observatories, it will be possible to understand the production mechanism of high energy particles and host properties of these sources up to very high redshifts ($z \sim 10$). As such sources will be rare --- only of the order of a few tens to hundreds in the whole observation period of LISA --- it will be crucial to have a network of telescopes operating in different frequency ranges that have a high sensitivity to detect counterparts to GW sources from far away redshifts and a rapid response system which can follow up any GW trigger from LISA. 

In the future, when the next-generation GW detectors such as CE and ET will be operational, we can reach redshifts beyond $z=1$ using BNS and NSBH. At that time, a few thousand GW sources per year with potential EM counterparts can be detected. A few of these sources can also be gravitationally lensed \cite{Takahashi:2003ix}. For a successful campaign of multi-messenger astronomy, we need to have a network of telescopes across the EM spectrum, and also operational neutrino and cosmic-ray detectors which can detect these sources.  Moreover, the joint operation of LISA and CE/ET can open up a new avenue of multi-band GW detection, where early warnings from LISA are possible for BBH sources in {AGN} discs, which are detectable in CE/ET after a few years. These sources can be detected along with EM counterparts over a range of wavelengths. Thus, the study of multi-band sources using multi-messenger observations will be feasible in the future if telescopes with high cadence, high sensitivity, and capability to cover large sky areas are operational.



\subsection{Multi-Messenger Astroparticle Physics with Neutrinos and Photons}

\subsubsection{Introduction}

The last century has made ground-breaking progress in understanding the origin of high-energy CRs. Because CRs are charged particles, it is difficult to point back to the source since charged particles with energies $E \lesssim 10^{16}$ GeV get deflected in the galactic magnetic field, and the information about the direction of their source is lost. On the other hand, high-energy gamma rays and neutrinos can serve as effective probes for sources of CRs. Over the last two decades, high-energy and very high-energy gamma-ray astrophysics have delivered a wealth of data and have discovered hundreds of high-energy gamma-ray sources. The first detection of high-energy extragalactic neutrinos by the IceCube neutrino observatory in 2013 opened a new window to the high-energy Universe. However, identification of a neutrino source had remained elusive till the observation of a high-energy neutrino from a blazar TXS0506+056 in 2018 along with observations of the same blazar in EM bands~\cite{Icecube:2018}. Electromagnetic data can also deliver information on where and when to expect a neutrino signal. In the same way, neutrino data can tell us when and where to look for an EM counterpart, providing improved power to discriminate between proposed source models. Hence combining neutrino data with EM data is the key to identifying the hadronic accelerators and hence sources of CRs~\cite{Gao2019,Cerruti2019,Padovani2019}. 

 In fact, the multi-messenger opportunity has already been established and exploited for low-energy neutrino astrophysics, in particular in the context of the MeV neutrinos from  \ac{CC-SNe}. The MeV neutrinos emitted during the shock breakout from the core of the collapsing star  bring information from the deepest parts of the star. These neutrinos can also influence the \ac{SN} mechanism, {and hence their observation as well as the study of their flavor conversions inside the star \cite{Raffelt:1992bs, Mirizzi:2015eza,Capozzi:2022slf} is crucial for a thorough understanding of this class of events.} The gravitational waves from the shock breakout deep inside the core and the EM signal counterparts from the shock breakout at the outer mantle open up the multi-messenger opportunity. The exciting correlation between the neutrino and EM counterparts in the context of SN1987A \cite{PhysRevLett.58.1490} was one of the finest and earliest examples of the multi-messenger reach of neutrino astrophysics. 
 
\begin{figure}
    \centering
    \includegraphics[width=0.49\linewidth]{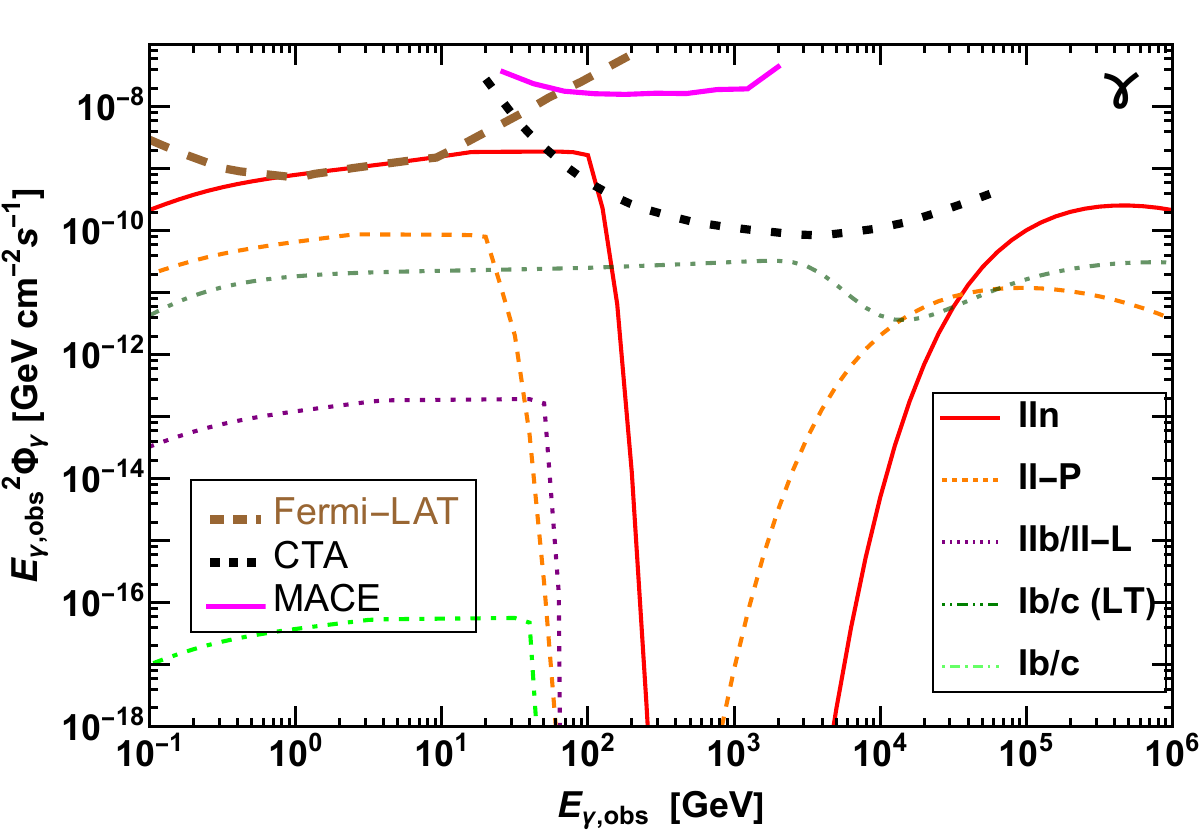}
    \includegraphics[width=0.49\linewidth]{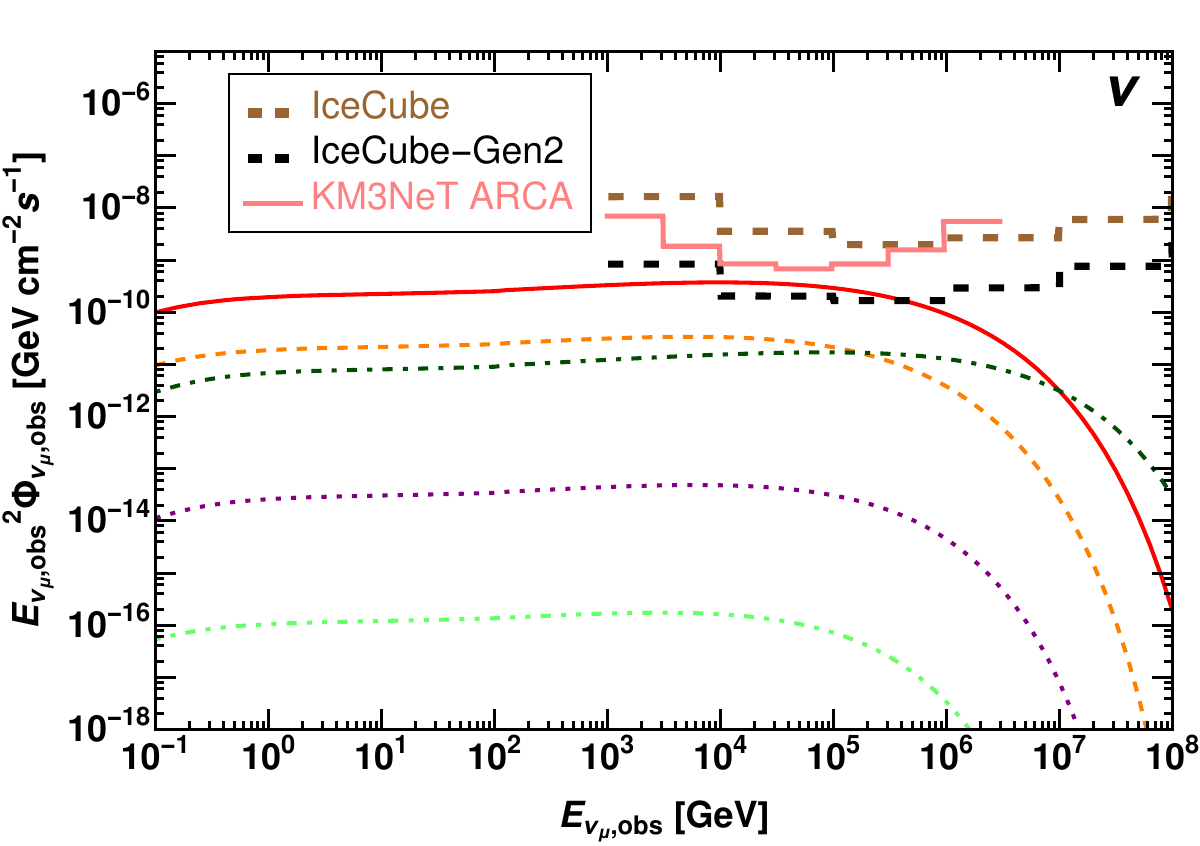}
    \caption{Typical flux estimates of gamma-rays (left) and high energy neutrinos (right) for various YSN types: IIn, II-P, IIb/II-L, Ib/c, and Ib/c LT. The sensitivities of current and upcoming telescopes are also depicted.}
    \label{fig:YSNe-point-sources}
\end{figure}

\begin{figure}
    \centering
    \includegraphics[width=0.9\linewidth]{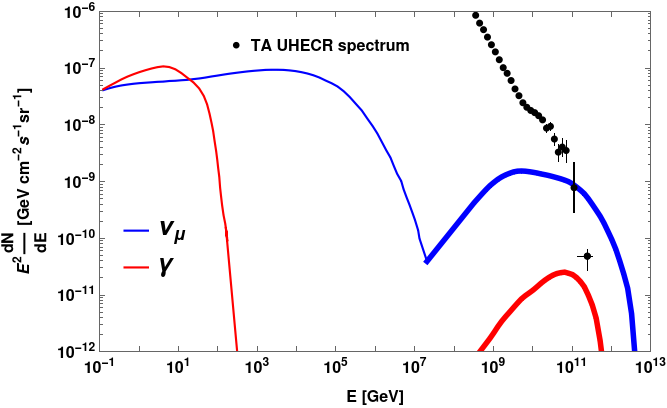}
    \caption{High energy neutrinos and gamma-rays from YSNe (thin) and UHECRs (thick).}
    \label{fig:HENG}
\end{figure}

\subsubsection{Current Status of Joint multi-messenger astrophysics}

\begin{itemize}
\item {\bf MeV neutrinos, photons and gravitational waves from core collapse SN} :
CC-SNe are huge explosions marking the violent death of massive stars ($M\gtrsim 8$~$\textrm{M}_{\odot}$), as they become unstable in late phases of their evolution \cite{Janka:2006fh}. These stellar explosions are driven by shock waves, which eventually eject the outer mantle of these collapsing stars. Indeed,  SNe are one of the most energetic events in the universe, liberating gravitational binding energy $E_B \simeq 3 \times 10^{53}$~erg. Remarkably, 99$\%$ of this binding 
energy is carried away by (anti)neutrinos of all three flavors, making the core-collapse supernova one of the 
most powerful source of (anti)neutrinos.

Neutrinos emitted during such a process are a crucial tool to probe both mass and mixing properties and the SN mechanism. These neutrinos are sensitive to many details of the SN physics, like temperature, density, and progenitor mass. Deep inside the core, neutrinos start undergoing flavor conversions due to the so-called ``collective effects'', arising from the neutrino-neutrino interactions \cite{Dasgupta:2016dbv}. Further, while passing through the mantle of the star and the stellar envelope, they are affected by matter effects on their flavor conversions \cite{Dighe:1999bi}.
Flavor changes inside an SN can also be really dramatic as the neutrinos pass through a dense turbulent
the environment during their propagation. As a consequence of all these effects, the SN neutrino fluxes reaching the detectors would be profoundly modified with respect to the initial ones, carrying intriguing signatures of oscillation effects occurring in the deepest supernova regions that are not accessible to the EM observations. These signatures would also give interesting information regarding the currently unknown neutrino mass ordering. Thus, detecting the astrophysical neutrinos from a future galactic SN event would prove to be pivotal in reaching a new frontier in low-energy neutrino physics and astrophysics. 

For the neutrino detectors \ac{SK} \cite{Super-Kamiokande:2002weg} and IceCube, inverse beta decay is the golden channel for $\bar{\nu}_e$ detection. 
However, SK and its upgrade \ac{HK} \cite{Hyper-Kamiokande:2018ofw} will also be capable of detecting electron neutrino signal through the elastic scattering $\nu_x+e^-\to e^-+\nu_x$ \cite{Kachelriess:2004ds}. On the other hand, the DUNE liquid Ar detector will have the unique opportunity to detect $\nu_e$'s by charged-current interactions on ${}^{40}\rm{Ar}$ with a very low-energy threshold. 
The total neutrino flux can also be measured at SK/HK from different neutral-current processes which are sensitive to all neutrino flavors. 
Overall, the large size of these detectors would allow huge statistics and spectral studies in time as well as energy domains. Additionally, the detection complementarity between $\nu_e$ and $\bar{\nu}_e$ in the different detectors would assure a unique way of probing the SN explosion and the intrinsic neutrino properties. 

Correlated GW detection from CC-SNe is expected to be possible at LIGO, Virgo, and {KAGRA} for CC-SNe up to a few kpc from the Earth, while future detectors such as the {ET} can reach the entire Milky Way. The onset of the 
GW and neutrino emission will be useful for understanding several fundamental properties in particle physics and gravity. 

The neutrino directionality information \cite{SNEWS} will be of crucial help in \ac{SBO} detection for the telescopes \cite{ASAS-SN}.
Indeed, the detection of galactic SN neutrinos in the current or future detectors will also be able to point to the location of the core collapse event in the sky with an uncertainty of up to about ten degrees. Such a pointing would be crucial for EM signal detectors trying to observe the SBO emission. The SBO of the progenitor surface is expected to be delayed by several hours from the core collapse triggering the neutrino and GW emission. Precise detection of the SBO, which would last for an extremely short duration, will help us understand various properties of the CC-SNe, such as the progenitor radius.  In general, the neutrino detectors will help in the 
pursuit of a precise early light curve from the photon telescopes, resulting in a better understanding of the progenitor properties and the explosion.

\item {\bf High-energy neutrinos and photons from young supernovae} : Observations of CC-SNe across various wavelengths have revealed intense emissions, indicating the presence of dense, hydrogen-rich  \ac{CSM} enveloping massive stars in their final stages~\cite{smith2010circumstellarmaterialevolvedmassive,Smith:2014txa}. This bright emission is thought to result from the photo-ionization of the CSM~\cite{Jacobson-Galan:2023ohh}. The dense CSM is created by excessive mass loss of the progenitor years prior to explosion~\cite{Smith:2014txa}.  
During SN explosions, high-velocity ejecta colliding with the CSM can generate a powerful forward shock.  This shock accelerates  protons to extremely high energies while sweeping out the CSM~\cite{Chevalier:2001at,Murase:2010cu,Murase:2013kda,Murase:2019tjj,Sarmah:2022vra,Sarmah:2023sds,Sarmah:2023xrm}. The shock acceleration can occur within an hour or a few weeks after explosion and can last up to days or years depending on the ejecta velocity, location and density of the CSM. Hence, these SN events are also termed as young SNe (YSNe)~\cite{Sarmah:2022vra}.  The maximum energy of the accelerated protons can be as large as $(1-100)$~PeV depending on the nature of the SN~\cite{Sarmah:2022vra}.  Subsequent collisions between these accelerated protons and unshocked CSM protons produce charged and neutral pions, which decay into secondary particles, including high-energy neutrinos and photons~\cite{Kelner:2006tc}. The flux of these particles depends on various properties of SN and its environment such as the density of CSM, shock kinetic energy, magnetic field, and acceleration efficiency. These properties can generally be associated with brightness and vary for different SN types such as IIn, II-P, IIb, II-L, Ib/c, and Ib/c late time~\cite{Murase:2017pfe,Sarmah:2022vra}. The type IIn class SNe, being the brightest ones, are expected to produce the largest flux of high energy neutrinos and photons,  followed by other SN types: Ib/c late time, II-P, IIb/II-L, and Ib/c, respectively. The flux from each SN type has large uncertainties due to the loosely constrained CSM interaction properties.  
{Recently, LHAASO detected the brightest GRB 221009A \cite{LHAASO:2023}, which was confirmed to originate from a type Ic supernova through JWST observations \cite{Blanchard:2023-JWST}. These very high-energy photons were subsequently used to constrain Lorentz-violating theories of gravity \cite{LHAASO:2024-LIV}.  }

Detection of these high-energy secondaries is essential to probe CSM interactions in YSNe. However, no high-energy neutrinos and gamma-rays have yet been detected from such SN events~\cite{IceCube:2023amf}. Indeed, the expected fluxes from the past SNe are below the sensitivities of the current high-energy neutrino (IceCube) and gamma-ray (Fermi-LAT) telescopes. The flux estimates of gamma rays (left) and high energy neutrinos (right) for various YSNe are shown in Fig.~\ref{fig:YSNe-point-sources}.    Studies~\cite{Murase:2017pfe,Sarmah:2022vra,Sarmah:2023xrm,Petropoulou:2017ymv} have shown that, for the best detection possibility, a type IIn SN should lie within a distance of $10$~Mpc. For other SN types, the detectable distances are much shorter.  Thus, we need to hope for closer events that would be detectable  with the currently available telescope facilities. The upcoming telescopes such as the IceCube upgrade (the proposed IceCube-Gen2) for neutrinos, and MACE, \ac{CTA} for gamma rays  may be able to chase such events beyond $10$~Mpc depending on the nature of the SNe~\cite{Sarmah:2022vra}.  It is worthwhile to note that the high energy photon flux gets severely attenuated due to pair production losses on source thermal photons, resulting in a dip in the photon spectrum above $\sim 100$~GeV.

While the ability to directly observe CSM interactions in YSNe relies on the chance of observing nearby events, a secondary approach involves analyzing  the past SNe across the Universe. The YSNe from the past could create diffuse backgrounds of high-energy neutrinos and photons. These diffused backgrounds depend on the   properties  of individual YSN type as well as their rate of occurrence~\cite{Sarmah:2022vra}. Type II-P SNe have the highest occurrence rate ($\sim 48\%$), followed by Ib/c ($\sim 26\%$, the subclass Ib/c late time is about 10$\%$), IIb/II-L ($\sim 17\%$) and IIn ($\sim 9\%$).   Although the abundance of type IIn is the smallest, they contribute the largest to the diffuse backgrounds as they produce the largest high energy neutrino and photon flux~\cite{Sarmah:2022vra}. Type Ib/c late time and II-P can also have a significant contribution to the diffuse backgrounds, whereas the contributions of the remaining YSN types   are negligible. An estimate  of these diffuse fluxes of neutrinos (thin blue) and gamma-rays (thin red) from \cite{Sarmah:2022vra} are depicted in Fig.~\ref{fig:HENG}.   The diffuse photon flux gets completely attenuated above $\sim 100$~GeV due to pair production on extra-galactic background light (EBL).   The total diffuse flux of high-energy neutrinos from YSNe is found to be capable of explaining a significant amount of the high-energy diffuse neutrino flux detected in IceCube. Whereas, the corresponding diffuse photon flux remains significantly below the Isotropic Gamma-Ray Background (IGRB) detected by Fermi-LAT.

\item {\bf The \ac{GZK} process and \ac{UHECRs}} : The CR spectrum at ultra-high energies, i.e., above $10^{18}$~eV, has several unanswered questions regarding their sources and the production mechanism~\cite{Bhattacharjee:1999mup,DUrso:2014vgv}. These CRs are certainly not produced by conventional astrophysical accelerators (SNRs, Pulsars)  in the Milky Way as the magnetic fields therein are not strong enough to confine CRs for a time sufficient to accelerate to such extremely high energies. Therefore, the UHECRs are more likely to originate in external galaxies, unless they are produced by exotic processes like dark matter decay or annihilation in the Milky Way. 

The UHECRs above energy $\sim 4 \times 10^{19}$~eV (GZK cutoff) can interact with the \ac{CMB} photons substantially if they propagate from sources at distances $10$~Mpc or beyond, via the so called  GZK process ($\text{CR} + \gamma_{\rm CMB} \to N + \pi^{+} +\pi^{-} +\pi^0 +....  $)~\cite{PhysRevLett.16.748,Zatsepin:1966jv}. As a result of this process, the UHECR spectrum is expected to be heavily suppressed above the GZK cutoff which is evident from different UHECR observations (\ac{TA}, \ac{PAO}, HiRes, AGASA)~\cite{Abbasi:2023swr,PierreAuger:2021hun,HiRes:2002uqv,HiRes:2007lra,Hayashida:2000zr}. However, there are a bunch of UHECR events that have been detected at energies up to $\sim 3 \times 10^{20}$~eV detected by these experiments at different times. The sources of these events are poorly understood to date. In fact, the second-most energetic event (known as the Amaterasu) recently detected in TA pointed back to a void in the local Universe~\cite{TelescopeArray:2023sbd}.  

The major problem of probing sources of UHECRs through their direct detection is that they being charged particles lose their directionality significantly due to interaction with the \ac{EGMF} during propagation. Therefore, this problem necessitates the need for other sources of information.   The  UHECR sources can also be probed via detection and analysis of multi-messenger signals such as \ac{UHE} neutrinos and UHE photons associated with the UHECRs. The emission of UHE neutrino and UHE photon (also referred to as GZK neutrino and GZK photon) occurs as a result of the decay of pions produced in the GZK interaction. As these secondaries are neutral particles, they do not suffer deflection on EGMF and are suitable for back-tracing. However, the UHE photons can undergo energy losses due to pair production  on different low-energy photon backgrounds like CMB, the  \ac{ERB}, and the EBL. As a result, the GZK photon flux can get heavily attenuated. Note that the neutrinos propagate unattenuated due to their weakly interacting nature.

While the attenuation of GZK photons due to CMB can be well estimated as the CMB spectrum is fairly understood, the ERB spectrum spanning over a broad frequency range (KHz-GHz) has large uncertainties. The ERB uncertainties give rise to  large uncertainties in the estimation of GZK photon flux. The upcoming radio telescope, the \ac{SKA}, can play a crucial role in understanding the ERB spectrum and resolving its features. On the other hand, the detection of GZK neutrinos and GZK photons will help in understanding both UHECR source properties and ERB absorption. If our current understanding of the GZK neutrino flux is reasonably correct, there is a high chance of detection of GZK neutrinos in IceCube-Gen2 or Giant Radio Array for Neutrino Detection. Detection of GZK neutrinos using these  will put strong constraints on the GZK photon flux.

\item {\bf High-energy neutrinos from  AGN, GRBs and other sources}: AGNs (e.g., blazars, quasars) and \ac{GRBs} are potential sources of high energy neutrinos and photons~\cite{Murase:2022feu,IceCube:2023jds,Kimura:2017kan,Kimura:2022zyg,Tamborra:2015qza}. CRs  accelerated to very high energies  in the powerful jets of these sources can interact with  different low-energy targets such as protons ($pp$ processes)  and photons ($p\gamma$ processes), producing high-energy neutrinos and photons. Observations of gamma rays from these sources have provided us with some evidence of CR acceleration. Detection of the neutrino counterpart can establish CR acceleration in these sources. 
In fact, high energy  neutrino detection in the IceCube experiment and follow-up gamma-ray detection in Fermi-LAT from  the direction of the blazar TXS 0506+056 is the best multi-messenger evidence of CR acceleration in blazars~\cite{IceCube:2018cha}. Several other tentative detections and associations with blazars have also been reported~\cite{Giommi2020}\cite{Padovani2022}\cite{Prince2024}. IceCube has also detected a bunch of neutrinos from the active galaxy NGC 1068, possibly produced in its AGN~\cite{IceCube:2022der}.

However, on the other hand, no neutrinos have ever been detected from GRBs. Even  the brightest GRB of all time, GRB 22109A, was not detected in IceCube~\cite{IceCube:2023uab}. Given the energetics and large distances to the GRBs observed to date, the flux of high-energy neutrinos may be too low for detection in current experimental facilities such as IceCube~\cite{Murase:2022vqf}. Nevertheless, one may expect to observe neutrinos if a GRB occurs closer than  a few hundreds of Mpc. 

High-energy neutrinos and photons from AGNs and GRBs across the Universe can produce diffuse backgrounds. There are several works in the literature that advocates these sources as potential contributors to the observed diffuse neutrino flux in IceCube  and gamma-ray flux in Fermi-LAT~\cite{Murase:2018iyl,Oikonomou:2022gtz,Murase:2022feu,Tamborra:2015qza,Liu:2012pf,Zhu:2021mqc}.   It has already been realized that the gamma-rays from blazars can explain the IGRB of Fermi-LAT, except for a portion above $100$~GeV~\cite{Fermi-LAT:2014ryh,PhysRevLett.116.151105}.

In order to firmly establish  the contribution of these sources to high-energy neutrinos and CRs, we need to understand the various processes occurring inside AGNs and GRBs. 
For both these, the flux of neutrinos and gamma rays generally depends on parameters such as shock velocity, magnetic field, density of accelerated CRs, and density of target particles. Due to a lack of sufficient data, these parameters  are poorly constrained. 
Therefore,  the detection of point sources in both high-energy neutrinos and gamma-rays  becomes crucial to constrain these parameters.   Unlike gamma rays which are impacted by pair production losses,  neutrinos can escape the dense source environment easily and are expected to arrive at the Earth much before the gamma rays.  Therefore, neutrino telescopes such as IceCube and KM3NeT are crucial players that can alert gamma-ray telescopes across the globe to prepare for monitoring signals coming from specific directions.  Apart from these two kinds of sources,  super-luminous SNe or hypernovae~\cite{Chakraborty:2015sta}, galaxy mergers~\cite{Bouri:2024ctc}, neutron star mergers~\cite{Kimura:2018vvz} and radio galaxies~\cite{Blanco:2017bgl,BeckerTjus:2014uyv} are also potential sources of high energy neutrinos and photons.

\item{\bf Interdependencies of diffuse neutrino, CRs, and gamma-ray backgrounds: } UHECRs have been detected by various experiments such as HiRes, AGASA, PAO, and TA. 
However, only 95\% C.L. upper limits on the integrated photon flux above 10$^{18}$ eV and 90\% C.L. upper limits on the normalization of a single flavor neutrino diffuse
flux above 10$^{17}$ eV have been reported by PAO. Similarly, IceCube has also reported  a 90\% C.L.  upper limit  on neutrinos with energies between 3 PeV and 100 EeV. The non-detection of GZK neutrinos by IceCube has already ruled out (at 95\% CL) models that predict event rates of $\sim 1$ neutrino/yr or more and has started to constrain the fraction of protons in the UHECR composition in a model-independent way without relying on hadronic interaction models.

\end{itemize} 

\subsubsection{Future Prospects}
\begin{itemize}
    \item \textbf{Supernova neutrinos:} Future  neutrino detector programs ({e.g., with a deuterated} liquid scintillator detector \cite{Chauhan:2021snf}) and neutral current-based detectors will be important for SN neutrino flavor identification. Correlation between the detection of GW waves and the neutrinos from neutronization burst will be important for analyzing a wide variety of physics models based on the precise measurement of the SN timing. 
\item \textbf{Young supernovae:}  The detectors such as IceCube-Gen2 and CTA have great potential of discovering nearby ($\sim 10$ Mpc) young supernovae. Participation in these experimental collaborations will help in probing the CR acceleration mechanism and YSNe properties. The energy ranges of MACE and GRAPES are also quite suitable for probing such YSNe. 

\item \textbf{Blazars:} From Fermi-LAT observations we know that blazars are the most abundant
extragalactic gamma-ray sources and constitute approximately 80\% of the entire extragalactic source population. However, current studies based on stacking catalog searches performed
with IceCube data estimate that neutrinos emitted by blazars, in the range between around
10 TeV and 2 PeV, can only contribute up to $\sim$ 27\% to the total diffuse neutrino flux. 
Hence, further multimessenger observations with next-generation experiments are required to test these estimates, and therefore provide valuable information to understand the particle
production and acceleration in blazars.

\item \textbf{UHECRs and GZK secondaries:} The extra-galactic radio background (ERB) plays a crucial role in the propagation of GZK photons. The SKA will be able to measure the ERB in a wide frequency range and will help in resolving the uncertainties in the ERB. This in turn will help in a precise prediction of the GZK photon flux. 
Cosmogenic neutrinos are produced in photo-hadronic interactions of cosmic ray protons with the CMB. The neutrino production rate can be constrained through the accompanying electrons, positrons, and gamma-rays that quickly cascade on the CMB and intergalactic magnetic fields to lower energies and generate a gamma-ray background in the GeV–TeV region. Bethe–Heitler pair production by protons also contributes to the cascade and can tighten the neutrino constraints in various models. This issue can be thoroughly investigated in light of the recent Fermi-LAT measurements of the diffuse extragalactic gamma-ray background observed by Fermi-LAT.  
Together with this, the detection of GZK neutrinos in future experiments like IceCube-Gen2 and Askaryan Radio Array will allow us to understand the origin of UHECRs and their composition. 

\end{itemize}

\subsection{Multi-Messenger Study of  Tidal Disruption Events}
\subsubsection{Introduction}  \ac{TDEs} are interesting astrophysical phenomena that arise when a compact object such as a star gets tidally disrupted due to a heavier object such as a massive black hole \cite{1988Natur.333..523R}. Several theoretical studies suggest such scenarios can give rise to multi-messenger signals leading to the emission of high energy neutrino, and CRs, along with the EM emission \cite{Farrar:2008ex, Hayasaki:2021jem}. GW emission from such systems can be present but is unlikely to be detected with the currently ongoing detectors. In the future, even with the space-based GW detector such as LISA,  GW emission is feasible to be detected only for a fraction of nearby sources \cite{Pfister:2021ton}. 

\subsubsection{Current Status}  
Currently, with the help of the \ac{ZTF} survey, observation of TDEs is becoming possible with a cadence of typically every 3-4 weeks, making it possible to open up an avenue for multi-messenger science by following up with other messengers such as high-energy neutrinos and CRs. Such two potential multi-messenger observations of TDE candidates are AT20129dsg \cite{Stein:2020xhk} and AT2019fdr \cite{Reusch:2021ztx} which are likely to be related to the neutrino events IC191001A and IC200530A respectively. These observations agree with the currently known models of high energy neutrino emission from TDEs \cite{Winter:2020ptf}. Until now there have been no GW detections with LVK corresponding to any TDE, as expected from theoretical models.

\subsubsection{Required Synergy between Messengers and Future Prospects}
To develop a complete physical understanding of different mechanisms involved in TDEs, it is important to have a multi-messenger observation using EM, CRs, and neutrinos and in the future, when LISA will be operational, using GW. Such    
joint observations will be useful in understanding the accretion mechanism, jet, and also the mechanisms of high energy particle acceleration.  

\vspace{1cm}

\noindent\textbf{\Large{Long-term science goals:}}

\subsection{Time variation of cosmic ray anisotropy}
\subsubsection{Introduction}
To date, there have been only a few studies conducted on the time variation of the  \ac{CRA}. They have been performed using the IceCube \citep{2010ApJ...718L.194A, IceCube:2017whz} and AMANDA neutrino observatories \citep{2023AAS...24120704P, Zhou_2010, mcnally2023cosmic, 2013ICRC...33..510D}, and data from the Tibet air shower array \citep{Amenomori_2004}. The origins of CRs are not as primitive as the Big Bang. So, contrary to the {CMB}, the last relics of the Big Bang, which shows almost no time variation, one expects to observe time variations in the cosmic ray anisotropies. Sources of CRs can range from cataclysmic events like supernovae, NSBH, and BNS mergers, as well as not-so-violent events like burning or shock-driven processes in a star. It can even come from very exotic processes like Hawking radiation from a black hole. Thus we expect the CR distribution throughout the sky to be non-stationary, i.e. it must vary over time scales of the individual source mechanisms. 

\subsubsection{Current Status}
Anisotropy in the CR has been observed by IceCube, AMANDA, etc. but its time variance has not been studied full-fledgedly. So, it must be understood and studied to a better level. A thorough observation and a detailed analysis must be conducted on it to reveal any exciting new phenomena in the depths of the cosmos.

\subsubsection{Future Prospects}
The study of CRA can help us understand the universe better. It could potentially help generate triggers for other observations. As an example, the sudden increase in the amount of cosmic ray flux from a particular direction in the sky could tell other EM telescopes to follow, since the highly energetic cosmic ray particles may appear much faster than some EM frequencies suffering from high dispersion in the intermediate medium. On the other side, GW signals can help generate triggers for a CRA study. For example,
 a core-collapse supernova GW emission or BNS GW emission can be the harbinger of a high flux of CRs from a particular direction. If the sky localization error for the direction of such an anisotropy is small, then using a cross-matching technique with the existing galaxy catalog can help us pinpoint the host galaxy. Many similar observations can reveal  some properties of the host galaxy, such as the star-formation rate, metallicity, etc.

\newpage

\section{Cosmology}
\label{cosmology}
\small \emph{Contributors: Samsuzzaman Afroz, Pratik Majumdar, Harsh Mehta, Arunava Mukherjee, Suvodip Mukherjee, Mohit Raj Sah}\\ \normalsize
\small \emph{Editors: Varsha Chitnis, Sanjeev Dhurandhar, Amol Dighe, Suvodip Mukherjee,  Tirthankar Roy Choudhury, Shriharsh P. Tendulkar} \\ \normalsize

\noindent\textbf{\Large{Short-term science goals:}}

\subsection{Cosmology and Testing Gravity through Multi-Messenger Observations}
\subsubsection{Introduction}

The \ac{GR}, formulated by Albert Einstein, has been the cornerstone of modern gravitational physics, providing a comprehensive framework for understanding how gravity operates by describing the warping of spacetime by massive objects. GR's predictions have been confirmed across a range of scenarios, from planetary orbits to the bending of light. However, the theory encounters limitations when faced with extreme conditions such as the environments near black holes or during the early moments of the universe\cite{thorne1995gravitational,sathyaprakash2009physics}. Also, at the largest scales, evidence of the accelerating Universe questions the foundation of GR at the cosmological scales and brings into the picture alternative models of gravity as possible scenarios to explain this phenomenon. Despite several alternative models of GR and dark energy, our understanding of cosmic acceleration has not improved significantly since its discovery. Furthermore, the mismatch in the measurement of the Hubble constant from low-redshift cosmological probes and high-redshift cosmological probes has grown into a crisis (popularly known as the Hubble tension), resulting in the doubting of our understanding of both cosmological models as well as of  cosmological observations. The recent detection of GWs by the LVK collaboration has opened a new observation window for testing the predictions of GR under extreme gravitational conditions and also for mapping the cosmic expansion history without requiring any distance calibrator \cite{abbott2016observation, abbott2016gw151226, scientific2017gw170104, abbott2017gw170817, goldstein2017ordinary}. 

\begin{figure}[ht]
    \centering
    \includegraphics[height=7.5cm, width=16cm]{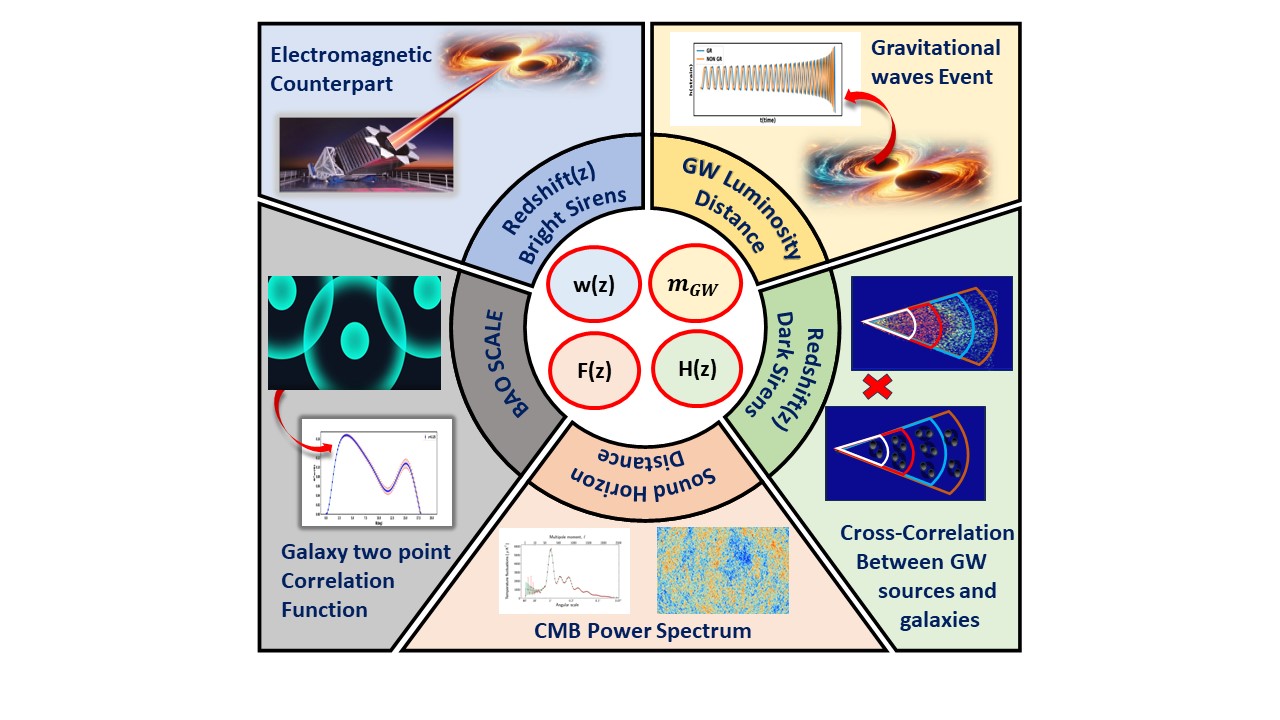}
    \caption{This diagram illustrates the methodology for exploring the frictional term in the context of modified gravity ($\mathcal{F}(z)$), the Hubble parameter (H(z)), the dark energy equation of state (w(z)), and the mass of the graviton ($m_{GW}$) by measuring three critical length scales: the sound horizon distance, the BAO scale, and the GW luminosity distance and redshift for both bright and dark sirens. These scales are derived from distinct observational methods, including CMB measurements, galaxy correlation studies, and detections of GWs with redshift using EM counterparts for bright sirens or through cross-correlations for dark sirens. By utilizing the distance duality relation, these measurements enable an empirical, redshift-dependent verification of GR.}
    \label{fig:Motivation}
\end{figure}

GWs, the ripples in spacetime caused by some of the most violent and energetic processes in the Universe, provide a direct way to study black holes and neutron stars. Observations of systems like BNS, NSBH, and BBH have allowed us to probe the strong-field as well as the weak-field regime of gravity. These observations, such as variations in the propagation speed of GWs, the mass of the graviton, variation of the Planck mass (also known as frictional term), or any forms of anisotropy in the spacetime fabric, {are critical not just for testing GR} but also for exploring any unmodelled physics related to gravity. Several alternative theories of gravity predict deviations from GR under extreme conditions. Although these modified gravity theories have yet to find empirical support against GR, the increasing sensitivity of GW detectors holds promise for potentially revealing new physics~\cite{deffayet2007probing, saltas2014anisotropic, nishizawa2018generalized, belgacem2018gravitational, belgacem2018modified, lombriser2016breaking, lombriser2017challenges}. 

Moreover, GW astronomy provides a novel method to measure cosmic distances independently of traditional EM observations. This capability is particularly important in addressing the Hubble tension, the discrepancy between the value of the Hubble constant obtained from local measurements and that derived from the CMB data~\cite{Abdalla:2022yfr}. By using GW sources, commonly known as ``standard sirens'', and measuring their distances without relying on the cosmic distance ladder, we can potentially offer a direct resolution to this tension. Additionally, the study of GWs contributes to our understanding of dark energy. The current standard model of cosmology ($\mathrm{\Lambda}$CDM), which includes dark energy characterized by an equation of state with a parameter {$\mathrm{w} = -1$},  could be further scrutinized through GW measurements. These can provide both model-dependent and independent ways to examine the equation of state of dark energy, thereby enhancing our understanding of the accelerating expansion of the universe.

Figure  \ref{fig:Motivation} presents a schematic diagram illustrating measurements of the variation of Planck mass ($\mathcal{F}$(z)), the dark energy equation of state (w(z)), the Hubble parameter (H(z)), and the mass of gravitons ($m_{GW}$) from the propagation of GWs. This diagram compares three significant cosmological lengths: the GW luminosity distance, the Baryonic Acoustic Oscillation (BAO) scale derived from the two-point angular correlation function, and the sound horizon distance, which can be estimated by analyzing CMB acoustic oscillations through missions like WMAP \citep{hinshaw2013nine}, Planck \citep{aghanim2020planck}, ACTPol \citep{thornton2016atacama}, and SPT-3G \citep{sobrin2022design}, along with the redshift which is inferred from EM counterparts for bright sirens (sources with EM counterparts) and using cross-correlation of galaxy surveys with GW sources for dark sirens (sources lacking an EM counterpart). By integrating these diverse observational methods, the schematic underscores the potential for a comprehensive and precise understanding of fundamental cosmological parameters.

\subsubsection{Current Status}
\begin{itemize}
    \item \textbf{Measurements of the Hubble Constant:} The Hubble constant ($\mathrm{H_0}$), which quantifies the expansion rate of the universe, has been the subject of intense debate due to the so-called Hubble tension, the discrepancy between measurements from the early and late universe. In the standard $\mathrm{\Lambda CDM}$ cosmological model, observations of the CMB by the Planck yield a value of $\mathrm{H_0 = 67.4 \pm 0.5 \, km \, s^{-1}Mpc^{-1}}$ \cite{Planck:2018vyg}. Comparable results have been reported by the Atacama Cosmology Telescope (ACT)\cite{ACT:2023kun} and the South Pole Telescope (SPT)~\cite{SPT-3G:2022hvq}. However, local measurements from Type Ia supernovae, calibrated by Cepheid stars, estimate $\mathrm{H_0 \approx 73 \, km \,s^{-1}Mpc^{-1}}$, showing a significant 5$\sigma$ tension with the CMB-based values \cite{2022ApJ...934L...7R}. Recent data from \ac{DESI} \ac{BAO}, {which has provided} a value of $\mathrm{H_0 = 68.52 \pm 0.62 \, km \, s^{-1}Mpc^{-1}}$ that is consistent with CMB results~\cite{DESI:2024mwx}, {has further complicated the picture and has intensified} the debate over the Hubble tension.

\item \texttt{Dark Energy Equation of state:} In addition to testing gravitational theories, GW astronomy is increasingly focused on exploring dark energy, one of the mysterious components of our universe and its equation of state. In the standard $\mathrm{\Lambda}$CDM model, the dark energy equation of state is characterized by $\mathrm{w = -1}$. However, various theoretical models suggest different impacts of dark energy on cosmic expansion, and observations of GWs offer a novel approach to assessing these effects. Significantly, there are several model-dependent parametrizations of the dark energy equation of state detailed in the literature, each proposing unique characteristics and implications for the universe's expansion. One notable model is the ($\mathrm{w_0, w_a}$) model, where the equation of state parameter $\mathrm{w}$ evolves with time, defined as $\mathrm{w(a) = w_0 + w_a(1-a)}$, with $\mathrm{w_0}$ representing the present-day value and $\mathrm{w_a}$ describing its evolution~\cite{belgacem2019testing}. Studying the dynamics of dark energy is critical to understanding the expansion of the universe, and GW observations offer a promising method for probing these effects.

Recent results from DESI provide significant constraints on the equation of state of the dark energy. DESI alone reports values of $\mathrm{w_0 = -0.55^{+0.39}_{-0.21}}$ and $\mathrm{w_a < -1.32}$, suggesting a mild deviation from the $\Lambda$CDM model, although no strong statistical preference for evolving dark energy is found. When DESI data is combined with CMB measurements, the constraints become tighter, with $\mathrm{w_0 = -0.45^{+0.34}_{-0.21}}$ and $\mathrm{w_a = -1.79^{+0.48}_{-1.0}}$, indicating a preference for an evolving dark energy equation of state at a 2.6$\sigma$ significance level. Further refinements come from combining DESI BAO and CMB data with supernova datasets. Using the PantheonPlus supernova sample, the equation of state parameters is constrained to $\mathrm{w_0 = -0.827 \pm 0.063}$ and $\mathrm{w_a = -0.75^{+0.29}_{-0.25}}$. Union3 data provides slightly different values: $\mathrm{w_0 = -0.64 \pm 0.11}$ and $\mathrm{w_a = -1.27^{+0.40}_{-0.34}}$. These combined datasets offer increasing statistical support for deviations from the standard $\Lambda$CDM model, with significance levels reaching as high as 3.9$\sigma$~\cite{DESI:2024mwx}.

\item \texttt{Modified Gravity:} The current status of GW astronomy is marked by intensive efforts to test GR through the detection of potential non-GR parameters. Although no significant deviations from GR have been observed yet, the field remains highly active with upcoming observational campaigns by LIGO and Virgo set to probe GR with unprecedented precision. These initiatives are supported by the development of both ground-based and future space-based observatories. Theoretically, a variety of alternative gravity theories that predict deviations from GR under extreme conditions are being rigorously tested. These theories include scalar-tensor theories and those proposing large extra dimensions, which speculate on variations in GW propagation and black hole dynamics. One key prediction of such theories is the change in the GW luminosity distance with respect to the EM luminosity distance due to the running of the effective Planck mass with redshift, commonly known as the frictional term \cite{Kobayashi:2019hrl}. Various model-dependent parametrizations of this frictional term are present in the literature, with one example being the ($\mathrm{\Xi_0, n}$) parametrization~\cite{belgacem2019testing}. The inference of this parametrization has been demonstrated for both the LVK network \cite{Mukherjee:2020mha,leyde2022current, Finke:2021aom, Chen:2023wpj} and LISA \cite{Baker:2020apq}. However, such parametric models are limited to capturing only power-law deviations. To overcome this limitation, a model-independent reconstruction of the frictional term $\mathrm{\mathcal{F}(z)}$ using a completely data-driven approach has been proposed \cite{Afroz:2023ndy, Afroz:2024oui, Afroz:2024joi}. This method utilizes both bright and dark standard sirens and applies to current and upcoming ground- and space-based detectors.

\item \texttt{Mass of Gravitons:} The question of whether gravitons, the hypothetical quantum particles that mediate the force of gravity, possess mass is a profound topic in theoretical physics. A massless graviton is a cornerstone of GR, but several alternative theories suggest that gravitons could have a small but nonzero mass. Recent measurements have provided stringent constraints on the mass of the graviton. The prompt detection of the EM counterpart of the BNS event GW170817, occurring approximately 1.7 seconds after the GW signal, has been pivotal in this context. This observation has enabled exceedingly precise constraints on the speed of GW propagation, which in turn impose limits on the graviton mass. Specifically, the near-simultaneous arrival of GWs and gamma rays from GW170817 implies that any difference in speed between these two signals is extremely small. This translates into an upper limit on the graviton mass, which is currently constrained to be less than $5.0 \times 10^{-23}$ eV/c$^2$\cite{abbott2017gw170817}.

\item \texttt{GR Insights from GW Polarizations:} Measuring the polarization of GW offers a powerful tool for probing fundamental properties of gravity. While GR predicts only two polarization states, alternative theories including modified gravity models allow up to six \cite{Will:2014kxa,Isi:2017fbj,Maggiore:2007ulw}. Detecting additional polarizations would challenge the predictions of GR and potentially reveal new physics. Observing GW polarization not only enhances the localization of sources, facilitating more precise EM follow-up observations, but also improves our understanding of GW propagation over cosmological distances \cite{LIGOScientific:2017ync,LISACosmologyWorkingGroup:2019mwx,Cutler:2009qv,Arai:2017hxj,Chen:2017wpg}.

\end{itemize}

\begin{figure}
\centering
\includegraphics[height=8.0cm, width=16cm]{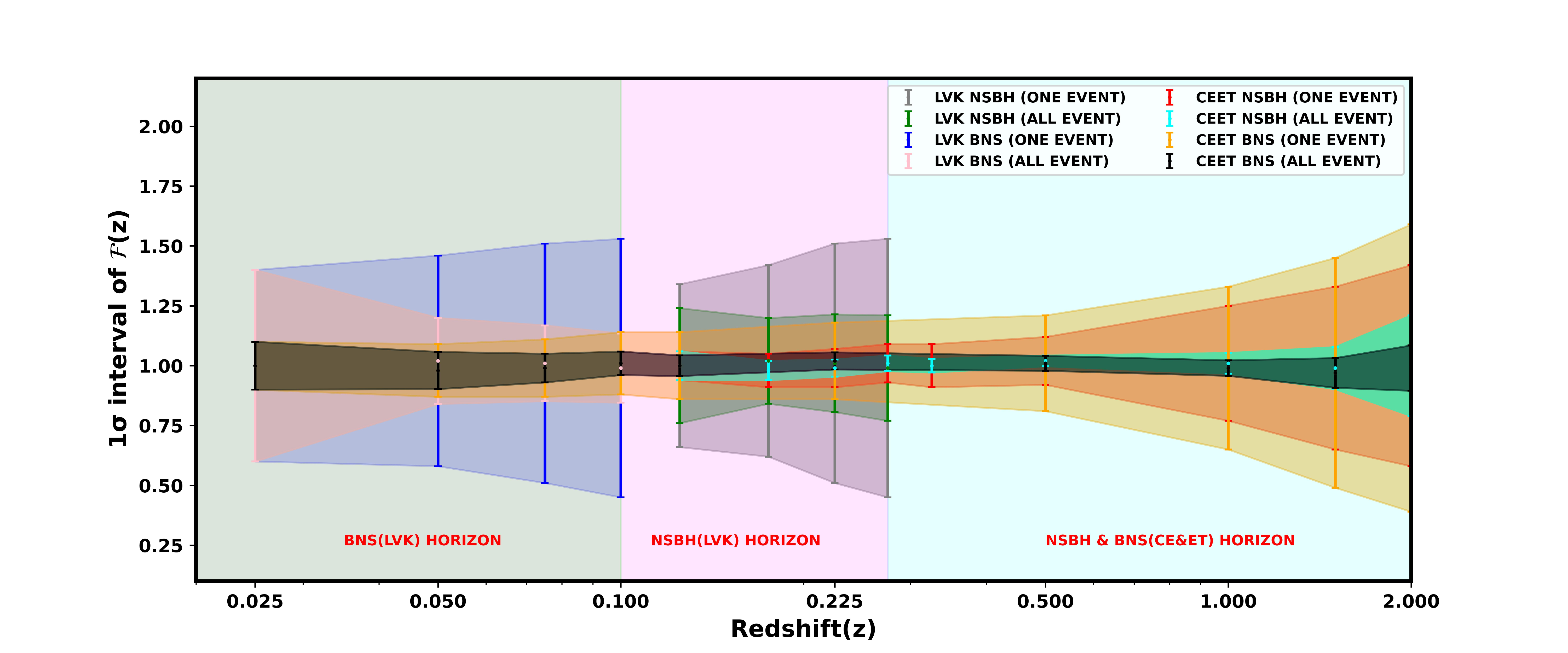}
\caption{This plot showcases the precision achievable in 
reconstructing the redshift variation of the Planck mass captured by $\mathrm{\mathcal{F}(z)}$ for both NSBH and BNS systems, employing the current LVK detector with 5 years of observation time and the future CE \& ET detectors with 1 year of observation time. The plot illustrates the enhanced accuracy in measuring the function $\mathrm{\mathcal{F}(z)}$ using future detectors compared to our current instruments in the redshift range $z=0.2$ to $z=1$, which is an interesting cosmic epoch when the acceleration of the universe starts becoming important.  This plot is taken from \cite{Afroz:2023ndy}.}
\label{fig:LVKFZ}
\end{figure}

\begin{figure}[ht]
\centering
\includegraphics[height=8.0cm, width=16cm]{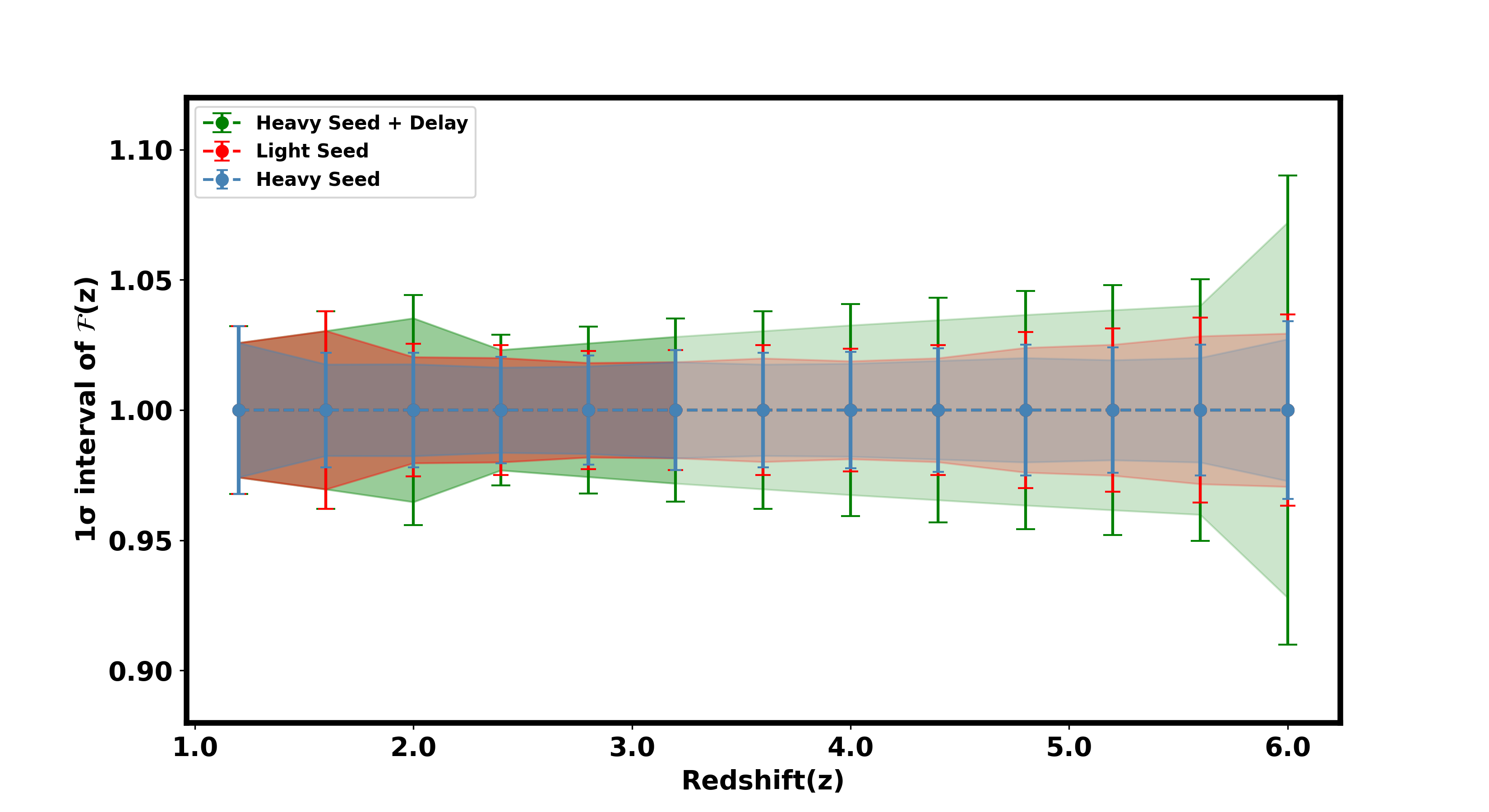}
\caption{This plot illustrates the precision in reconstructing the redshift evolution of the Planck mass, represented by $\mathrm{\mathcal{F}(z)}$, for different black hole seed models \cite{2018MNRAS.481.3278R,2024ApJ...966..176Y} the Light Seed, Heavy Seed, and Heavy Seed+delay models  using the upcoming space-based LISA observatory. The observation time is 4 years for the Light Seed and Heavy Seed models and 10 years for the Heavy Seed+delay model. The filled regions represent the $1\sigma$ intervals for the assumption of $75\%$ EM counterpart detection, while the error bars with lines indicate the $1\sigma$ intervals for the assumption of $50\%$ EM counterpart detection. The plot highlights LISA's accuracy in measuring the function $\mathrm{\mathcal{F}(z)}$ up to a redshift of 6. Beyond a redshift of 3.0, a fainter color is used to indicate that the BAO scale cannot be measured beyond this redshift with the currently planned galaxy surveys. The details of the analysis to obtain this plot can be found in \cite{Afroz:2024oui}.}
\label{fig:LISAFZ}
\end{figure}

\begin{figure}
\centering
\includegraphics[height=8.0cm, width=16cm]{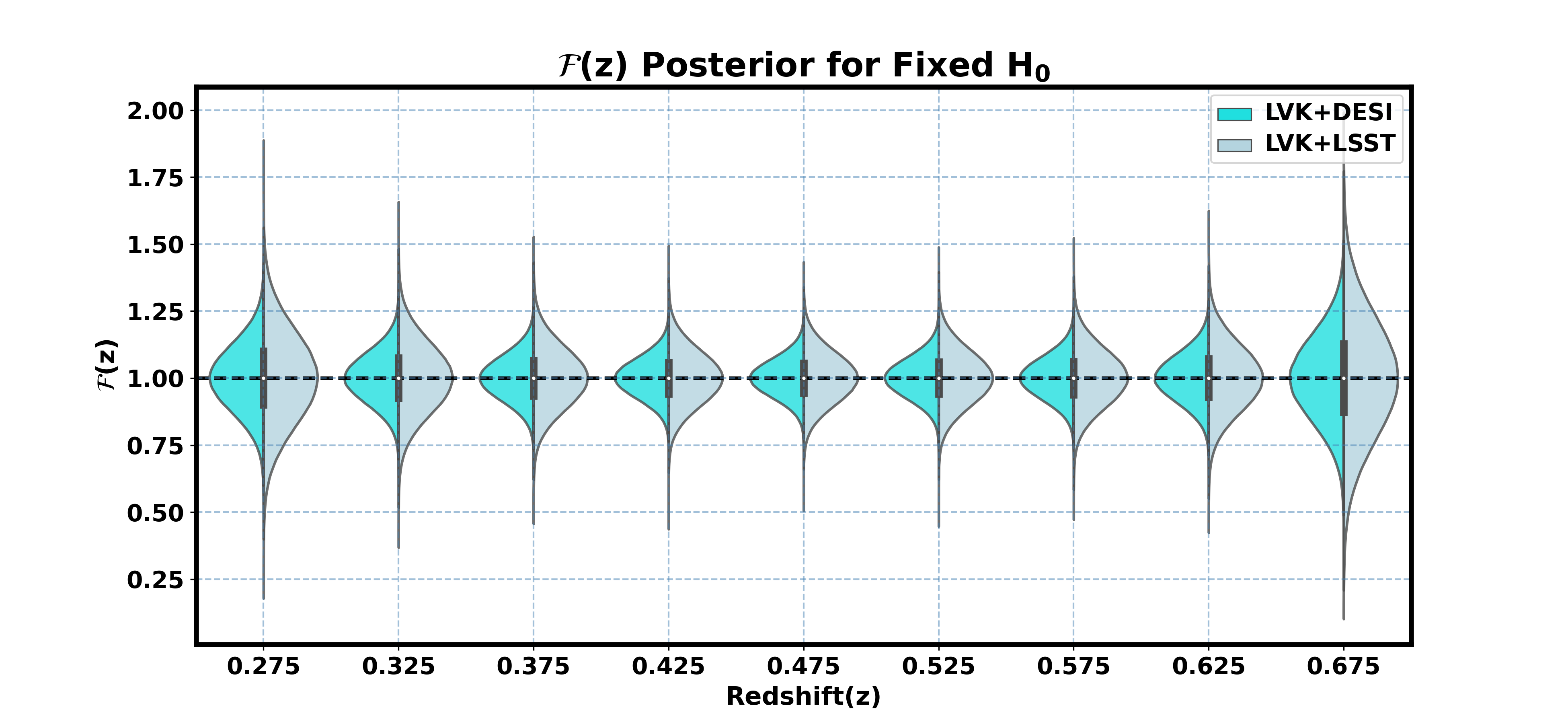}
\caption{This ``violin''-plot illustrates the posterior on the reconstruction of the non-GR parameter $\mathrm{\mathcal{F}(z)}$ as a function of cosmic redshift {from} the LVK system, considering both the galaxy survey DESI (LVK+DESI) and {LSST} (LVK+LSST), for BBHs mergers up to a redshift of $\rm{z=0.675}$. The plot presents results under fixed Hubble constant scenarios. This plot highlights the potential for achieving more accurate measurements of $\mathrm{\mathcal{F}(z)}$ at significantly deeper cosmic redshifts for current GW detectors when combined with galaxy surveys like DESI and LSST. Additionally, it suggests that DESI provides better measurements than LSST, as expected at lower redshifts. The details of the analysis behind this plot can be found in \cite{Afroz:2024joi}.}
\label{fig:DarkSirenLVK}
\end{figure}

\begin{figure}[ht]
\centering
\includegraphics[height=8.0cm, width=16cm]{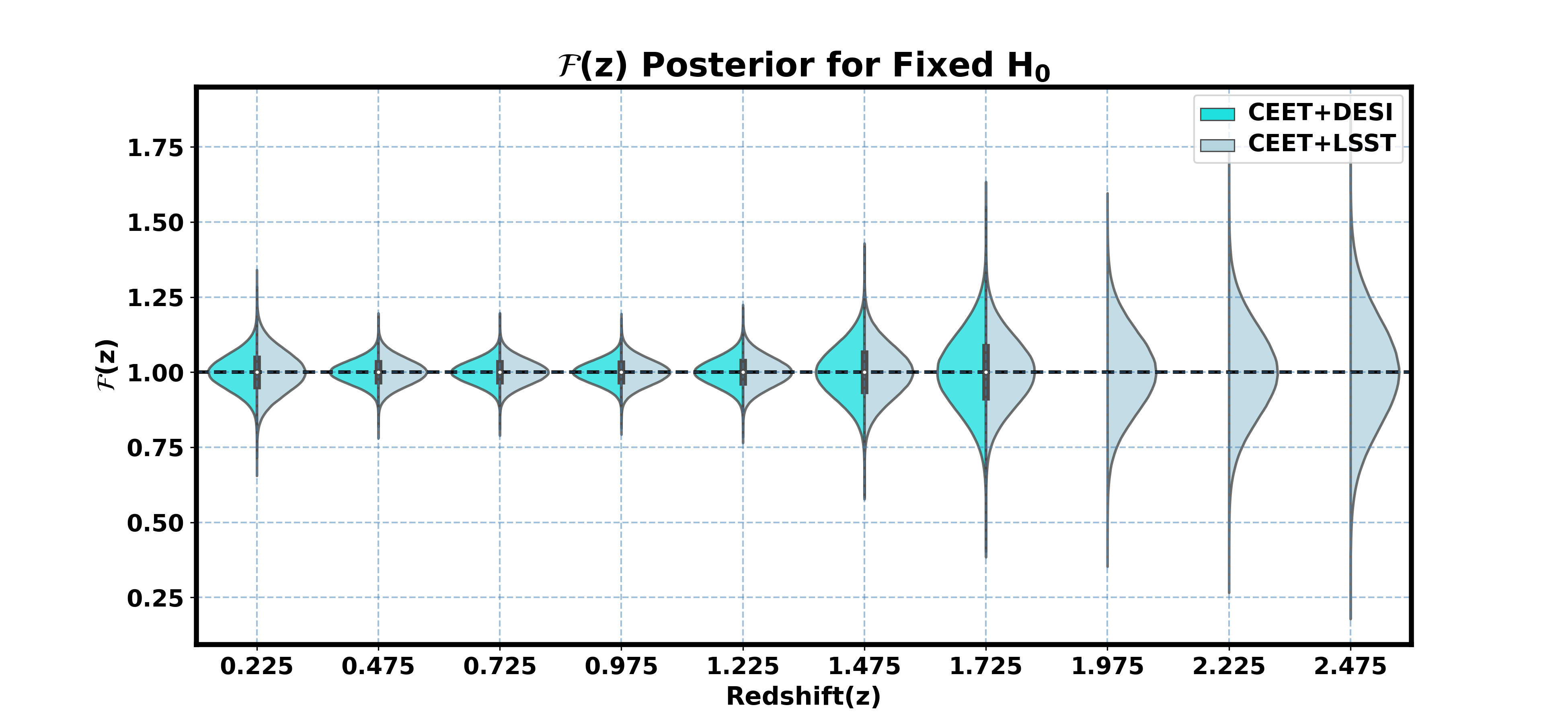}
\caption{This violin plot illustrates the posterior on the reconstruction of the non-GR parameter $\mathrm{\mathcal{F}(z)}$ as a function of cosmic redshift for CE and ET systems, considering both the galaxy survey DESI (CEET+DESI) and LSST (CEET+LSST), for BBH mergers up to a redshift of $\rm{z=2.5}$. The plot presents results under fixed Hubble constant scenarios. This plot highlights the potential for achieving more accurate measurements of $\mathrm{\mathcal{F}(z)}$ at significantly deeper cosmic redshifts for future GW detectors like CE and ET when combined with galaxy surveys like DESI and LSST. It also indicates that DESI gives better measurements at low redshifts, while LSST provides better measurements at high redshifts. This is expected because DESI is a spectroscopic survey, which excels at low redshifts where individual galaxy spectra can be obtained for precise redshift measurements. In contrast, LSST is a photometric survey, which is more effective at higher redshifts where obtaining individual galaxy spectra becomes more challenging. This plot is taken from \cite{Afroz:2024joi}.}
\label{fig:DarkSirenCEET}
\end{figure}

\begin{figure}
\centering
\includegraphics[height=8.0cm, width=16cm]{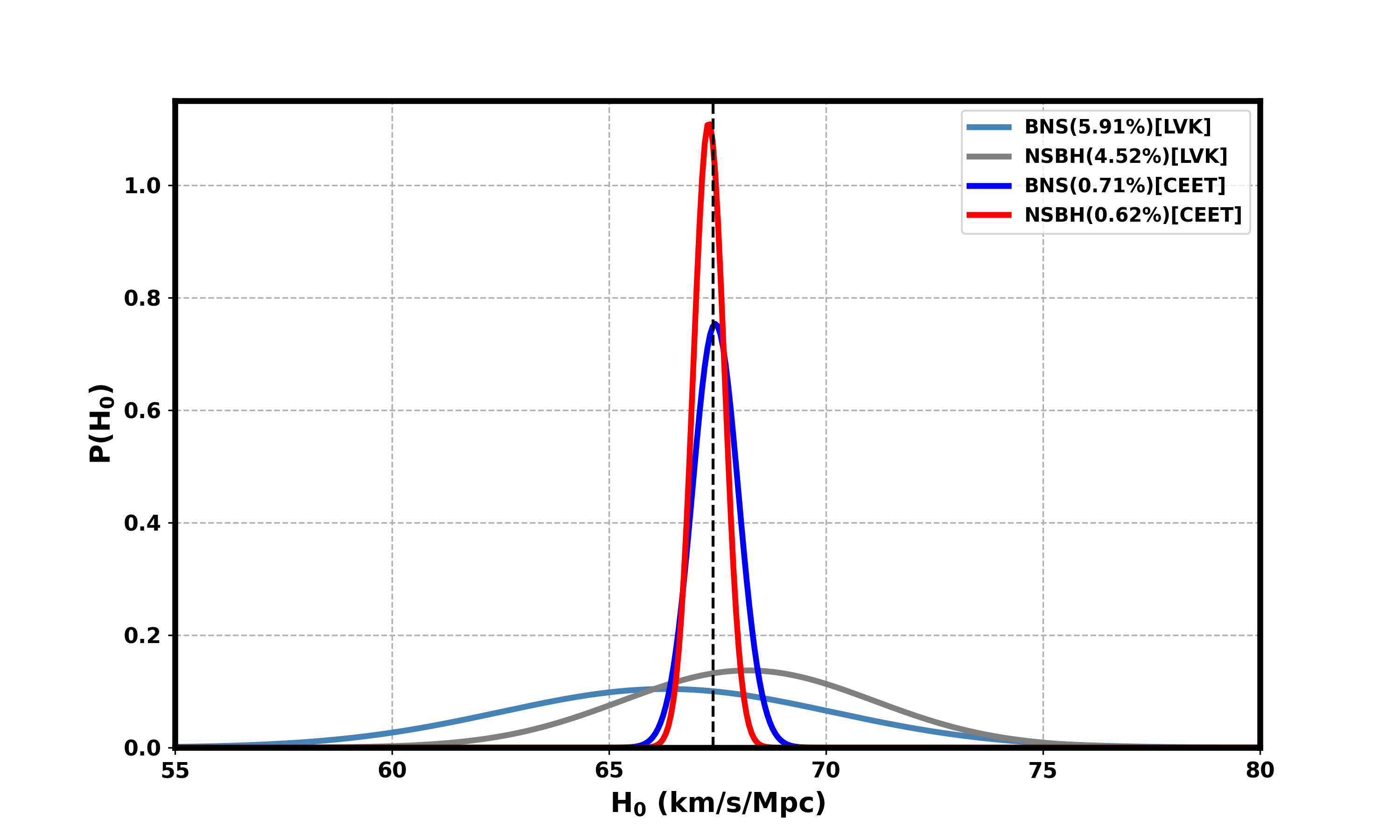}
\caption{The plot illustrates the posterior distribution of the Hubble constant, $\mathrm{H_0}$, for bright siren events detected by the LVK GW detector and the CE and ET detectors. The analysis includes BNS and NSBH mergers. The LVK analysis assumes 5 years of observation time, while the CE and ET detectors assume 1 year of observation time. The errors associated with each combination are indicated in parentheses in the legend.}
\label{fig:BrightSirenH0}
\end{figure}

\begin{figure}
\centering
\includegraphics[height=8.0cm, width=16cm]{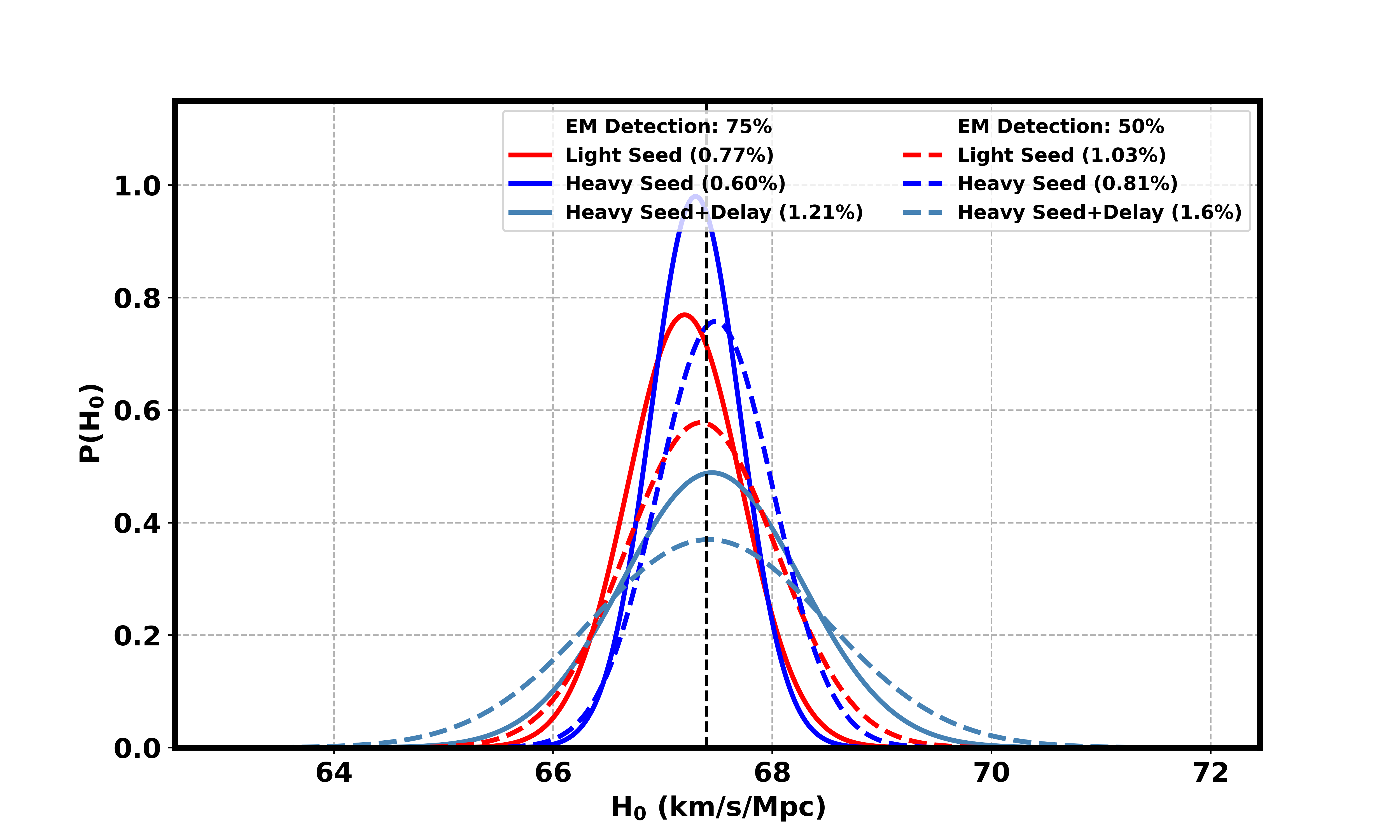}
\caption{This plot shows the posterior distribution of the Hubble constant, $\mathrm{H_0}$, for all events up to a redshift of 6, based on three models: Light Seed, Heavy Seed, and Heavy Seed+delay, using the upcoming space-based LISA observatory. The observation periods are 4 years for both the Light Seed and Heavy Seed models and 10 years for the Heavy Seed+delay model. The plot includes results assuming that EM counterparts are detected for 75\% of the events (solid lines) and 50\% of the events (dashed lines). The 1$\mathrm{\sigma}$ uncertainties in the measurements are provided in parentheses in the legend,   highlighting the varying degrees of uncertainty depending on the survey and detector combination. The details of the analysis behind this plot can be found in \cite{Afroz:2024oui}.}
\label{fig:LISAH0}
\end{figure}

\begin{figure}
\centering
\includegraphics[height=8.0cm, width=16cm]{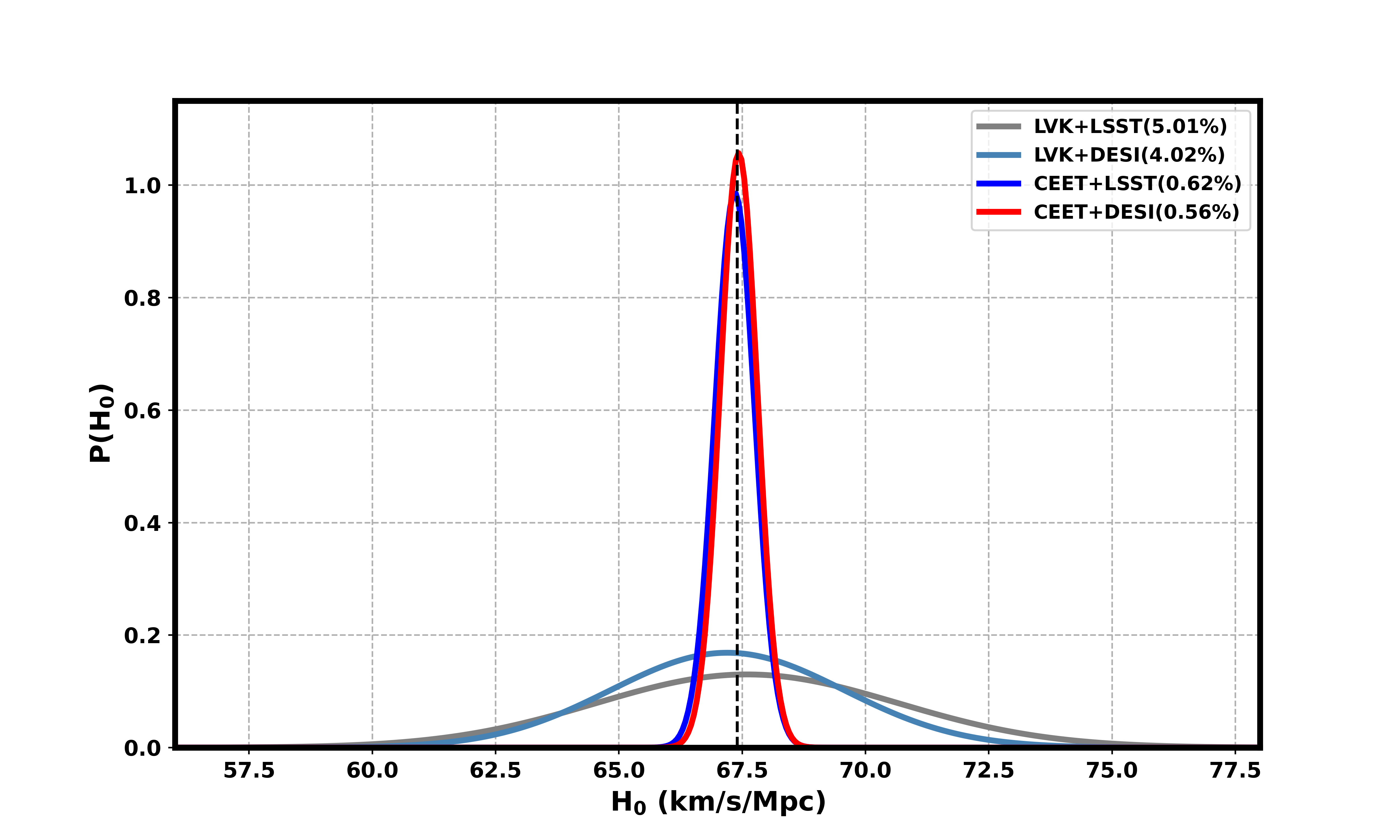}
\caption{The plot illustrates the posterior distribution of the Hubble constant, $H_0$, for combined sources from the LVK GW detector with galaxy surveys DESI (LVK+DESI) and LSST (LVK+LSST), as well as the CE and ET detectors with DESI (CEET+DESI) and LSST (CEET+LSST) surveys. The analysis covers BBH mergers up to a redshift of $\rm{z=2.5}$ for CEET and up to $\rm{z=0.725}$ for LVK, considering five years of observational time with a 75\% duty cycle. The error associated with each combination is indicated in parentheses in the legend. The details of the work behind this plot can be found in \cite{Afroz:2024oui}.}
\label{fig:DarkSirenH0}
\end{figure}

\subsubsection{Required Synergy between Messengers}
\begin{itemize}
    \item \texttt{EM}:  The concept of luminosity distance is pivotal in astronomy for understanding the universe's expansion. Luminosity distance can be measured using different methods, each with its implications for cosmological studies. Standard sirens such as GWs, and standard candles {such as} supernovae, are two primary ways of measuring this distance. Standard sirens measure distance based on the GWs emitted by events like BNS or BBH mergers, while standard candles rely on the observed brightness of objects, such as Type Ia supernovae. Additionally, other EM probes like the CMB and BAO offer alternative measurements of cosmic distances. These different measurements can show discrepancies due to the unique ways {in which} they interact with the contents and structure of the universe. Combining these various measurements provides a more clear understanding of dark energy and its equation of state. By comparing the luminosity distances from GWs and EM phenomena like supernovae, we can refine their models of how dark energy influences the  acceleration of the universe. This comparison is also crucial for addressing the Hubble tension. By comparing distances measured from GWs with those from EM observations, unique tests of GR can be conducted. Additionally, measuring the time delay between these signals provides an opportunity to infer the mass of gravitons and the speed of GWs, offering valuable insights into the fundamental laws of gravity.

\item \texttt{GW}: GW observations provide a unique tool for independently determining distances to cosmic events, a method known as standard sirens. This approach is essential for addressing the Hubble tension and could potentially reveal whether new physics is required to explain these discrepancies. Additionally, GW astronomy plays a significant role in investigating the equation of state of dark energy. By analyzing how the expansion of the universe affects the propagation of GWs, insights can be gained into the nature of dark energy, which drives the acceleration of cosmic expansion. Furthermore, GW data allow for the testing of non-GR signatures by examining the strain observed in GW detectors.  By placing constraints on these parameters, GW astronomy can challenge and refine modified theories of gravity, providing a stringent test of GR under extreme conditions.

\end{itemize}

\subsubsection{Future Prospects}

The future of GW astronomy is set to be transformative with the development of more sensitive detectors and the launch of space-based observatories. The evolution of ground-based GW observatories, such as LIGO \citep{aasi2015advanced}, Virgo \citep{acernese2014advanced}, and KAGRA \citep{akutsu2021overview}, along with upcoming advancements like LIGO-Aundha \citep{Unnikrishnan:2013qwa,saleem2021science}, CE \citep{reitze2019cosmic}, and ET \citep{punturo2010einstein}, has significantly enhanced our ability to detect these waves, primarily focusing on the hertz to kilo-hertz frequency range. The upcoming space-based observatory, LISA \cite{amaro2017laser}, is set to revolutionize our understanding of GWs by observing massive binary black holes over unprecedented cosmological distances, exploring the millihertz frequency range. This ability to probe such diverse astrophysical phenomena is crucial because ground-based detectors are limited by unavoidable noise artifacts, preventing them from accessing this lower frequency domain. The combination of advancements in both ground-based and space-based detectors promises enhanced sensitivity and deeper exploration of the cosmos. These different frequency-band explorations will also facilitate multi-band searches, allowing for the detection of GWs from a variety of events such as mergers involving intermediate-mass black holes, and potentially uncovering new astrophysical phenomena.

In Figure \ref{fig:LVKFZ}, we illustrate the measurements of the frictional term for bright standard sirens using our current and upcoming ground-based detectors. This figure demonstrates how the new detectors will help us test GR with percent to sub-percent level accuracy. Figure \ref{fig:LISAFZ} showcases the precision with which the upcoming space-based LISA observatory can reconstruct the redshift evolution of the Planck mass, denoted by $\mathcal{F}(z)$. Together, these plots illustrate the complementary capabilities of ground-based and space-based detectors in measuring the frictional term for bright standard sirens.

In Figures \ref{fig:DarkSirenLVK} and \ref{fig:DarkSirenCEET}, the reconstruction of the non-GR parameter $\mathcal{F}(z)$ {for dark sirens} across cosmic redshifts is illustrated for different GW detector systems and galaxy surveys. Figure \ref{fig:DarkSirenLVK} presents results for the LVK system combined with the DESI and \ac{LSST} galaxy surveys, focusing on BBH mergers up to $z = 0.675$. This plot demonstrates that DESI provides more accurate measurements than LSST, particularly at lower redshifts. Figure \ref{fig:DarkSirenCEET} extends this analysis to future detectors, CE and ET, in conjunction with DESI and LSST, for BBH mergers up to $z = 2.5$. It shows that DESI is superior at low redshifts, while LSST offers better precision at higher redshifts, highlighting the complementary strengths of these surveys. Together, these figures illustrate the improved accuracy in measuring $\mathcal{F}(z)$ for dark sirens when combining GW detectors with galaxy surveys, with DESI and LSST providing complementary capabilities across different redshift ranges.

Figure \ref{fig:BrightSirenH0} displays the posterior distribution of the Hubble constant, $\mathrm{H_0}$, for all bright siren events up to a redshift of 2, utilizing the LVK, CE, and ET detectors. The analysis includes both BNS and NSBH mergers. 
Figure \ref{fig:LISAH0} displays the posterior distribution of the Hubble constant, $\mathrm{H_0}$, for all events up to a redshift of 6, utilizing the upcoming space-based LISA observatory. In Figure \ref{fig:DarkSirenH0}, we illustrate the posterior distribution of $\rm{H_0}$, obtained from the analysis of all sources using the LVK detectors with galaxy surveys DESI (LVK+DESI) and LSST (LVK+LSST), as well as the CE and ET detectors with DESI (CEET+DESI) and LSST (CEET+LSST) surveys. These measurements are expected to provide significant insights into the Hubble tension. The upcoming detectors, CE, ET, and LISA, will cover a wide range of cosmic redshifts, from low-redshift to the deep high-redshift universe. By observing various types of sources across this extensive redshift range, these observatories will offer a comprehensive view of the cosmic expansion. 

Several theoretical models predict that gravitons could exhibit mass in frequency bands other than those currently accessible to the LVK collaboration. Future GW observatories, which will operate in different frequency ranges, will provide opportunities to test these models. As these new observational windows open, we anticipate further refinements in the constraints on the graviton mass, or potentially the discovery of a nonzero mass for the graviton. Such a discovery would have profound implications for our understanding of gravity and fundamental physics. These next-generation detectors will provide more precise measurements, facilitating a detailed exploration of cosmological models. This will not only aid in testing GR and its alternatives but will also deepen our understanding of fundamental cosmological parameters, offering new insights into the nature of gravity and dark energy. Moreover, the integration of GW detectors with projects like DESI, LSST, and Euclid \cite{Amendola:2016saw} for BAO measurements will substantially enhance the precision testing of GR, the dark energy equation of state, and the Hubble tension. This synergy between different observational platforms and methodologies will allow for a more thorough examination of the expansion rate of the universe, potentially resolving inconsistencies observed with current methods.

\subsection{Dark Matter Searches}
\subsubsection{Introduction}
There is overwhelming evidence of the presence of dark matter over all length scales in our Universe. Over the past several decades,  \ac{WIMPs} have generally been considered the leading class of candidates for the dark matter of our Universe~\cite{Leszek2018}. To detect and understand the particle nature of dark matter, several experiments have been designed and carried out to detect the interactions of dark matter particles with atoms (direct detection), to produce particles of dark matter in collider environments, and to detect the products of dark matter annihilations or decays (indirect detection). The direct search of dark matter is not discussed in the white paper as it is outside the scope of multi-messenger astrophysics.

\subsubsection{Current Status}
Indirect searches for dark matter include efforts to detect the gamma rays, antiprotons, positrons, neutrinos, and other particles that are produced in the annihilations or decays of this substance~\cite{Gaskins2016}. The abundance of dark matter that emerged from the early universe is set by its self-annihilation cross-section. It can be demonstrated that a stable particle species with a thermally averaged annihilation cross-section of
$\sigma_{v}$ $\sim$ 3 $\times$ 10$^{-26}$ cm$^{3}$/s {would} freeze out of thermal equilibrium with an abundance equal to the measured cosmological density of dark matter.  In many models, the dark matter is predicted to annihilate with a
similar cross-section in the current universe, thus making indirect detection searches very appealing. In particular, gamma-ray and cosmic-ray searches for dark matter annihilation products have recently become sensitive to dark matter masses of around 100 GeV and higher. 

Searches for dark matter using gamma-ray telescopes benefit from the fact that these particles are not deflected by magnetic fields and are negligibly attenuated over galactic distance scales, making it possible to acquire both spectral and spatial information. Using a combination of space-based and ground-based gamma-ray telescopes, strong constraints on the annihilation cross-sections of dark matter masses up to a few hundred GeV have been obtained over the last few years~\cite{McDaniel2024,MAGIC2016}. In the future, these constraints are expected to improve due to (i) the growing data set from Fermi-LAT (and future satellite-based and ground-based gamma-ray telescopes) and (ii) the discovery of new ultra-faint dwarf galaxies that are expected from LSST and
other surveys. 

On the other hand, the cosmic-ray spectrum is dominated by protons and nuclei. 
It has long been appreciated that if dark matter particles are annihilating or decaying in the halo of the Milky Way, such processes would produce equal amounts of matter and antimatter, leading to an excess of antimatter relative to that predicted by standard astrophysical mechanisms. In this sense, searches for cosmic-ray antimatter, such as antiprotons, positrons, or anti-deuterons become very important ingredients for dark matter searches. 
In 2008, the PAMELA satellite reported  that the cosmic-ray positron fraction (the ratio of positrons to positrons-plus-electrons) rises between approximately 10 GeV and 100
GeV~\cite{PAMELA2008}. This rise is in stark contrast to the behavior expected for a positron spectrum dominated by secondary particles produced during cosmic-ray propagation. Within this context, the possibility that annihilating dark matter might be responsible for this signal generated a great deal of interest in the community, although it was also pointed out that nearby pulsars or the acceleration of secondary positrons in supernova remnants could potentially account for the excess positrons. This data was later combined with data from the AMS-02 experiment which hinted at the presence of a dark matter particle at TeV energies~\cite{Heros2020}. 

\subsubsection{Future Prospects}
In the future, we expect indirect searches for dark matter to be bolstered by a range of new experiments and
observations. The CTA is an array of ground-based gamma-ray telescopes that will be soon 
entering into the construction phase in 2025
and will be offering unprecedented sensitivity to the very high-energy gamma-ray sky above 100 GeV. At lower gamma-ray energies, several proposals are designed to be significantly more sensitive than Fermi at MeV-GeV energies.  Searches for gamma rays from dark matter annihilation in dwarf galaxies will be further enhanced by {LSST}, which is expected to discover many new dwarf spheroidals. Additionally, future measurements of AMS-02 \cite{PhysRevLett.110.141102} and the GAPS balloon experiment \cite{2020NIMPA.95862201O}, in particular in regards to their search for anti-deuterium and anti-helium in the cosmic-ray spectrum, will provide us with valuable clues to the dark matter searches.

Apart from these probes, neutron stars can be potential astronomical objects to accrete dark matter in abundance at their cores. In case the accreted amount of dark matter in the interior of the star is substantially large, it will affect the stellar structural properties, e.g. maximum mass, stellar radius, and tidal deformability parameters, as compared to the standard neutron stars without their presence~\cite{PhysRevD.88.123505, PhysRevD.110.043006, PhysRevLett.122.071102}. These properties can then be inferred and constrained with GW observations~\cite{McKeen2018,PhysRevD.109.023021, Husain2022a,Shirke2023b,Shirke2024} with upcoming second- and third-generation ground-based detectors.

\subsection{Cosmology of Supermassive Black Hole with Multi-Messenger Astronomy}

\subsubsection{Introduction}

\ac{SMBHs} are extremely massive objects in the universe, with masses millions to billions of times greater than the Sun. These black holes reside in the centers of most of the massive galaxies, including our Milky Way, and play a pivotal role in the evolution of their host galaxy. Despite their importance, their formation and evolution remain an open question in cosmology. Multi-messenger study of these objects, which combines information from GWs, EM waves, and other probes, can offer a powerful approach to understanding these processes. 


Detecting \ac{nHz} GWs from \ac{SMBHBs} via \ac{PTAs} \citep{verbiest2022pulsar} offers a promising method to study SMBH formation over cosmic time \citep{sesana2013insights,ravi2015prospects}. The detection of individual inspiraling binaries, coupled with the identification of their host galaxies through EM counterparts, can break the mass-distance degeneracy as well as help us understand the environment of the SMBHBs.  Periodic variability in an AGN is a potential indicator of the presence of SMBHBs, as the binary is expected to periodically perturb the circumbinary disk, leading to periodic flares \citep{burke2019astrophysics,dey2018authenticating}. If both SMBHs are active, their combined influence could produce a distinct periodic light curve \cite{o2022unanticipated}. However, the detection of an individual binary will only be possible for low-redshift sources. The imprint of the evolution of the SMBHB population with cosmic time will be encoded on the GW background.

The GW background generated by the superposition of GWs from individual SMBHBs is expected to be anisotropic. Moreover, since these binaries reside in the centers of galaxies, the distribution of the anisotropic signal is expected to follow the galaxy distribution in the universe. Therefore, the study of the angular power spectrum of the \ac{SGWB} and its cross-correlation with galaxy distribution presents a promising avenue for investigating the cosmic evolution of SMBHs.

\begin{figure}
    \centering
    \includegraphics[width=\textwidth]{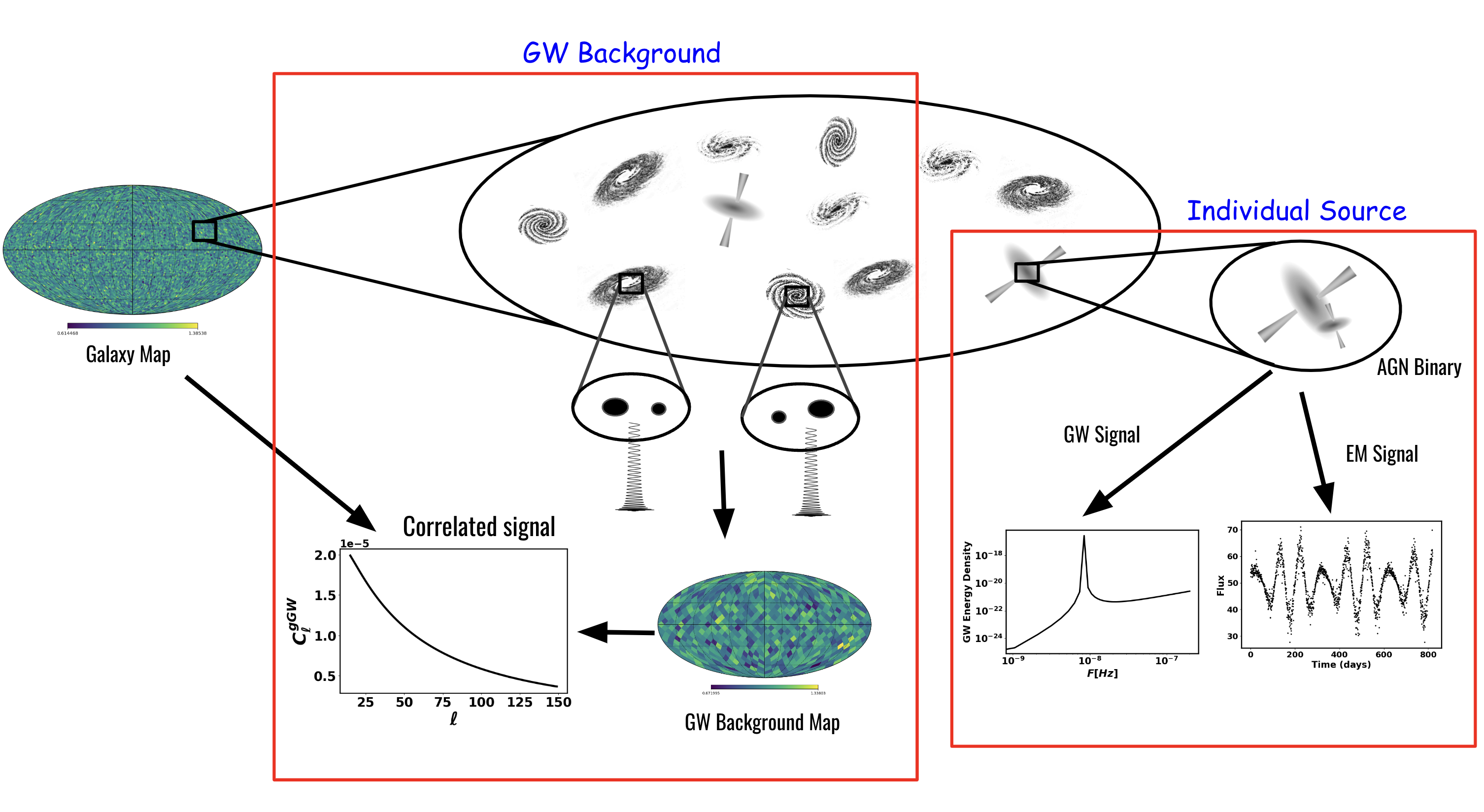}
    \caption{A schematic diagram illustrating the multi-messenger study of SMBHBs. The SGWB generated by SMBHBs, which reside at the centers of galaxies, is expected to be spatially correlated with the large-scale galaxy distribution in the Universe. Furthermore, the detection of GW  from individual SMBHBs may be accompanied by an EM counterpart if at least one of the BHs in the binary is active.}
    \label{Motv}
\end{figure}

Of particular interest is the cross-correlation between nHz GWs and galaxy surveys, which promises to unlock new insights into the redshift evolution of SMBHBs \citep{sah2024discovering}. The SGWB generated by unresolved SMBHBs in the nHz frequency range provides a unique probe into the coalescence history and spatial distribution of SMBHs across cosmic time. By correlating this background with the large-scale galaxy distribution, we can trace the formation and growth of SMBHs and their host galaxies, offering a comprehensive understanding of their evolution. A schematic diagram summarizing the multi-messenger study of SMBHs is shown in Fig. \ref{Motv}.

Complementing the nHz observations from PTAs, {LISA} will play a crucial role in the multi-messenger study of SMBHs by detecting GWs from binaries with masses between $10^{5} M_{\odot}$ and $10^{7}  M_{\odot}$ in the millihertz (mHz) range \citep{mangiagli2022massive}. LISA sensitivity to these intermediate-mass BHs and lower-mass SMBHs will bridge the observational gap between PTAs and ground-based detectors, allowing us to explore a wider range of SMBH populations.

\subsubsection{Current Status}

Recent advances in PTA observations, including results from the \ac{NANOGrav} \citep{mclaughlin2013north}, \ac{EPTA} \citep{desvignes2016high}, \ac{InPTA} \citep{joshi2018precision}, \ac{PPTA} \citep{manchester2013parkes}, and \ac{CPTA} \citep{xu2023searching} have provided compelling evidence for the presence of an SGWB in the nHz range of GWs. The most promising source of this SGWB is believed to be the population of SMBHB. However, the current PTA configuration is not sensitive enough to provide any valuable information about the anisotropic distribution of the SGWB signal.

\begin{figure}
    \centering
    \includegraphics[width=12cm]{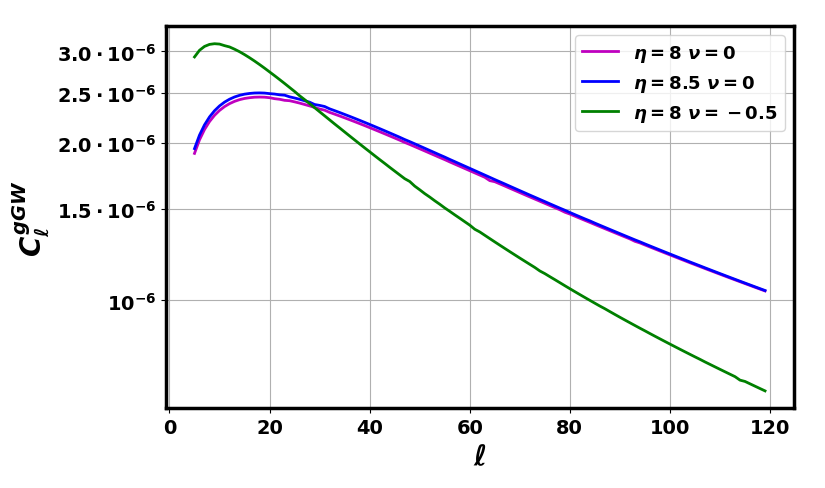}
    \caption{Theoretical angular cross-correlation between SGWB and the galaxy distribution ($C_{\ell}^{\rm gGW}$) for LSST-like surveys for different stellar mass and SMBH mass relation and its evolution.}
    \label{ClgGW_L}
\end{figure}

Despite this, significant progress has been made in developing methodologies to detect and interpret the anisotropies in the SGWB \citep{sato2023exploring,gardiner2024beyond,sah2024imprints}. The work done in \cite{sah2024discovering} demonstrates techniques to cross-correlate the nHz SGWB signal and the spatial distribution of galaxies from surveys like Rubin LSST \citep{ivezic2019lsst}. These studies have shown that the angular power spectrum of the SGWB  retains the imprint of the evolution of the SMBHB population with cosmic time and can be correlated with galaxy catalogs to infer the formation and evolution of the SMBHs (see Fig. \ref{ClgGW_L}).

\subsubsection{Future Prospects}

The future of studying SMBH evolution through multi-messenger astronomy is promising, especially with upcoming facilities such as the {SKA} \citep{smits2009pulsar} and the availability of the LSST and other surveys. The sensitivity of the SKA to nHz GWs, combined with extensive galaxy catalogs, will provide an unprecedented opportunity to study the cross-correlation signal between the SGWB and galaxy distributions. 

This approach will enable us to probe the evolution of SMBH across a wide range of redshifts, significantly beyond what is currently possible. In particular, the SKA is expected to monitor thousands of millisecond pulsars, increasing the angular resolution of the SGWB map and allowing for more detailed cross-correlation studies. This will enable us to disentangle the complex relationship between SMBHs and their host galaxies, including how SMBH masses and growth rates evolve with cosmic time. 

In summary, the integration of GW with traditional EM observations as galaxy catalogs in the next decade will transform our understanding of SMBHs. By studying the cross-correlation between the SGWB and galaxy distributions, we will gain new insights into the formation and evolution of SMBHs, paving the way for a deeper understanding of the high-redshift universe and the role of SMBHs in shaping the universe. In addition to the SKA, the LISA will play a critical role in the study of the lower-mass SMBH population.

\newpage

\noindent\textbf{\Large{Long-term science goals:}}

\subsection{Cosmic Neutrino Background}
\subsubsection{Introduction}
Neutrinos had decoupled from the matter-photon fluid in the first seconds of the Big Bang (when the temperature of the universe was around $10^9$ K) \cite{akita2020precision,hannestad1995neutrino} and serve as the oldest cosmic messengers. These neutrinos have been traveling unobstructed due to their low interaction cross-section. These neutrinos contributed to the radiation at the earlier cosmic epoch when they were relativistic, and contribute to the matter density of the universe at the current epoch. The temperature of these neutrinos is estimated to be about a bit lower compared to the CMB at about 1.95 K \cite{Dodelson:2003ft}. These neutrinos can give us information about the gravitational and magnetic field fluctuations in the early universe. Moreover, the EM and gravitational effects can be used to look for warm dark matter candidates in the form of sterile neutrinos \cite{Raffelt:1996wa}. 

\subsubsection{Current Status}
The {cosmic} neutrinos rarely interact and currently have low energies, hence are very difficult to detect. Many of these neutrinos are non-relativistic, which further reduces their detectability. The direct detection experiments are mainly ground-based and rely on the interaction of neutrinos with Tritium or other elements \cite{Dodelson:2003ft}.  Indirect evidence of these primordial neutrinos comes from their effect on the large-scale structure, Big Bang nucleosynthesis, anisotropies in the CMB, etc. The tightest bounds on the number of effective neutrino species at the time of decoupling is about 3.6 (up to 95$\%$ confidence interval) from Planck \cite{ade2016planck}.

\subsubsection{Future Prospects}
Neutrino astronomy has started establishing itself as one of the frontiers in science and detection of the CNB is a difficult task, but can bring 
deep insights into early Universe physics. PTOLEMY \cite{PTOLEMY:2018jst} is one such proposed experiment that may detect it. Signatures of the primordial neutrinos on various other probes (EM, GW, and CRs) currently serve as the best way to study them. Thus cooperation between various sectors is required, if we are to be able to study the early universe. Neutrino astronomy still has a long way to go and will provide information that is not yet accessible to us.

\newpage

\section{Fundamental physics}
\label{fundamental-physics}

\small \emph{Contributors: Soumya Bhattacharya, Debanjan Bose, Indranil Chakraborty, Debarati Chatterjee, Arpan Hait, Md Emanuel Hoque, Subhendra Mohanty, Arunava Mukherjee, Dhruv Pathak, Swarnim Shirke}\\ \normalsize
\small \emph{Editors: Varsha Chitnis, Sanjeev Dhurandhar, Amol Dighe, Suvodip Mukherjee,  Tirthankar Roy Choudhury, Shriharsh P. Tendulkar} \\ \normalsize

\noindent\textbf{\Large{Short-term science goals:}}

\subsection{Extreme Matter Physics of Neutron Stars with Multi-Messenger Astronomy}
\subsubsection{Introduction}
The direct detection of GW in 2015 from mergers of binary compact objects~\cite{GW150914} revolutionized astronomy by opening a new window to the Universe. Further, the detection of GW signal from the binary neutron stars merger event GW170817 in 2017~\cite{GW170817}, along with its counterparts across the EM spectrum ushered in a new era of multi-messenger astronomy~\cite{GW170817_MMA, 2017ApJ_GW170817GRB}. These events underlined the importance of the synergy between the global array of GW detectors and the conventional ground-based and space-based EM telescopes, which allows us to collect a wealth of information about compact stars. Astronomical observations of this event have already been able to shed light on several areas of fundamental physics as well as astrophysical processes~\cite{eos_gw170817, 2019PRL_GW170817TGR, 2017Natur_GW170817_H0, 2020PRL_StandardSiren}. Specifically, it has been able to put significant constraints on the neutron star equation of state~\cite{2018PRL_SoumiDe_etal,eos_gw170817,Forbes:2019xaz,Annala:2017llu}.

\begin{figure}
    \centering
    \includegraphics[width=0.5\textwidth]{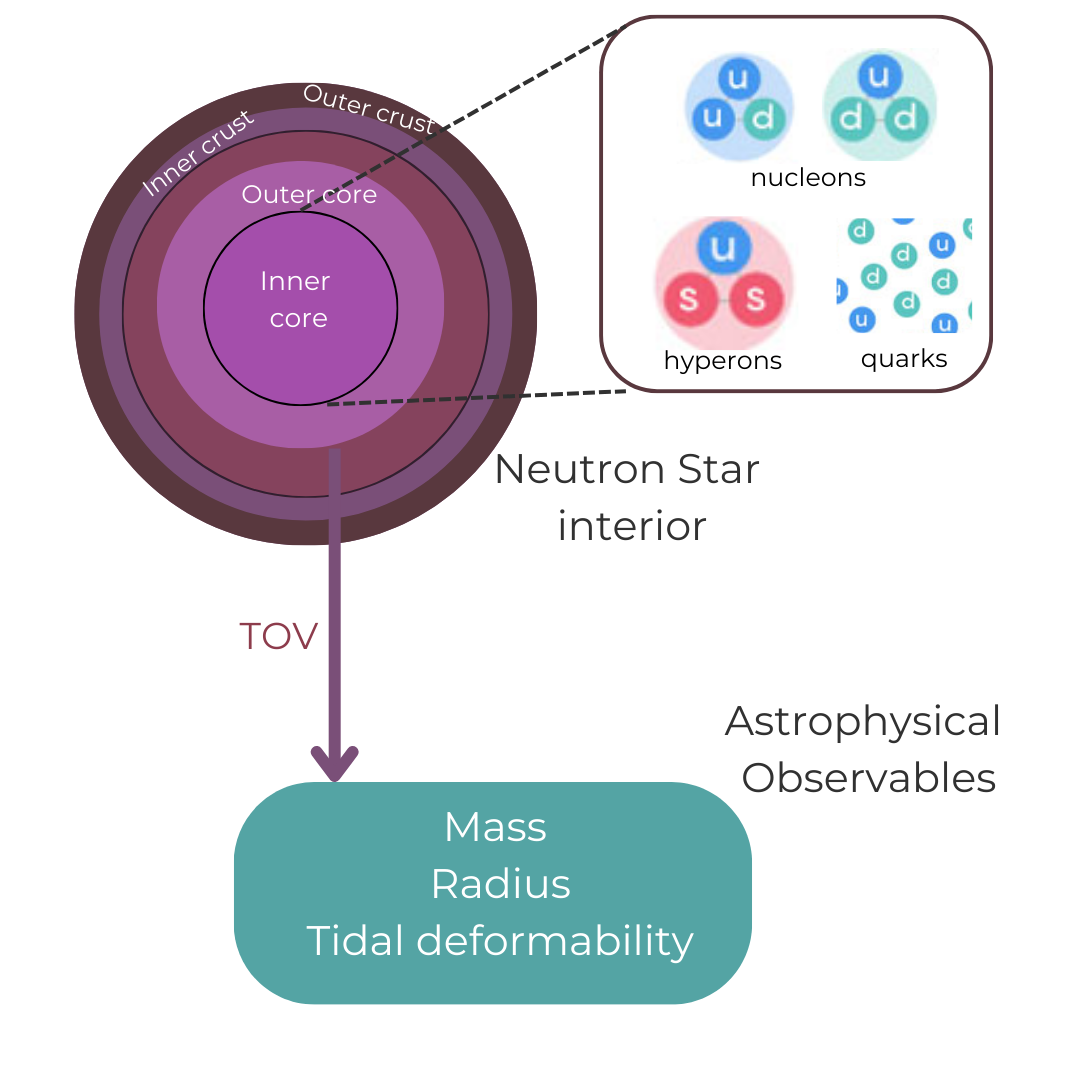}
    \caption{Schematic figure connecting neutron star interior composition, {EoS} and
astrophysical observables.}
    \label{fig:Deba-1}
\end{figure}

NSs provide clues about the unknown behavior of matter under extreme conditions of density, temperature, and magnetic fields, far from the conditions accessible to terrestrial experiments, such as nuclear laboratories and heavy-ion experiments in particle accelerators. Therefore, a description of the internal composition of neutron stars requires theoretical modeling to extrapolate the current knowledge derived from experiments to unknown regimes. 
{The equation of state}
encodes the information about the interior composition and connects it with global neutron star observables, such as its mass and radius, that can be calculated by solving the \ac{TOV}  hydrostatic equilibrium equations. The model predictions are then compared against astrophysical data, to constrain the parameter space of uncertainties in the models (see Fig.~\ref{fig:Deba-1}).

Currently, NSs are observed with the help of a large number of space-based and ground-based telescopes (e.g. Fermi, INTEGRAL, Chandra, VLT, VLA, NICER) across the EM spectrum, from gamma-rays and X-rays to radio frequencies. From the multi-wavelength data, one can measure the structure properties of neutron stars, such as their mass, radius, moment of inertia, or compactness. However, there are uncertainties in the astrophysical measurements that limit their accuracy and do not allow us to impose stringent constraints on the theoretical models.

\begin{figure}
    \centering
    \includegraphics[width=0.5\textwidth]{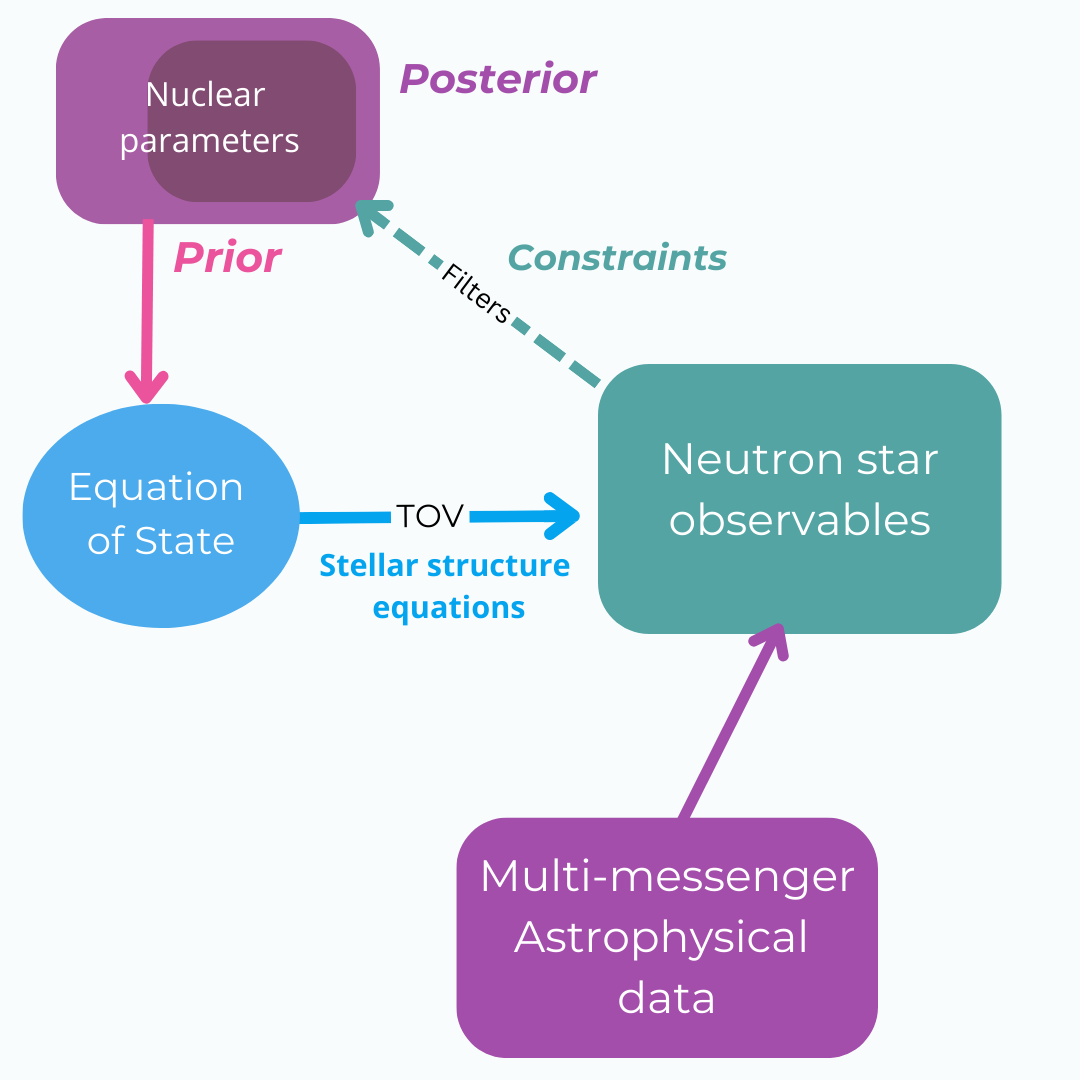}
    \caption{Flowchart demonstrating how nuclear parameters can be constrained using neutron star multi-messenger astrophysical data within a Bayesian scheme.}
    \label{fig:Deba-2}
\end{figure}

\subsubsection{Current Status}
Simultaneous measurements of the mass and radius of a neutron star are crucial in constraining its equation of state. The TOV equations map a given equation of state to a unique curve in the mass-radius plane. {Inversely, a precise mass-radius measurement can pin down the microscopic equation of state.} Similarly, neutron stars in the BNS (binary neutron star) system get tidally deformed by the external gravitational field of the companion star~\cite{Hinderer_APJ2008}. 

\textbf{X-ray timing measurement of neutron stars: } The NICER measurements of pulsars PSR J0030+0451 and PSR J0740+6620 have measured their radii down to a precision of 10\% (see~\cite{NICER0030_Miller,NICER0740_Miller,NICER0030_Riley,NICER0740_Riley}), leading to new constraints on the \ac{EoS}. A recent observation of a compact supernova remnant HESS J1731-347~\cite{Doroshenko2022} has reported a low mass and low radius hinting towards the possibility of exotic matter in neutron stars.

\textbf{GW signal from coalescing BNS:} GW170817, a GW event of a BNS merger, was the first joint detection of gravitational and EM radiation from a single source. A long GW signal was observed along with observations of short GRB (GRB170817) and an optical transient (AT2017gfo) in the host galaxy NGC 4993. The source was detected across the EM spectrum, in the X-ray, ultraviolet, optical, infrared, and radio bands, over hours, days, and weeks~\cite{Abbott2017b}.

The binary neutron star merger event GW170817 provided a novel method to constrain the neutron star EoS. A typical neutron star merger comprises three phases: an inspiral, the coalescence, and the post-merger (long/short-lived) remnant which may collapse into a black hole eventually, depending on the remnant mass, spin, and the equation of state. During the inspiral phase, the tidal deformability of one neutron star due to the strong gravitational field of the other is sensitive to the EoS. The detection of the tidal deformation from the event GW170817 therefore provided important constraints on the EoS and restricted the possible range of the neutron star radii~\cite{Abbott2018}.

Information from the GW event was combined in subsequent analyses with data from EM observations (e.g. kilonova AT2017gfo, GRB170817 and NICER measurements~\cite{GW170817_MMA}) to impose even further restrictions on the dense matter EoS~\cite{Raaijmakers2021}. Several studies also showed that neutron star multi-messenger astrophysical data at high densities can be combined with multi-disciplinary physics at other densities, e.g. with microscopic nuclear theory calculations (Chiral Effective Field theory) at low densities, heavy-ion data at intermediate densities, or perturbative QCD at very high densities, within a Bayesian scheme to improve our understanding of the neutron star EoS (see~\cite{Huth2021, GhoshEPJA2022, Shirke2023}). The scheme is demonstrated in Fig.~\ref{fig:Deba-2}, with the help of a flowchart.

\textbf{\ac{CGW} signal from spinning neutron stars:} Rapidly spinning neutron stars with non-axisymmetric deformations~\cite{1998PhRvD_JKS, 2023LRR_KeithRiles} or oscillating isolated NSs can produce continuous (i.e., persistent) nearly-monochromatic gravitational waves. Although only a few sources are currently within the sensitivity of the current generation of detectors, targeted and directed searches are being performed for possible candidates. We have not yet observed any CGW signals with the GW detectors. The intrinsic strengths of such signals are expected to be orders of magnitude weaker than those originating from compact binary merger events, regularly observed with the LIGO and Virgo detectors to date. Accreting neutron stars in \ac{LMXB} systems are an excellent class of objects to harbor such highly spinning, deformed neutron stars~\cite{2015PRD_LeaciPrix, 2018PhRvD_MMR, 2023PhRvD_BinaryWeave}. Even a non-detection of continuous GWs imposes limits on the maximum possible deformation in NSs~\cite{Riles2023}. Nevertheless, many recent studies have been dedicated to looking for signatures of the neutron star's internal composition on unstable oscillation modes and GW emission (e.g.~\cite{Pradhan2022,Pradhan2023, Tran2023}).

\subsubsection{Required Synergy between Messengers}

\begin{itemize}
\item {\bf EM:} Synergies between gravitational waves and EM observations are critical to extracting significant benefits in this area of research. For the follow-up of EM counterparts of GW triggers, a lot of effort has been invested internationally. {Capturing} the nature of EM signals, for example, {detecting} the spectrum and chemical abundances in the kilonova events corresponding to a BNS event with good SNR (signal-to-noise-ratio), can shed light on the r-process nucleosynthesis and productions of heavy elements in our universe. 
Hence, the Indian Multi-messenger community needs to come together for EM observations in case of the next BNS merger event. The uGMRT can be used for radio signal detections~\cite{ugmrt1, GMRT}, X-ray Polarimeter Satellite can be used for X-ray observations~\cite{xposat1,xposat2}, and Daksha should be used for the detection of associated GRB and other EM counterparts~\cite{dakshascience2024}. They can also select specific targets (such as potential host galaxies) and look for remnant EM observations.

Additionally, it will be highly rewarding to have coordinated X-ray and radio observations for several rapidly-spinning neutron stars/millisecond pulsars for the direct detection of a CGW signal (details are discussed in~\ref{coord_obs_cw_em}). Since the intrinsic amplitudes of the CGW signals are much weaker, it needs the observer to integrate the signals for a much longer observational duration~\cite{2015PRD_LeaciPrix}. However, a thorough search over a wide range of parameter space for any interesting source becomes computationally prohibitive while integrating the signal over a longer duration \cite{PhysRevD.61.082001}. Thus, a coordinated effort of planned X-ray and radio observations of a set of high-priority targets will be immensely beneficial to drastically narrow down the relevant parameter space that needs to be searched. This will significantly improve the scope of direct detection of CGW signal for current and future generation detectors~\cite{2023PhRvD_BinaryWeave}. 

\item {\bf GW:} In 2023, the LVK Collaboration published the  GWTC-3~\cite{GWTC3}, which summarises cumulative GW signal detections with LIGO and Virgo up to the end of their third observing run. In the three observing runs, 90 candidates for compact binary coalescences were identified, of which 83 were black hole mergers, five were confirmed to involve neutron stars (two binary neutron star mergers and three neutron star/black hole mergers), and two were identified as a merger between a black hole and another uncertain compact object (either an unusually massive neutron star or an unusually light black hole). Thus far, EM emission has been observed only in association with one event (GW170817). The LVK's fourth observing run (O4) started on 24 May 2023. Improvements in the sensitivity of the LVK interferometers allow us to explore a wider volume of the universe, with a large increase in the detection rate with respect to the O3 run. Simulations predict the detection of several EM counterparts to neutron star mergers, with certain input assumptions for merger rates and neutron star mass distributions (see e.g.~\cite{Colombo2022}). In May 2023, the Indian government gave the final nod to the construction of a GW detector on Indian soil.  Joining the global array of GW detectors, the LIGO-India project will improve the landscape of GW astronomy by virtue of its cutting-edge technology and its global position leading to a better localization of GW sources. LIGO-India is expected to complete construction and begin its science run by 2030 and join the global array of five detectors, possibly after the fifth observation run of LVK. 

Coordination between GW signals (both transient and persistent/continuous) and EM astronomy is critical to pursue simultaneous/near-simultaneous observations of multi-messenger signals involving neutron stars, either in a binary system or in isolation. Gravitational waves will generally provide an important and reliable estimate of certain source properties, gravitational masses, tidal deformability, moment-of-inertia, quasi-normal modes of fluid oscillations, etc., that are associated with coherent bulk-motion of the mass density or mass current density in the system. This signal, if suitably coupled and complemented with relevant information detectable in EM observations, {would yield} more comprehensive and effective constraints {on the} neutron star structure, composition, and equation of state. {These observations can also} potentially be applied to study cosmology (see the discussion in Sec. \ref{cosmology}). 

\item {\bf Neutrino:} The supernova SN1987A was the first observation of a neutrino pulse from a collapsing star that lasted for around 10 seconds. The measurement of neutrinos from supernovae put important bounds on the cooling mechanism via other exotic channels~\cite{RaffeltSeckel1988, FornalGrinstein2018, Husain2023}. Such supernova observations also hold the key to the formation mechanism of compact objects like neutron stars. As the neutron star interior is optically opaque, neutrinos provide an excellent alternative to probe the interior matter, {similar to the role played by the GWs.}  

BNS merger is also associated with the production of a large amount of MeV neutrinos since during the merger process, the beta-equilibrium condition gets disturbed. Depending on the final mass and spin of the merged object it can either form a black hole with prompt or delayed collapse, or a stable neutron star to live for a long duration. In the case of a stable neutron star as the final remnant object, it is expected that neutrino cooling will be an important component of the astrophysical process. However, these MeV neutrinos may not have enough flux to be detectable for a source distance of several 100 Mpc distances with the present generation detectors. Perhaps a more details theoretical analysis would be useful to explore any other observational scenario. 

{Plans of the erstwhile} proposed India-based Neutrino Observatory {also} included studies of astrophysical neutrinos, e.g. from supernovae~\cite{IndianNeutrinoObservatory}. {Such a future neutrino observatory in India}  would play a vital role in the Indian efforts of multi-messenger constraints on compact objects. A detector with sensitivity to MeV - GeV range neutrinos can further constrain DM models from indirect detections of DM~\cite{RaffeltSeckel1988}. Such constraints would be critical in answering questions concerning the admixture of DM in NSs~\cite{McKeen2018,Shirke2023b, Shirke2024}.

\item {\bf CRs:} The amount of different species of CRs produced in binary neutron star mergers and supernova explosions will be relatively small in amount to be able to detect on earth, for a majority of astrophysical sources. Therefore a follow-up  CR signal of a GW event is not guaranteed. However, in a rare case, we might be able to observe CRs coincident and/or (near-)simultaneous  with GW events.

\end{itemize}

\subsubsection{Future Prospects}

{Multi-messenger observations of events involving neutron stars} is an active area of research, particularly after the coveted simultaneous detection of GW and EM signals from the GW170817 event. Over the last few years, substantial progress has been made to constrain neutron star EOSs using multi-messenger observations of colliding neutron star sources both on the theoretical front~\cite{2020NatAs_Capano_etal, 2020Sci_Dietrich_etal}, and in the observational and data analysis sector~\cite{2024arXiv_McGinn_etal, 2024arXiv_Wouters_etal}. However, considering the complexity and richness of the problems, it requires additional inputs of theoretical calculations complemented by suitable numerical relativity simulations to provide more accurate descriptions of a large number of astronomically observable quantities. In the coming years, numerous scientific literature is expected to be devoted to these investigations, along with dedicated observational efforts that will be pursued internationally to detect and study these signals. 

As the current generation LVK detectors of GW near the end of their fourth run of observations, many more GW detectors are on the horizon. Several other future-generation detectors such as the CE in the US, the ET in Europe, and the space-based LISA are planned over the next decade. A dedicated high-frequency GW interferometer called  Neutron Star Extreme Matter Observatory \cite{Ackley:2020atn}, {designed to be sensitive to the high-frequency part of the GW spectrum,} to measure the fundamental properties of nuclear matter at extreme densities with GWs has also been proposed. A proposal for an Indian space-based GW interferometer at complementary frequencies is also being considered. With the advancement of space technologies, lunar GW missions such as the Lunar GW Antenna \cite{Ajith:2024-LGWA}, and Gravitational-wave Lunar Observatory for Cosmology (GLOC) \cite{Jani:2020gnz} are also being proposed.

Other disciplines in fundamental physics are also advancing steadily. The Compressed Baryonic Matter (\ac{CBM}) experiment \cite{Senger:2002xp} is being developed at the future \ac{FAIR} in Darmstadt, Germany to study the behavior of dense matter at 4-5 times nuclear densities. Several hyper-nuclear experiments have been proposed at future heavy-ion facilities to probe the physics of strange matter relevant to neutron star interior composition. Experiments such as PREX \cite{Mammei:2012uu} and CREX \cite{Kumar:2023bmb} aim to measure neutron skin thickness and other properties of neutron-rich nuclei relevant to neutron stars. On the other hand, rapid progress in theoretical techniques such as chiral perturbation theory and lattice QCD are providing additional information about the behavior of matter at extreme conditions. Information from future advances in theoretical calculations and experiments can be combined with future multi-messenger data to impose important constraints on the nuclear EoS and therefore better understand the behavior of dense nuclear matter. 

As the sensitivity of the ground-based detectors will affirmatively improve in the next few years to a couple of decades, we expect multiple simultaneous observations of gravitational waves and EM emissions from the same astronomical events, possibly accompanied by other observational channels, e.g., neutrinos and CRs. This area of astronomy and astrophysics has a reach potential to enhance our scientific knowledge on probing the strong nuclear interactions in dense and extreme environments of strong field gravity and high magnetic field regime. It will also be effective as a natural testbed of general relativity and several other modified theories of gravity (see the previous section  \ref{cosmology}).

\vspace{1cm}
\noindent\textbf{\Large{Long-term science goal:}}

\subsection{Gravitational Memory Signal from Self-Interacting Neutrinos in SNe}
\subsubsection{Introduction}

Large self-interactions in neutrinos due to their interaction with light scalars or vectors of mass $M_\phi \simeq 10$ MeV have been proposed to alleviate the Hubble tension and also to modify cosmological structure formation from the standard $\Lambda$CDM. Moreover,   
neutrino self-interaction has been recently proposed to increase the duration of the neutrino burst. In \cite{Bhattacharya:2023wzl}, the authors have shown that self-interactions in supernova neutrinos can produce distinct changes in the gravitational waveform accompanying the anisotropic neutrino burst in the supernova, which in principle can be detected from planned future detectors such as the Big Bang Observatory (BBO) \cite{4453737} and DECIGO \cite{Kawamura:2020pcg}. The detectable self-interaction scale for these GW signals is higher than the weak interaction scale.

\subsubsection{Current Status}

The issue of self-interaction among supernova neutrinos and their possible observational implications remains an open question. In \cite{Chang:2022}, it was demonstrated that significant self-interactions lead to an increased duration of the neutrino burst. Subsequent works, such as \cite{Fiorillo:2023-PRL} and \cite{Fiorillo:2023-PRD}, have shown that neutrinos propagate as a pressure wave with sonic velocity near the proto-neutron star surface, eventually reaching luminal velocity after traversing a newly defined length scale called the free-streaming radius. However, it was argued that the pressure wave would not significantly alter the duration of the neutrino burst. In \cite{Bhattacharya:2023wzl}, the authors contend that GW features due to the substantial self-interaction of neutrinos, as proposed in \cite{Fiorillo:2023-PRL} and \cite{Fiorillo:2023-PRD}, could potentially be observed. These GW bursts typically manifest as memory signals \cite{Mukhopadhyay:2021}. This work is part of broader efforts to detect memory signals, which have also been predicted to arise from binary mergers \cite{Hait:2022ukn, Mitman:2024} and have intriguing theoretical connections to low-energy physics \cite{Strominger:2014}. 
\begin{figure}
    \centering
    \includegraphics[width=.8\textwidth]{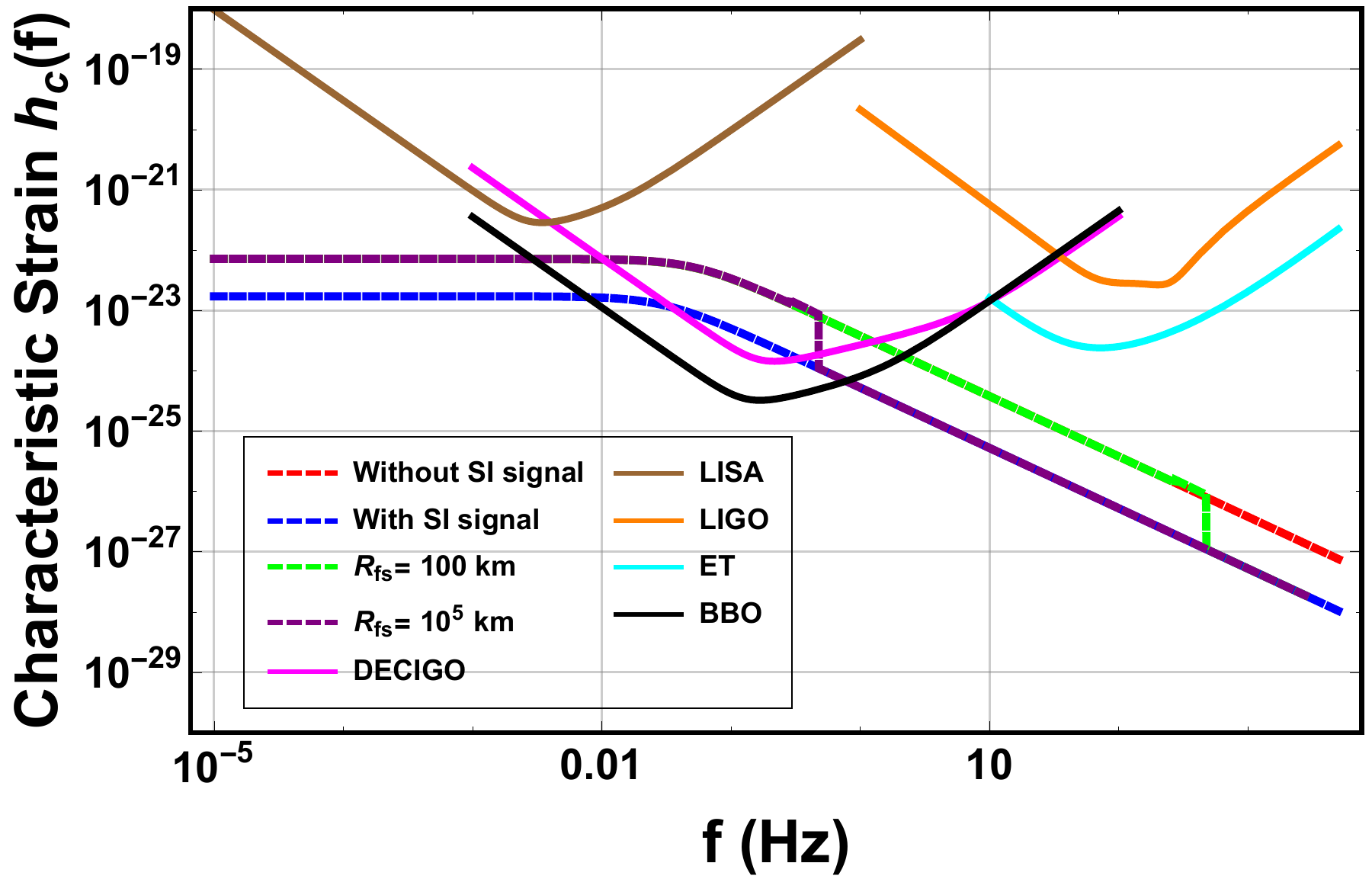}
    \caption{Characteristic strain for the memory signal is plotted along with the characteristic noise of the detectors. The plots show that only DECIGO and BBO can detect the memory signal. The transition occurring in $R_{fs}=10^5$ km can also be detected by BBO and DECIGO. For lower values of $R_{fs}$, {\em i.e.} smaller diffusion region, the transition occurs at a higher frequency value which is beyond the sensitivity of the current and future detectors.}
    \label{fig:slefintneutrino}
\end{figure}

\subsubsection{Required Synergy between Messengers}

\begin{itemize}
    
\item {\bf GW:} The signal predicted in \cite{Bhattacharya:2023wzl} will be in the deci-hertz band and hence detectors like BBO, DECIGO, and LGWA will be useful in observing this signal \cite{Ajith:2024-LGWA}. It will be interesting to see how gravitational waveforms for more realistic models of anisotropic neutrino emission incorporating the effects of self-interaction would behave. Phenomenological works involving no self-interaction have shown that memory signals from $\sim$ 1 Mpc  can be detected \cite{Mukhopadhyay:2021}.

\item {\bf Neutrino:} The GW wave signals coming from SN neutrino bursts present in the deci-hertz band 
could potentially give pointers for extragalactic supernova emissions.  In the kpc range, simultaneous observations from the neutrino signal can be observed at SK, and at HK {once it is operational}. Thus, knowing the total neutrino luminosity from the neutrino data and contrasting it with GW data for the neutrino memory signal, one can disentangle the neutrino memory contribution from the matter memory one, which occurs at higher frequencies.

\item {\bf EM:} The EM signal is delayed compared to the joint signal of the GW and neutrino bursts.  Thus, the joint signal can act as an alert for EM follow-ups. {Given} the source distance, EM astronomy can provide better source localization compared to GW detectors. This could in turn help in understanding the properties of the host galaxy and supernova explosion. 

\end{itemize}

\subsubsection{Future Prospects}
In \cite{Bhattacharya:2023wzl}, the authors have assumed constant anisotropy. However, realistic supernova simulation models have shown non-trivial time dependence of the anisotropy parameter. Thus, from these studies, one can understand the nature of anisotropy driving the neutrino burst. Moreover, it has been shown recently in \cite{Nagakura:2019} and \cite{Nagakura:2024} that asymmetric neutrino-assisted {proto-neutron star} kicks can get amplified using fast oscillations. This would produce memory signals. Accurately predicting the anisotropy might give clues to a better understanding of these flavor oscillations. 

\newpage

\section{Synergy With Other Research Areas}
\small \emph{Contributors: Debarati Chatterjee, Sourendu Gupta, Nilmani Mathur, Rishi Sharma}\\ \normalsize
\small \emph{Editors: Varsha Chitnis, Amol Dighe, Suvodip Mukherjee, Tirthankar Roy Choudhury, Shriharsh P. Tendulkar} \\ \normalsize

\subsubsection{Synergy with Nuclear Physics}
The Mega-Science vision document for Nuclear Physics \cite{MSVN-2035} recommends (1) continued Indian participation in collider studies of the phase diagram of strongly interacting matter and (2) development of new accelerator facilities in India for \ac{RIBs} and continued participation in similar international facilities. Both these goals are of relevance to multi-messenger astrophysics.

GW170817 is a typical example of this. The pre-merger dynamics of neutron stars give information about nuclear densities in the range of 1--6 times normal nuclear densities. This is exactly the range of densities that are to be probed by the {CBM}  experiment at GSI. The study of such dense matter is one of the core interests in nuclear physics. These will yield an increasing amount of information about the nature of nuclear matter and enable more detailed checks of gravitational waves from such objects.

Several recent works have tried to impose constraints on the allowed range of uncertainties in nuclear and hypernuclear empirical parameters using multi-disciplinary physics (nuclear theory, heavy-ion data, multi-messenger astrophysical observations) at different densities within a Bayesian approach~\cite{Huth2021, GhoshEPJA2022,Ghosh_2022_Frontiers,Gorda:2023usm}. These studies have been further combined with nuclear experimental data such as measurements of nuclear skin~\cite{Biswas_2020,Biswas_2021,Mondal_2023} which are in apparent tension with astrophysical data.  Further, some studies also demonstrated how the future detection of gravitational waves emitted by unstable oscillation modes can be used to constrain the nuclear equation of state and nuclear parameters effectively~\cite{Pradhan2023,Pradhan_2024_cost,Iacovelli_2023}. Tidal interactions during neutron star mergers or unstable oscillation modes may also be searched for potential signatures of strangeness~\cite{Ghosh_2024_tidal,Tran2023}. Nuclear experiments in the future, e.g. precise measurements of neutron skin thickness as well as hypernuclear experiments, may be able to provide necessary information on baryon-baryon interactions that are crucial inputs for neutron star equation of state models. {Using }combined information from the ongoing STAR/RHIC and HADES/GSI experiments, the future FAIR and NICA facilities becoming operational in the near future, and improved multi-messenger observations, the high-energy nuclear physics and astrophysics communities are expected to set stringent constraints on the equation of state of strongly interacting matter~\cite{Dexheimer_2021}.  

Optical and IR observations of the resulting kilonova are currently being interpreted in terms of the reaction network that follows from rapid neutron capture dynamics. The physics of these processes are validated at RIBs. As observations of post-merger ejecta become more precise and numerous, theoretical inputs validated by observations at RIBs will become more precise.

\subsubsection{Synergy with Lattice QCD}
The comprehensive study of neutrino-nucleus scattering is a key for uncovering various crucial aspects of nuclear and astrophysics, namely,
(i) determining neutrino masses and flavor mixing, 
(ii) investigating BSM interactions between neutrinos and ordinary matter 
(iii) understanding nuclear structures. 
One of the primary challenges in neutrino-nucleus scattering experiments is accurately reconstructing the incident neutrino energy, particularly since nuclear remnants are not directly detected. Achieving precision in these experiments relies on modeling nuclei, which presents a complex problem spanning a wide range of energies.

Presently, discrepancies between event generators used in
neutrino-nucleus scattering experiments and experimental data often necessitate adjustments through ad hoc modeling. However, accurately reproducing one process does not guarantee precise predictions for other processes or energies, highlighting the limitations of such
approaches.

Indeed, the challenge of unknown input parameters for constructing a consistent many-body theory can introduce unpredictability to the results. Numerical computations using first-principles Lattice QCD offers a unique solution in this endeavor. Lattice QCD calculations have the potential to provide the necessary input parameters, often leveraging their capability to
systematically improve precision with increased computing
resources. In recent years, exciting developments in Lattice QCD
calculations have emerged, particularly in various form factors of
hadrons. These include advancements in determining nucleon axial and tensor charges, parton distribution functions, and their generalizations, along with studying the properties of light
nuclei. Such progress highlights the significance of Lattice QCD in advancing our understanding of nuclear and particle physics. With the advent of exascale computing, it is anticipated that calculations in this field will become significantly more precise. This anticipated enhanced precision has the potential to constrain the parameter space of event generators used in neutrino-nucleus scattering experiments, leading to more precise scientific results.

Additionally, Lattice QCD calculations are crucial not only for light nuclei but also for providing input to construct consistent many-body effective field theories for heavier nuclei. NPLQCD~\cite{Wagman:2017tmp} and the HAL QCD~\cite{Inoue_2016,Inoue_2017,Inoue_2021} are two examples of approaches used to study few-baryon systems in Lattice QCD. In the HAL QCD approach, baryon–baryon interaction potentials can be extracted from Lattice QCD. Using the extracted potentials, properties of many-baryon systems, e.g. nuclei, hypernuclei, or infinite baryonic matter in the neutron star core~\cite{Tolos_2020}, can be obtained with many-body techniques. 

Lattice QCD also provides direct first-principles calculations of the QCD phase diagram in certain regimes which can be useful in informing us about the properties of dense baryonic matter, even though the regime of large ($\sim 1$ GeV) baryon chemical potentials and low temperatures ($\lesssim 20$ MeV) is currently not directly accessible using lattice techniques due to the sign problem. For example, calculations of the equation of state (for eg. see Ref.~\cite{Datta:2016ukp}) at temperatures of about $150$ MeV and small chemical potentials ($\lesssim 300$ MeV) can validate effective theories which are also applicable at lower temperatures and higher chemical potentials that are interesting for neutron star mergers. Similarly, neutron star matter features matter at high baryon chemical potential as well as iso-spin chemical potential, a regime difficult to access in lattice calculations. However, lattice calculations with only iso-spin chemical potentials have been performed~\cite{Bali:2016nqn,Brandt:2017oyy,Brandt:2022hwy} and can be used to constrain theories of neutron star matter. As mentioned above, once the few-nucleon interactions are known, quantum many-body techniques~\cite{Hebeler:2015hla,Tews:2024owl} can be used  to predict the thermodynamical properties of matter, including its equation of state properties of hadronic matter at densities higher than nuclear saturation densities. 

To advance in this direction, a coordinated effort involving
high-energy theory, lattice QCD, nuclear many-body theory, and event generators are necessary. This collaborative approach will facilitate progress in understanding neutrino-nucleus scattering and its implications for nuclear physics and astrophysics. Indian lattice gauge theorists in collaboration with other theorists can play a crucial role
in this program.


\chapter{Available and Required Facilities for Multi-Messenger Science}

\small \emph{Contributors: Dipankar Bhattacharya, Subir Bhattacharyya, Varun Bhalerao, Chinmay Borwanker, Ishwara Chandra C. H., Bitan Ghosal, Pratik Majumdar, Arunava Mukherjee, Suvodip Mukherjee, Krishna Kumar Singh, Gaurav Waratkar, Kuldeep Yadav}\\ \normalsize
\small \emph{Editors: Varsha Chitnis, Amol Dighe, Suvodip Mukherjee, Tirthankar Roy Choudhury, Shriharsh P. Tendulkar} \\ \normalsize


\section{Multi-Messenger Astrophysics with High Energy Neutrinos and Photons}
\subsubsection{Introduction}
The origin of high-energy CRs is one of the most outstanding problems in high-energy astroparticle physics. As described in the earlier section, high-energy gamma rays and neutrinos can serve as effective probes for sources of high-energy CRs. Over the last two decades, high energy and very high energy gamma-ray astrophysics have delivered a wealth of data and have discovered hundreds of high energy gamma-ray sources~\cite{Bose2022}. The first detection of high-energy extragalactic neutrinos by the IceCube neutrino observatory in 2013 opened a new window to the high-energy Universe. However, identification of a neutrino source {remained elusive}  till the observation of high energy neutrinos from a blazar TXS0506+056 in 2018, {followed by its} observations in the EM bands~\cite{Icecube:2018}. {Subsequent to} this observation, several tentative detections and associations with blazars have been reported~\cite{Giommi2020}\cite{Padovani2022}\cite{Prince2024}. 
{Along with the above-mentioned extragalactic sources, recently IceCube Observatory has detected high-energy neutrinos that originate within the Milky Way from possible astrophysical sources such as SNe \cite{IceCube:2023ame}.}


\subsubsection{Available telescopes/resources}

The real-time follow-up of neutrino events is a promising approach to search for astrophysical neutrino sources. It has already provided compelling evidence for a point source: the flaring gamma-ray blazar TXS 0506+056 observed in coincidence with the high-energy neutrino IceCube-170922A detected by IceCube. The detection of very-high-energy gamma rays (VHE, E $>$ 100 GeV) from this source helped establish the coincidence and constrained the modeling of the blazar emission at the time of the IceCube event. The major \ac{IACTs} around the world --- H.E.S.S., MAGIC, and VERITAS --- operate a vigorous and active follow-up program of \ac{ToO} observations of neutrino alerts sent by IceCube. It will be extremely important to coordinate these observations with the Indian gamma-ray telescope MACE located at Hanle along with the upcoming array of CTA which is slowly entering its construction phase. 

{The secondary electron-positron pairs produced from charged pion decays along with neutrinos and gamma rays produced from neutral pion decays, initiate EM cascade interactions inside the jet. The energies of the secondaries are redistributed from higher to lower energies in EM cascade interactions and typically may fall in the X-ray energy band (0.1 to 100 keV) and also MeV energies. Herein lies the importance of having very good X-ray observations over a broad energy range. The Indian multiwavelength satellite ASTROSAT which has several X-ray detectors on board over a very wide range of energy, is perfectly well-suited to carry out such multi-messenger observations on AGNs in association with neutrinos.}

{IceCube provides two alert channels: A "Gold" channel will issue alerts for neutrino candidates at least 50\% likely to be of astrophysical origin and is expected to deliver $\sim$ 10 alerts per year. Additionally, a more frequent "Bronze" channel will provide $\sim$ 20 alerts per year for neutrino candidates that are between 30\% and 50\% likely to be of astrophysical origin~\cite{Blaufuss2019, Abbasi2023}.
Once high-energy neutrinos are detected by IceCube, the information is disseminated through publicly issued alerts to the wider community, especially for follow-up observations by EM detectors and telescopes.
Simultaneous observations of these potential sources of neutrinos with ASTROSAT, and with very high-energy gamma-ray telescopes will be a valuable addition to the program and will be crucial in unraveling the nature of high-energy emission in the sources.}

{Clearly, optical follow-up observations of neutrino events will yield a significant discovery potential and will also be useful for future improvements in our understanding of highest-energy astrophysical sources. They are important not only for blazars but also for supernovae and TDEs. As recently reported by ZTF, a high-energy neutrino was identified in spatial correlation with the supernova SN AT2019pqh. An automated trigger system SuperNova Early Warning System for neutrino observations is available \cite{SNEWS:2021ewj}.}
{Regular follow-ups of neutrino events using the optical telescopes available in the country are being pursued in India, at HCT and 1.3 mt telescope at ARIES. All these need to be done in a more coordinated manner along with the observatories available outside India. In the future, it will be extremely important to carry out multimessenger observations with IceCube-Gen2 and CTA, in combination with the Indian observatories.}

\subsubsection{Requirement for synergies between different observatories}

There exist promising synergies among very high-energy messengers such as CRs, gamma rays, and neutrinos as spatial and temporal correlation of high-energy gamma rays and neutrinos can provide clues to the identification of the sources of high-energy CRs. The intensity of gamma rays, neutrinos, and UHECRs {has been shown to be comparable (see Figure \ref{fig:HENG})}, suggesting that they may be powered by the same sources. As blazars are the most abundant extragalactic sources, they are by default the most promising candidates. However, blazars do not seem to fit the bill as they are subdominant in the high-energy neutrino flux, and we may need to look at other candidates such as starburst galaxies and TDEs.

It has been recently claimed that the diffuse gamma-ray flux detected by Fermi-LAT is actually dominated by star-forming galaxies. At the same time, the data collected in the Pierre Auger Observatory has shown an indication of anisotropy in the arrival direction of UHECRs, showing an excess from the direction of nearby starburst galaxies, claiming that star-forming galaxies can be the cause of $\sim$ 10\% of the ultra-high-energy CR flux. It has been also shown that the contribution of the starburst galaxies to the neutrino diffuse flux is sub-dominant and constrained to be at the level of $\sim$ 10\%. Therefore, it remains unanswered for now whether there is a dominant type of source accelerating all cosmic messengers. Hence the need for detailed multi-messenger observations and follow-up analysis programs between different observatories. 

Finally, in our own galaxy, it has been shown that a sub-PeV diffuse galactic gamma-ray emission exists and the galactic neutrino contribution should constitute roughly 5–10\% of the IceCube diffuse flux. Also, in the 10–100 TeV range, the expected galactic neutrino flux should be comparable to the total neutrino diffuse flux. If so, the next-generation neutrino telescopes should be sensitive enough to detect it. It is possible that part of the measured gamma-ray diffuse flux,
for example, the {very-high-energy} sources detected by HAWC and LHAASO, comes from individual sources. This can be further confirmed by combined multi-messenger observations.

\subsubsection{Trigger/ Alerts}

The real-time alerts (track-like events) from IceCube are distributed as \ac{GCN} Notices. 
These can then be followed up by the EM telescopes.
For the public alerts, approximately 10 Gold alerts per year are issued and about 20 Bronze alerts per year are issued. 
Moreover, IceCube also sends out alerts of neutrino flares (called multiplets) from a list of pre-selected sources once a predefined detection threshold is reached.
This threshold is set based on the theoretical modeling of the source using available multi-wavelength information. Once the threshold is crossed, an alert will be sent out via GCN. 

The data from all-sky detectors such as Fermi-LAT is immediately publicly available. However, the data from ground-based Cerenkov telescopes is propitiatory in nature and can be accessed through effective collaboration.


\subsubsection{Recommendations}
Follow-up observations of all Gold alerts after applying a suitable angular resolution cut --- the angular resolution of track-like events varies from a few arc minutes to a couple of degrees --- by gamma-ray telescopes (MACE, MAGIC, etc.), X-ray observatory (ASTROSAT, Swift, XMM-Newton etc.) and optical telescopes (HCT, telescopes at ARIES etc.) is imperative. It is extremely important to form a consortium and then derive a mechanism to have a dialog with the IceCube partners regarding several interesting alerts that may be taken up for future joint publications. It would be further useful for the neutrino observatories to coordinate with GW events to send out alerts on a regular basis even for sub-threshold events.  


\section{Multi-Messenger Observations of Binary Compact Objects}
\subsubsection{Introduction}
The multi-messenger observations of the GW sources such as the coalescing binary compact objects open a discovery space in broad research areas, which include astrophysics, cosmology, and fundamental physics as described in the previous chapter. However, to achieve these science goals, coordinated observations of the GW sources with EM, CRs, and neutrino observations will be crucial. The multi-messenger observation and the optimized strategy depend on the nature of GW sources and in which frequency band of GW they are observed. 

At the very high-frequency end of the GW spectrum (above 10 Hz), a rapid follow-up within a few seconds in the gamma-ray band to a few tens of hours to days time scale in the radio is required for EM-bright sources such as BNS and NSBH \cite{KAGRA:2013rdx}. As described in previous chapters, these sources are also going to have neutrino and cosmic-ray emission, and observation of these sources will be important to understand the internal structure of the neutron star. Also, for the milli-hertz GW signal accessible from 
LISA \cite{Baker:2019nct}, the sources can be seen for a time scale of hours to years depending on the masses of the BBHs. The EM signal from these sources is expected across a broad range of spectrum in EM, from X-ray to Radio, due to the interactions of the baryonic matter around the SMBHs. The EM counterpart can be present from hours before mergers to days after mergers. On the extreme low-frequency GW signal in the nHz range, signals from coalescing SMBHs are going to contribute to the stochastic background as well as a few well-detected events which are bright and nearby \cite{burke2019astrophysics, sah2024discovering}. 

These sources are likely to be embedded at the center of the galaxies that have recently gone through mergers. These sources can also have AGN discs around them \cite{Chen:2023xrm, Bogdanovic:2021aav, Baker:2019nct}. A potential follow-up of these sources through EM telescopes will make it possible to identify the host galaxies and study their properties and morphology. These are also potential sources of CRs and neutrinos. So a joint detection of three messengers from a source will be extremely valuable to learn the physics of the source.

\subsubsection{Available telescopes/resources} 

The existing and upcoming Indian facilities that can contribute to the multi-messenger observations span the areas of GW, EM, and CRs. We describe below the salient aspects of these facilities and the bands that can be covered for these messengers.

\begin{itemize}

\item {\bf GW detectors:} The two frequency bands of GWs that can be explored by the existing/upcoming detectors are the high-frequency band from LIGO-India detector \cite{Unnikrishnan:2013qwa,saleem2021science} to be constructed at Aundha and the low-frequency band from GMRT \cite{ugmrt1} to search for the nHz signal. It is likely that LIGO-Aundha will join the network of other GW detectors LIGO-Hanford, LIGO-Livingston, Virgo, and KAGRA from 2030 onwards.  The operation of LIGO-Aundha jointly with other GW detectors will make significant advancements in improving the sky-localization of the GW sources. As a result, the multi-messenger observation in the follow-up to a GW source will be easier when LIGO-Aundha joins the network. On the nHz band of GW signal, upgraded GMRT plays a pivotal role in accessing the low-frequency radio band as a part of InPTA \cite{joshi2018precision} which in collaboration with another network of radio antennas such as Nanograv\cite{NANOGrav:2023gor}, EPTA\cite{EPTA:2023sfo}, PPTA \cite{Reardon:2023gzh} helps in detecting the signal. In the future, SKA  \cite{Janssen:2014dka, ChandraJoshi:2022etw}, of which India is part, will be able to access this signal with higher statistical significance and make discoveries in both astrophysics and cosmology frontier of nHz GW signal. 

\item \textbf{EM Telescopes: } \\
\begin{itemize}
\item \textbf{MACE: } MACE (Major Atmospheric Cherenkov Experiment) is a third-generation large-size IACT for very high energy gamma-ray observations from the Earth's surface. It pushes the frontier of ground-based gamma-ray astronomy in India \cite{Singh2021, Singh2022}, following the idea of high altitude IACTs with low threshold energy. Situated at an altitude of $\sim$ 4.3 km above mean sea level, MACE has the distinction of being the highest among the present and future IACTs in the world. The geographical location of MACE at Hanle (32$^\circ$ 46$^\prime$ 46$^{\prime\prime}$N, 78$^\circ$ 58$^\prime$ 35$^{\prime\prime}$E) in the Ladakh region \cite{Yadav2022}, appropriately reduces an important longitudinal gap among the major instruments around the globe. It is equipped with a 21 m diameter quasi-parabolic optical reflector 
with an f-number of $\sim$ 1.2 and has an alt-azimuth mount. Such a large light collecting surface has been achieved by using a tessellated structure comprising 1424 small square-shaped spherical metallic mirror facets of size 0.488 m$\times$0.488 m each having varying focal lengths between 25.0 m and 26.25 m \cite{Dhar2022}. This offers a total light collection area of ~339 m$^2$ and an effective focal length of $\sim$25 m. The imaging camera of the telescope, placed at the focal plane, consists of an array of 1088 photomultiplier tubes with a uniform pixel resolution of 0.125$^\circ$. It provides a total optical field of view of $\sim~ 4.36^\circ \times 4.03^\circ$. 
Moreover, the telescope has been designed for detecting the gamma-ray photons of lower energies close to 20 GeV and above with very high point source flux sensitivity. This can be achieved by employing advanced machine algorithms for more effective background rejection at energies below 100 GeV. The energy resolution is estimated as $\sim 34\%$ at 100 GeV which further improves to $\sim 20\%$ at 1 TeV. 

Regular science observations with MACE have been ongoing since 2021. The measured 50-hour integral sensitivity of MACE is $\sim 9.6\%$ of the Crab Nebula flux in the energy range above 80 GeV \cite{Borwankar2024}. It is in good agreement with the predictions from the Monte-Carlo simulation studies using the CORSIKA software package. 
Thus, the MACE telescope, as a national facility, is fully capable of playing a pivotal role in exploring the highest energy end of the EM spectrum in the current era of multi-messenger astronomy.     

\item \textbf{AstroSat: } AstroSat is an observatory-class space mission operated by the Indian Space Research Organisation. Launched in 2015, the mission has been active for nearly a decade in the area of Ultraviolet and X-ray Astronomy \cite{AstroSat:2017Agrawal}.  With four co-pointed instruments --  \ac{UVIT} (130-550 nm), \ac{SXT} (0.3-8 keV), \ac{LAXPC} (3-80 keV) and \ac{CZTI} (20-150 keV) -- the mission provides a simultaneous broad spectral coverage with fast timing capability, highly valuable for variable sources \cite{AstroSat:2014Singh+}.  In addition, the CZTI instrument acts as an open all-sky detector in the 100-500keV band and is able to detect many high-energy rapid transients such as GRBs even if the satellite is not pointed at them \cite{CZTI:2017Bhalerao+}.  The CZTI also has the ability to measure polarization above 100 keV for bright sources.  For the study of important transients \cite{CZTIPol:2015Vadawale+}, AstroSat can trigger ToO pointed observations with a typical response time of a few days.  Additionally, CZTI data can be examined for possible detection of high-energy counterparts coincident with the original event. Such searches are being routinely carried out for many announced transients, including all the detections reported by the LVK collaboration. The AstroSat policy allows high-priority ToO execution for the follow-up of EM counterparts of GW sources, with a reduced turn-around time.

\item  \textbf{GROWTH-India: } GROWTH (Global Relay of Observatories Watching Transients Happen) -India \cite{Kumar:2022svq} is a 0.7 m diameter robotic telescope situated at Hanle, Ladakh near the MACE telescope about 4.5 km above the mean sea level. It is operational in the optical band (it has the ugriz filters) of the EM spectrum and is capable of performing follow-up observations of GW sources. It is a part of the international GROWTH network \cite{Kasliwal:2019pyx}. The telescope has a maximum slew rate of 50 degrees/sec, a pointing accuracy of 10 arc-seconds, and a tracking accuracy of 1 arc-second for an exposure of about 1 second. The sensitivity of the telescope in the g and r band is typically up to 20.5-21 magnitude for ten minutes of exposure and has a field of view of about 0.8 degrees. The geographical advantage of the telescope and a good seeing quality from the site along with these telescope specifications make it capable of performing fast follow-up of the transients in response to the alerts systems for transients such as GCNs. This telescope, in coordination with the GROWTH network, is expected to detect the optical counterpart expected from BNS and NSBH events in the low redshift Universe probed by the network of GW detectors.

\item \textbf{uGMRT: }
The GMRT (Giant Metrewave Radio Telescope) \cite{1991CSci...60...95S} consists of thirty 45 m diameter antennas spread over a 25 km region, about 80 km north of Pune, India. Among these, 14 antennas are in a compact region of about 1 km area called the Central Square. The rest of the antennas are on the 3 arms of a “Y” array. Each arm of the array is about 14 km in length and has 5 to 6 antennas. The longest baseline is about 25 km and the shortest is about 200 m. The observing bands of the \ac{uGMRT} \cite{2017CSci..113..707G} since 2019 are 120-250 MHz (Band-2), 250-500 MHz (Band-3), 550-850 MHz (Band-4) and 1000-1460 MHz (Band-5). For the data acquisition, the GMRT Wideband Backend (GWB) is available with bandwidths of 400 MHz, 200 MHz, or 100 MHz in a mode with 2048, 4096, 8192, or 16384 channels. Both continuum (imaging and spectral line) and beamformer modes for pulsar observations with high time resolutions are possible.

The observing time with {uGMRT} is through proposal submission, in two cycles per year with deadlines on January 15 and July 15 each year.
The proposals are reviewed by the GMRT time allocation committee (GTAC).  ToO proposals can also be submitted to the GTAC, but any time allotted will be scheduled dynamically, as and when the specified trigger conditions are met.  Additionally,  there is a Director’s Discretionary Time (DDT) option, which is meant for special uses, such as (i) pilot observations/feasibility studies that might lead to future GTAC proposals, (ii) urgent confirmatory observations, (iii) ToO observations, primarily intended for short-lived or time-dependent astronomical phenomena.   Interferometric data from all standard observations is available to the public 18 months after the date of observation.  For all ToO / DDT observations, the corresponding period is 3 months. A \href{https://github.com/ruta-k/CAPTURE-CASA6}{CASA \cite{casa}-based Pipeline-cum-Toolkit} is now available for producing continuum images from the uGMRT.

The main advantage of uGMRT is the time delay. The transients generally peak in the radio window with a delay which makes it logistically easier to organise the observations. For uGMRT, at low radio frequencies, the delay is even longer, hence a larger time window to plan the observations. The observations can be done in either or both the continuum imaging and/or the time series
mode. This is useful for fast time-varying sources such as pulsars or Fast Radio Bursts, as well as for continuum sources such as GRBs. The angular resolution of GMRT at the highest band (band-5) is about 2 arc-seconds, which will be very useful to localize the sources. Radio sources as weak as a few tens of $\mu$Jy can be detected in a few hours of observation in Band-4 and Band-5 in imaging mode.

\end{itemize}
\end{itemize}
\subsubsection{Requirement for synergies between different observatories} 

Among the different kinds of GW signals, the requirement for the synergies with telescopes will be different for short and long-duration EM  signals. For the short-duration EM signals, a rapid follow-up of the GW sources will be required over different EM frequency bands starting from gamma-ray to radio. As a result, an automated follow-up technique for the GW triggers from LVK GRACEDB \cite{gracedb} by MACE, AstroSat, and GMRT will be essential for candidates such as BNS, NSBH, and BBHs. Furthermore, sub-threshold GW signals, which are not well-detected by the GW detectors due to the larger distance to the source, can have a detectable EM counterpart in gamma rays. As a result, follow-up of the sub-threshold events in gamma rays will be useful. To have a successful campaign of multi-messenger observations, both loud events, and sub-threshold events need to be followed up with gamma-ray telescopes and followed by telescopes in other EM bands in a coordinated way. Any observed transient signal in any of the messengers (GW, EM, neutrino, CRs) should immediately send alerts along with the sky position, magnitude, and light curve, if available. 

For a long-duration signal such as in radio, a dedicated follow-up with the uGMRT observation can be made if a transient signal is observed in the GW and gamma rays. The  detection of EM signal in radio is usually after a few hours after GW and gamma rays and can stay for years. Joint exploration of these signals will play a key role in exploring the science cases discussed in the previous chapters.

\subsubsection{Information/Data sharing:} The information about the GW super events is publicly available in GRACEDB, and GCN notices are announced for these events. For these sources, a limited amount of information on the source properties such as distance, sky localization, and  whether it includes a neutron star or not. These events need to be followed up by the EM telescopes as soon as a trigger is announced. If EM telescopes in the gamma-ray band can provide a sky map and magnitude of any transient signal along with the time-stamp, then a follow-up of it in low-frequency EM bands will be useful. 

\subsubsection{Triggers/alerts: } To follow-up signals from coalescing compact objects through multi-messenger probes, it is important to have automated alert systems such as GCN not only for the loud GW events but also for loud and sub-threshold events seen in gamma-ray which can be follow-up by other multi-messenger probes. An automatic alert system with the necessary information of the signal such as sky position, energy flux (or luminosity distance), and arrival time of the signal communicated as soon it is detected in GW, gamma-ray, and neutrino to other observatories will be useful for searches of the signals even if they are weak.   

\subsubsection{Roadmap: } 

A rapid follow-up system for super-threshold and sub-threshold GW events by the EM observatories in the country such as MACE, AstroSat, GROWTH, and uGMRT with the sharing of essential data products such as sky position, time of arrival of the signal, energy flux (or luminosity distance) for identification and classification of a multi-messenger signal. Along with these future ground and space-based missions from India on gamma-ray and X-ray will be useful for multi-messenger science. We briefly describe below a few promising Indian facilities that can benefit multi-messenger science. 

The stereoscopic MACE system, comprising two additional MACE-like telescopes, is proposed for very high-energy gamma-ray observations with better flux sensitivity and improved energy and angular resolutions. Enhancement in the flux sensitivity will help in the detection of faint gamma-ray sources whereas improvement in the angular resolution offers detection/discovery of point sources in galaxy clusters. An energy resolution of better than 15$\%$ is important in the exact spectral measurement of the gamma-ray emission. Installation and commissioning of the stereo MACE will be completed over a period of 7 years. 

A proposed space-based high-energy transient mission Daksha \cite{dakshatech2024, dakshascience2024} with a primary objective of studying EM counterparts of GW sources and GRBs. 
Daksha has been designed to be the most sensitive high-energy transient mission, with higher volumetric coverage than any current or proposed mission.
Daksha will consist of two satellites covering the entire sky from 1~keV to 1~MeV and will have a higher sensitivity than any other current or proposed international missions. On-board algorithms will detect, localize, and broadcast important data of these bursts for further rapid follow-up observations. 

When LIGO-Aundha joins the international GW detector network, Daksha can detect up to 8 joint short GRB and GW events every year. Daksha can also provide key insights into the X-ray counterparts of BBH mergers by probing a much fainter luminosity space than any other mission \citep{waratkar2024}. Further, Daksha is expected to detect 500 long GRBs and 50 short GRBs every year, with a large number of poorly understood high-redshift GRBs \citep{dakshascience2024}. The wide-band coverage of Daksha including all--sky soft X-ray detectors will create the first opportunity ever to obtain truly broad-band spectra, which can be used to accurately distinguish various components of prompt emission. These observational inputs can plugged into theoretical models to understand source physics including details like the neutron star Equation of State \citep{singhal2022}. The high sensitivity will enable finely time-resolved prompt emission studies, again giving an unprecedented window into source physics. Moreover, Daksha can measure the polarization of about five GRBs per year thanks to the open structure of the payload, large collecting area, and varied detector orientation \citep{dakshapol2023}. 

Over the next few years,  the SKA Observatory which is a next-generation radio in which India is playing a major role \cite{2023JApA...44...27G}. In the era of SKA, multi-messenger science in synergy with GW will be feasible with far better sensitivity, and possible synergy with five hecto-Hz GW observatories (LIGO-Hanford, LIGO-Livingston, Virgo, KAGRA, and LIGO-Aundha) and milli-hertz GW observatory (LISA) will be feasible \cite{Chandra:2016rvb}. Furthermore, SKA will make a paradigm shift in the nHz GW astronomy \cite{Janssen:2014dka}.    

\section{Multi-Messenger Observations of Continuous GW Signals}\label{coord_obs_cw_em}

\subsubsection{Introduction}

Spinning neutron stars are expected to emit {CGW} at various frequencies characteristic of their production mechanisms. Non-axisymmetric deformation in mass-quadrupole (and other higher-order) moments can generate persistent/long-duration gravitational waves~\cite{2023LRR_KeithRiles, 2023NatAs_HaskellBejger}. These CGW signals are quasi-monochromatic with plausible slow variations in their spin frequencies~\cite{1998PhRvD_JKS, 2009ASSL_Prix}. Rapidly spinning neutron stars have also been speculated to emit CGW signals corresponding to the quasi-normal oscillation modes of the star (for example, r-mode, g-mode, f-mode) depending on the nature of the fluid perturbations in their interior~\cite{1998ApJ_Andersson, 1998PhRvD_Owen_etal}. Frequencies of these oscillation modes carry information on the internal composition and structure of the neutron stars~\cite{2009PhRvL_ChukHorowitz}, as discussed in section~\ref{fundamental-physics}. 

All the realistic CGW signals are intrinsically much weaker in amplitude as compared to the presently detected ones in Advanced-LIGO/Advanced-Virgo detectors originating from compact binary coalescence events~\cite{1998PhRvD_JKS, 2023LRR_KeithRiles}. Detecting CGW signals, therefore, requires an integration of a large amount of observational data, often covering the entire science/observational runs for the ground-based aLIGO/aVIRGO detectors, which can be more than a year long~\cite{2023LRR_KeithRiles}. Thus, searching for a CGW signal becomes computationally expensive~\cite{1998PhRvD_PBrady_etal} in comparison to the transient CBC/burst signals present only for a small duration (fraction of a second to a few minutes).

\subsubsection{Available telescopes/resources}

The LMXB systems are a promising class of candidates for observing CGW signals in the present and future generation GW detectors~\cite{1998ApJ_Bildsten}. It has been proposed in the literature that mass accretion rate can play an important role in producing non-axisymmetric deformations of the neutron stars in the LMXB systems~\cite{1998ApJ_Bildsten, 2008AIPC_Deepto}. 

Several highly promising candidates for the detection of CGW signals in LMXB systems are visible with X-ray telescopes that are sensitive to ~1-20 keV photon energy band~\cite{2018PhRvD_MMR}. At present, AstroSat is one of the relevant facilities~\cite{2014SPIE_ASTROSATmission} that can perform such observations in X-ray bands with fast timing capabilities for a set of accreting neutron stars in LMXB systems. {LAXPC} onboard AstroSat will be a useful instrument to detect the rapid periodicity of millisecond time scales, along with a scanning sky monitor for the long-term monitoring of mass accretion rates from those sources. In the future, XSPECT and POLIX instruments onboard XPoSat (X-ray Polarimeter Satellite)~\cite{XPoSat} with their fast timing capabilities, along with the Daksha ~\cite{dakshascience2024} X-ray satellite, can also be utilized for this purpose. 

Glitching radio pulsars are another class of neutron star populations that are also high-value targets for detecting transient {CGW} signals~\cite{2011PhRvD_Prix_etal, 2016PhRvD_Keitel, 2019PhRvD_Keitel_etal}. An independent knowledge of frequency shifts, glitch timing, and other ephemeral parameters from observations with radio telescopes will be beneficial to performing more optimized searches producing higher detection sensitivities~\cite{2023PhRvD_BinaryWeave}. \ac{ORT} ~\cite{OotyRadioTelescope} and  uGMRT~\cite{GMRT} are the two relevant facilities in India that can be utilized for this purpose. Specifically, ORT and uGMRT have been observing many pulsars for a long duration over the past decade and are particularly well suited for long-term monitoring of these glitching pulsars~\cite{ORT_uGMRT_monitoring2024}. In the future, SKA can play an important role in this regard.

\begin{figure*}[t]
  \hspace{-0.55cm}
  \includegraphics[width=0.55\columnwidth]{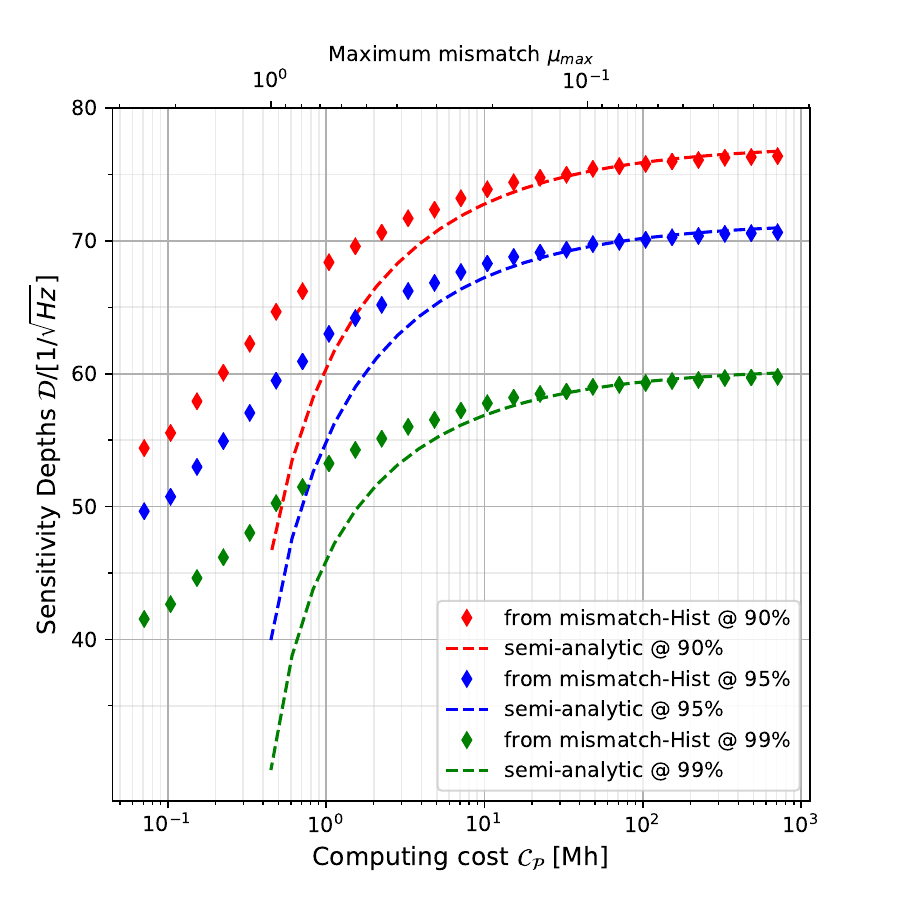}
  \hspace{-0.55cm}
  \includegraphics[width=0.55\columnwidth]{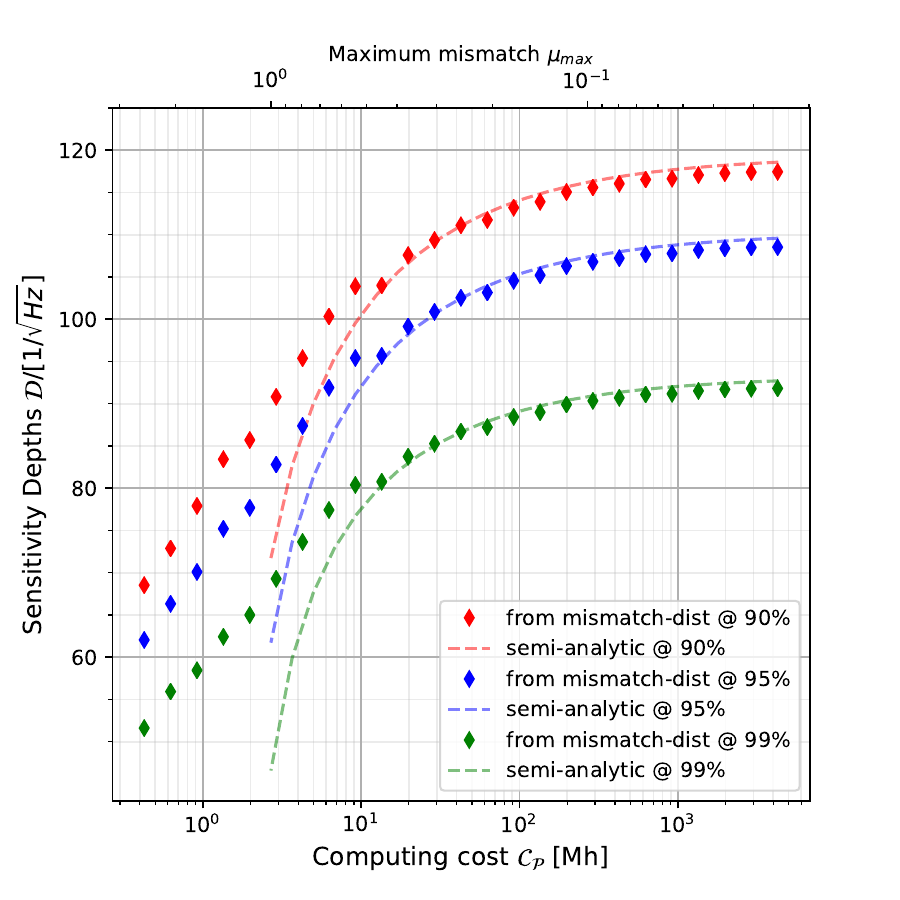}\\
  \caption{Sensitivity depth $\mathcal{D}$ as a function of (4D) computing cost $\mathcal{C}_{\mathcal{P}}$ (measured in million core hours [Mh] of a single-core \textsc{CPU}) for varying maximum mismatch $\mu_{max}$ at fixed number and length of segments, assuming Sco X-1 search parameter space of \textsc{Table-II} from~\citep{2023PhRvD_BinaryWeave}. Sensitivity depth is estimated for a fixed (per-template) false-alarm of \texttt{pfa} $= 10^{-10}$ and different confidence levels of \texttt{pdet} = 90\% (top), \texttt{pdet} = 95\% (middle) and \texttt{pdet} = 99\% (bottom). For illustration, (i) a total of 180 segments of coherence length 1 day each (left plot) for six months of gap-less observation, and (ii) a total of 120 coherence length 3 days each (right plot) for one year of gap-less observation, corresponding to \textsc{setup-I} and \textsc{setup-II}, respectively (see~\textsc{Table-II} in~\citep{2023PhRvD_BinaryWeave}) are shown here. The dashed lines correspond to the sensitivity estimate assuming a theoretical $A_{n}^{*}$ lattice mismatch distribution, while the diamond markers correspond to using the measured \textsc{BinaryWeave} mismatch distributions. 
  }
  \label{fig:sensdepth_computing_cost}
\end{figure*}

\subsubsection{Requirement for synergies between different observatories}

As outlined above, it is widely proposed in the literature that many promising targets for detecting CGW signals, e.g. accreting neutron stars in LMXB systems, glitching pulsars, and rapidly spinning radio pulsars with high spin-down parameters, may also emit EM radiations. Although some of these interesting targets have been observed in either X-ray or radio bands in the EM spectrum, the source parameters are poorly constrained by respective observations. This decreases the detection efficiency of the search pipelines for the CGW signal due to limitations in computational resources and statistical effects of noise artefacts~\cite{2023PhRvD_BinaryWeave}. 

Any prior knowledge of source parameters of important targets, obtained using coordinated EM and GW observations, will significantly reduce the computational cost for a sensitive search. This will increase the detection probability, and hence the sensitivity depth. Specifically, the knowledge of ephemerides of a neutron star in a binary system will drastically enhance the scope of detection for several high-value targets, e.g., Sco X-1. A detailed analysis for the detection of a CGW signal in Sco X-1-like LMXB systems has demonstrated the strong dependence of sensitivity depths on the prior knowledge of ephemerides \cite{2023PhRvD_BinaryWeave}. In addition, if the spin frequency and its time derivatives can be observationally determined then the search sensitivity of the CGW signal can be further improved significantly. Thus, multi-messenger observations of a set of high-value targets  will be significantly beneficial for the potential discovery of continuous GW signals.

\subsubsection{Information/Data sharing}

A list of interesting/high-value targets, for the CGW along with their sky positions, needs to be provided to the EM observers. In addition, the specific epoch during which the CGW signal is planned to be searched may also be communicated to suitable EM observers well in advance. 

For glitching pulsars, the list of sources needs to be monitored at regular intervals with {ORT}/uGMRT-like radio telescopes. The set of glitches observed or inferred from their timing solutions (for example, {\texttt Tempo2}-type data/information) needs to be provided back to the CGW search community. This information can then be utilized to perform optimized searches to increase the detection sensitivity for any putative CGW signal present in the data. 

In the case of accreting neutron stars in LMXB systems, the list of important targets needs to be shared with the X-ray, optical, and UV astronomical observers. Multi-wavelength observations in the EM channel can be used to measure the concurrent ephemerides and orbital parameters — at least partially — for those binary systems. For the accreting millisecond X-ray pulsars (AMXPs), precise spin frequency and its time derivatives will be additionally useful. This goal can be achieved in the country with LAXPC onboard AstroSat, and with the Daksha X-ray satellite, using pointed observation of these sources.

\subsubsection{Triggers / alerts}
The CGW are expected to be persistently emitting the GW signals in general. However, many of the neutron star LMXB sources are transient in the X-ray or other EM bands and thus need to be observed during their active/bright phases in the respective EM bands. 
On the other hand, glitching radio pulsars are persistent sources but their glitch phenomena, which are of interest for detecting transient CGW signals, can occur at random intervals without any prior warnings and can occur at any epoch. Thus, these types of sources need to be monitored at regular intervals. 

\subsubsection{Recommendations}
In the case of accreting neutron star systems in LMXBs, the sources should be observed a few times during the entire observation period, covering the  span of observation at some regular intervals, for example once in 3-6 months depending on the source properties. More specifically, one planned observation at/near the middle of the science/observation runs of the GW detectors would be the most important requirement. In addition, for some of the high-value targets, at least two additional observations --- one close to the beginning and the other close to the end of observing/science runs of the GW detectors --- will be highly desirable. 

For the case of glitching radio pulsars, a long-term monitoring campaign for the list of important/interesting candidate pulsars will be necessary. Over the past few years, {ORT} has observed a list of glitching pulsars and reported the ephemerides. A similar exercise during observation runs of GW detectors will be useful. 

\chapter{Conclusion and Future Directions}

The frontier of multi-messenger science discussed in this white paper highlights some of the key goals in the immediate to long-term future. This can bring a paradigm shift in our understanding of the Universe by bringing new insights into the laws of physics and the constituents of the Universe across different energy scales over most of the cosmic epoch. We mention in the table below the key science cases which are observable from different messengers.
\vspace{1 cm}
\begin{center}
\begin{tabular}{ |p{4cm}|p{2.5cm}|p{2.5cm}|p{2.5cm}| p{2.5cm}| }
 \hline
 \multicolumn{5}{|c|}{\textbf{Science Summary Chart Achievable from Ongoing/Proposed Experiments}} \\
 \hline
 \textbf{Topics} & \textbf{CRs} & \textbf{EM}& \textbf{GW} &\textbf{Neutrino}\\
 \hline
    Active Galactic Nucleus & UHECRs&  All bands &  nano-hertz to milli-hertz &Above TeV\\
 \hline
 Cosmic Expansion History & ----& All bands & Above milli-hertz &----\\
 \hline
  Neutron star Physics  & UHECRs & All bands & kilo-hertz &Above MeV\\
 \hline
  SMBHBs Physics &---- &  All bands &  nano-hertz to milli-hertz &----\\
 \hline
   Supernovae  & UHECRs&  All bands &  hecto-hertz to kilo-hertz &Above MeV\\
 \hline
  Tidal Disruption Events  & UHECRs& All bands & ---- & Above GeV\\
 \hline
   Testing Fundamental Physics  & ----&  All bands & All bands &Above MeV\\
 \hline
 
\end{tabular}
    
\end{center}

\vspace{1 cm}
We list down a few crucial points that are important for exploring multi-messenger science by synergizing between different observatories spanning different cosmic messengers. 

\begin{itemize}
\item Science goals
\begin{enumerate}
\item Astrophysical properties of compact objects and physical mechanisms involved in their evolution can be explored in great detail with the help of multi-messenger observation using GW, EM signals (including CRs), and neutrinos, which will span a vast range of energy scales for sources, both transient and persistent in nature.
\item The cosmic evolution of the Universe, its constituents including dark matter and dark energy, and fundamental laws of physics governing the Universe at the largest scales can be explored using GW signal in synergy with EM observations.
\item The internal structure of neutron stars and the fundamental laws of physics in extreme conditions can be explored from the multi-messenger observations of BNS, NSBH mergers, and continuous GW signals.
\end{enumerate}
\item Coordination between telescopes and observatories
\begin{enumerate}
\item A rapid follow-up of the GW super-events and sub-threshold events in the EM signals and neutrino, and possible measurement of the variation in the flux with time, will be essential for multi-messenger studies. Such observations followed by the identification of the host galaxy will be crucial for several science cases.
\item Sharing of data such as the sky position, time of arrival of the signal, and flux (or distance to the source) between different observations will be essential for joint studies.
\item Measurement of the multi-messenger counterparts of BNS and NSBHs contributing to the GW signal in the astrophysical stochastic background can be interesting to explore the events at high redshift.
\item Multi-messenger observations should be carried out for BBH events as well as for the sources present in AGN discs. Similarly in the future multi-messenger observations of the {SMBHBs} will be beneficial.
\end{enumerate}
\item Key requirements of the multi-messenger scientific community in India
\begin{enumerate}
\item Building national telescope facilities in GW, EM bands (including CRs), and neutrino will be important for prompt and continuous observations of multi-messenger signals.  
\item A Coordinated network of multi-messenger observation between different observatories (national and international) and sharing data products, with a clearly defined chain of communications, is crucial. 
\item High performance and high throughput computing facilities and storage for data reduction, data analysis, and theoretical calculations will be essential for multi-messenger data analysis. 
\item The training of undergraduate students, graduate students, post-doctoral fellows, scientists, and engineers in the cross-disciplinary areas of multi-messenger astronomy will pay rich dividends in the long term future. 
\end{enumerate}
\end{itemize}

In conclusion, the paradigm of multi-messenger science using different cosmic probes opens up unexplored scientific territories and discovery space. The impacts of these discoveries extend to various cornerstones of physics ranging from nuclear physics to cosmology and have far-reaching consequences in shaping our understanding of the Universe. To achieve this scientific milestone, coordinated observations between different messengers and joint scientific analysis between these are pertinent. Such a scientific venture will require a dedicated facility 
with people from different research areas to collaborate and coordinate in performing observations, data analysis, and theoretical calculations. Along with the coordination with current observation facilities in the country, the development of new observation facilities and joining international observation facilities will be useful. A multi-messenger science consortium of India with people from different research institutes, universities, and national laboratories will play a crucial platform in scientific studies and in building human and technical resources for the sustainability of multi-messenger scientific research in the country along with the rapidly growing global community. 

\newpage

\section*{Acknowledgement}
This white paper has been initiated following the GW-EM-Nu-2023 conference held at the Tata Institute of Fundamental Research (TIFR), Mumbai from November 20-24, 2023. We acknowledge the support from TIFR, the National Center for Radio Astrophysics (NCRA)-TIFR, and the Department of Atomic Energy, Government of India for this meeting.  A part of the research works presented in this white paper is carried out at the \texttt{⟨data|theory⟩ Universe-Lab} supported by TIFR and the Department of Atomic Energy, Government of India. We acknowledge the support of the Department of Atomic Energy,
Government of India,
under Project Identification No. RTI 4002. AH acknowledges support from the MHRD, Government of India for a research fellowship.  S.P.T. is a CIFAR Azrieli Global Scholar in the Gravity and Extreme Universe Program. S.K.A. acknowledges the support of the Department of Atomic Energy (DAE), Government of India, under Project Identification No. RIO 4001. S.K.A. acknowledges the financial support from the Swarnajayanti Fellowship (sanction order No. DST/SJF/PSA-05/2019-20) provided by the Department of Science and Technology (DST), Government of India, and the Research Grant (sanction order No. SB/SJF/2020-21/21) provided by the Science and Engineering Research Board (SERB), Government of India, under the Swarnajayanti Fellowship project.
 SM also thank Yashwant Gupta for his valuable suggestions on the white paper. The authors are thankful to the members from national facilities, institutes, and universities for their contributions to this white paper.

\bibliographystyle{unsrtnat}
\bibliography{WPtheorydata.bib}


\end{document}